\documentclass[12pt]{article} 

\usepackage[letterpaper,left=1in, right=1in, bottom=1.5in, top=1in]{geometry} 
\usepackage[margin=1cm]{caption}
\usepackage{subfigure,amsmath,amsthm,amsfonts,amssymb,graphicx,bbm,bm, enumitem,slashed,xcolor}
\usepackage{cite}
\usepackage{setspace}
\usepackage[colorlinks]{hyperref}
\hypersetup{
    citecolor = {blue}
}

\def\RR{\mathbb R}

\def\1{\mathbbm{1}}
\def\2{\mathbbm{2}}
\def\C{\mathsf{C}}
\def\c{\mathsf{c}}
\def\G{\mathsf{G}}
\def\g{\mathsf{g}}

\def\dc{d_{\mathsf{c}}}
\def\dC{d_{\mathsf{C}}}
\def\cupc{\cup_{\mathsf{c}}}
\def\calC{\mathcal{C}}
\def\calO{\mathcal{O}}

\newcommand{\ZZ}{\mathbb{Z}}

\usepackage[margin=1.5cm,font=small,labelfont=bf]{caption}

\title{Gauging C on the Lattice}
\author{Theodore Jacobson$^1$\footnote{tjacobson@physics.ucla.edu} \\ \\
{\it\small $^1$Mani L. Bhaumik Institute for Theoretical Physics, Department of Physics and Astronomy,}\\
{\it\small University of California, Los Angeles, CA 90095, USA} 
}
\date{\today}							

\begin{document}
\maketitle
\thispagestyle{empty}

\begin{abstract}

We discuss general aspects of charge conjugation symmetry in Euclidean lattice field theories, including its dynamical gauging. Our main focus is $O(2) = U(1)\rtimes \ZZ_2 $ gauge theory, which we construct using a non-abelian generalization of the Villain formulation via gauging the charge conjugation symmetry of pure $U(1)$ gauge theory. We describe how to construct gauge-invariant non-local operators in a theory with gauged charge conjugation symmetry, and use it to define Wilson and 't Hooft lines as well as non-invertible symmetry operators. Our lattice discretization preserves the higher-group and non-invertible symmetries of $O(2)$ gauge theory, which we explore in detail. In particular, these symmetries give rise to selection rules for extended operators and their junctions, and constrain the properties of the worldvolume degrees of freedom on twist vortices (also known as Alice or Cheshire strings). We propose a phase diagram of the theory coupled to dynamical magnetic monopoles and twist vortices, where the various generalized symmetries are typically only emergent. 

\end{abstract} 

\newpage
\begingroup
	
	\hypersetup{linkcolor=.,linktoc=all}
	\tableofcontents

\endgroup

\section{Introduction}

Lattice gauge theory has historically served as an important testing ground for novel concepts in quantum field theory, ranging from confinement and other strong coupling phenomena to kinematic features such as symmetries and dualities. In part inspired by developments in condensed matter physics concerning topological phases of matter, more recently the lattice has proven to be a fruitful setting to explore generalized global symmetries and their anomalies~\cite{Gukov:2013zka,Kapustin:2013qsa,Kapustin:2013uxa,Kapustin:2014gua,Gaiotto:2014kfa}. Notably, various `modern' concepts such as higher-form symmetry appear quite naturally on the lattice, and some of their features were appreciated in early studies of lattice gauge theories, for instance Refs.~\cite{Wegner:1984qt,Wilson:1974sk,Fradkin:1978dv,Mack:1979gb,Polchinski:1981nq} (see also \cite{Greensite:2016pfc} and references therein). 

\smallskip
Given some target continuum quantum field theory of interest, it is an important but challenging task to find a lattice discretization which preserves as many continuum features as possible. Global symmetries, together with their 't Hooft anomalies, constitute some of the most basic features of a quantum field theory, and provide non-perturbative constraints on the dynamics and possible long distance behavior of lattice and continuum theories alike. As the landscape of generalized symmetries continues to expand (see Refs.~\cite{Sharpe:2015mja,Cordova:2022ruw,McGreevy:2022oyu,brennan2023introduction,Schafer-Nameki:2023jdn,luo2023lecture,Bhardwaj:2023kri,Shao:2023gho} for reviews), it is also important to establish concrete realizations of novel symmetry structures. The lattice is a useful arena for providing such realizations and exploring their implications. Indeed, there is a large collection of works exploring generalized symmetries in Euclidean spacetime lattice models~\cite{Kapustin:2013uxa,Aasen:2016dop,Aasen:2020jwb,Koide:2021zxj,Nguyen:2021yld,Nguyen:2021naa,Cherman:2021nox,Choi:2021kmx,Hayashi:2022fkw,Cherman:2022eml,Apte:2022xtu,Kan:2023yhz,Abe:2023ncy,Inamura:2023qzl,Cordova:2023bja,Berenstein:2023ric,Berkowitz:2023pnz,Honda:2024yte} as well as Hamiltonian lattice systems~\cite{Inamura:2021szw,Barkeshli:2022wuz,Barkeshli:2023bta,Barkeshli:2022edm,Moradi:2023dan,Seiberg:2023cdc,Inamura:2023qzl,Seifnashri:2023dpa,Seiberg:2024gek,Seifnashri:2024dsd,Delcamp:2023kew,Fechisin:2023dkj,Sinha:2023hum,Lu:2024ytl,Bhardwaj:2024wlr,Bhardwaj:2024kvy,Chatterjee:2024ych,Choi:2024rjm,Hsin:2024aqb,Cao:2024qjj}, including higher-group~\cite{Kapustin:2013uxa,Kan:2023yhz,Barkeshli:2022wuz,Barkeshli:2023bta,Barkeshli:2022edm,Abe:2023ncy,Moradi:2023dan} and non-invertible symmetries~\cite{Aasen:2016dop,Aasen:2020jwb,Inamura:2021szw,Koide:2021zxj,Nguyen:2021yld,Nguyen:2021naa,Cherman:2021nox,Choi:2021kmx,Hayashi:2022fkw,Apte:2022xtu,Honda:2024yte,Cordova:2023bja,Berenstein:2023ric,Inamura:2023qzl,Delcamp:2023kew,Fechisin:2023dkj,Sinha:2023hum,Lu:2024ytl,Bhardwaj:2024wlr,Bhardwaj:2024kvy,Hsin:2024aqb,Seiberg:2023cdc,Seifnashri:2023dpa,Seiberg:2024gek,Seifnashri:2024dsd,Chatterjee:2024ych,Choi:2024rjm,Hsin:2024aqb,Cao:2024qjj}. 

\smallskip
In this paper, we use the lattice to study a familiar global symmetry, namely charge conjugation, through the lens of generalized symmetry. What makes charge conjugation special is that it is a 0-form symmetry (i.e. it is generated by a codimension-1 operator) that nonetheless has an intrinsic action on extended operators such as lines and surfaces in addition to local operators.\footnote{See e.g. Ref.~\cite{Roumpedakis:2022aik} for a treatment of charge conjugation and other 0-form symmetries in topological theories \emph{without} local operators. } It is this property that sets charge conjugation apart from more conventional 0-form internal symmetries which (intrinsically) only act on local operators.\footnote{A generic 0-form symmetry can also act projectively on extended operators --- this is known as symmetry fractionalization.} The lattice is particularly well-suited for studying gauge theories of charge conjugation: first, discrete gauge theories are very natural on the lattice, and second, the action on extended operators is inherited in a straightforward way from the transformation properties of fields living on higher-dimensional cells of the lattice, i.e. links, plaquettes, cubes, etc.

\smallskip
Promoting the charge conjugation symmetry of a gauge theory to a local gauge redundancy results in a disconnected, non-abelian gauge group. Gauge theories with disconnected gauge groups were first studied in Refs.~\cite{Kiskis:1978ed,Schwarz:1982ec} and further explored in the '90s, see for instance Refs.~\cite{Alford:1989ch,Alford:1990mk,Alford:1990ur,Preskill:1990bm,Bucher:1991qhl,Bucher:1991bc,Alford:1992yx}, and more recently in Refs.~\cite{Barkeshli:2009fu,Teo:2015xla,Argyres:2016yzz,Cordova:2017vab,Bourget:2018ond,Nguyen:2021yld,Henning:2021ctv,Heidenreich:2021xpr,Bhardwaj:2022scy,Arias-Tamargo:2023duo}. Much of the earlier literature on disconnected gauge groups was focused on the peculiar properties of non-abelian vortices dubbed `Alice' or `Cheshire' strings. The simplest example is furnished by $O(2)= U(1) \rtimes \ZZ_2$ gauge theory, which can be obtained by gauging the charge conjugation symmetry of $U(1)$ gauge theory. Alternatively, in the continuum $O(2)$ gauge theory can be realized by Higgsing $SO(3)$ using a Higgs field in the spin-2 representation~\cite{Schwarz:1982ec}.\footnote{If the UV gauge group is taken instead to be $SU(2)$, the resulting gauge group in the IR is a double cover of $O(2)$ called $Pin^-(2)$~\cite{Preskill:1990bm}. We discuss the differences between these two theories on the lattice in Appendix~\ref{app:pin2}. } The Alice strings of $O(2)$ gauge theory in the continuum arise as semiclassical vortex solutions characterized by the fact that a charged particle encircling the vortex experiences a reflection, flipping the sign of its charge.

\smallskip
In a conventional Wilson-type formulation of $O(2)$ lattice gauge theory, one integrates over $O(2)$-valued link fields in the path integral using an appropriate Haar measure, and Wilson loops are given by traces (in some representation) of path-ordered products of $O(2)$ link fields. Despite its simplicity, this formulation obscures certain global properties, in particular the role of topological excitations. Instead, in this paper we introduce a novel, non-abelian generalization of the Villain formulation~\cite{Villain:1974ir}: we start with a non-compact $\RR$ gauge theory, gauge a $\ZZ$ 1-form symmetry to obtain $U(1) = \RR/2\pi\ZZ$ gauge theory, and then gauge charge conjugation symmetry to land on the non-abelian $O(2) = U(1)\rtimes \ZZ_2$ theory. On the one hand, in this formulation it takes more work to construct gauge-invariant operators such as Wilson lines --- our method for defining gauge-invariant operators refines a construction in Ref.~\cite{Alford:1992yx} with projection operators, or `condensation defects.' However, the advantage of this formulation is that we can easily identify and control magnetic monopoles and twist vortices. The former are familiar from $U(1)$ lattice gauge theories, while the latter are specific to disconnected gauge groups like $O(2)$ --- they are nothing but the Alice strings mentioned above, i.e. the codimension-2 defects which induce a holonomy for reflections. 

\smallskip
In order to endow our lattice theory with the global symmetries of the continuum $O(2)$ theory, we employ the modified Villain formulation~\cite{Sulejmanpasic:2019ytl,Gorantla:2021svj} and completely suppress dynamical monopoles and twist vortices (see Refs.~\cite{Chen:2019mjw,Sulejmanpasic:2019ytl,Gorantla:2021svj,Gorantla:2021bda,Gorantla:2022eem,Gorantla:2022ssr,Anosova:2021akr,Anosova:2022yqx,Anosova:2022cjm,Jacobson:2023cmr,Berkowitz:2023pnz,Cherman:2024exo,Fazza:2022fss,Yoneda:2022qpj,Cheng:2022sgb,Thorngren:2023ple,Jacobson:2024hov,Nguyen:2024ikq,Chen:2024ddr} for the development and subsequent applications of this formalism). In particular, this enables us to preserve the higher-group~\cite{Hsin:2019fhf} and non-invertible~\cite{Nguyen:2021yld,Heidenreich:2021xpr} symmetries of $O(2)$ gauge theory at finite lattice spacing. As mentioned above, one of the main motivations of this work is to study the implications of these generalized symmetries using a concrete lattice realization. 

\smallskip
The higher-group symmetry involves the $\ZZ_{2,e}^{(1)}$ and $\ZZ_{2,m}^{(d-3)}$ electric and magnetic symmetries (the invertible subgroups of $U(1)_e^{(1)}$ and $U(1)_m^{(d-3)}$ in the $U(1)$ theory which commute with the action of charge conjugation) as well as the dual, or quantum, symmetry $\ZZ_{2,v}^{(d-2)}$ arising from the suppression of dynamical twist vortices. The higher-group arises from the fact that $\ZZ_{2,e}^{(1)}$ and $\ZZ_{2,m}^{(d-3)}$ are consistent symmetries on their own, but one cannot turn on backgrounds for both symmetries simultaneously without also activating $\ZZ_{2,v}^{(d-2)}$. As we discuss in Sections~\ref{sec:higher_group} and \ref{sec:twists}, the higher-group has important implications for the physics of twist vortices. In particular, the twist vortex must support charged degrees of freedom in order to match anomaly inflow from the bulk, and the higher-group constrains how these degrees of freedom couple to the bulk electric and magnetic symmetries. Our analysis of twist vortices clarifies the connection between the notion of `Cheshire charges' in lattice models of condensed matter systems~\cite{Else:2017yqj,gaiotto2019condensations,Barkeshli:2022wuz,Tantivasadakarn:2023zov} to the treatment of Alice strings in continuum high-energy physics~\cite{Alford:1990ur,Alford:1990mk}. 

\smallskip
The non-invertible symmetries we study here are among the simplest --- they arise from gauging a discrete 0-form symmetry which acts non-trivially on invertible symmetries. This class of symmetries has been studied in a number of works, for instance Refs.~\cite{Teo:2015xla,Nguyen:2021yld,Heidenreich:2021xpr,Thorngren:2021yso,Sharpe:2021srf,Bhardwaj:2022yxj,Arias-Tamargo:2022nlf,Bhardwaj:2022scy,Antinucci:2022eat,Schafer-Nameki:2023jdn,Damia:2023gtc,Gutperle:2024vyp,Hsin:2024aqb,Hofman:2024oze}, and we follow Ref.~\cite{Hsin:2024aqb} and refer to them as `coset' non-invertible symmetries. In the context of $O(2)$ gauge theory, we have electric and magnetic coset non-invertible symmetries which we denote by $(O(2)/\ZZ_2)_e^{(1)}$ and $(O(2)/\ZZ_2)_m^{(d-2)}$ --- these are the non-invertible counterparts of  $U(1)_e^{(1)}$ and $U(1)_m^{(d-3)}$ in the $U(1)$ theory. In our lattice construction, the non-invertibility arises from the need to dress the symmetry operators with condensation defects which effectively `ungauge' charge conjugation on the defect. The non-invertible symmetries give rise to (sometimes subtle) selection rules, and their realization can be used to label phases of the lattice theory. For instance, near the continuum limit the non-invertible electric and magnetic symmetries are spontaneously broken, and this phase is robust to the introduction of dynamical magnetic monopoles and twist vortices --- we present a possible phase diagram in Section~\ref{sec:phases}.  

%

\smallskip
The remainder of the paper is organized as follows. We begin in Section~\ref{sec:generalities} with lattice preliminaries, and explain how to couple a generic Euclidean lattice theory to gauge fields for charge conjugation. In Section~\ref{sec:general_construction} we describe how to use projection operators, or condensation defects, to construct gauge-invariant extended operators in a theory where charge conjugation symmetry is gauged. The construction is also used to define twisted-sector \emph{extended} operators. We apply these methods to $O(2)$ gauge theory in Section~\ref{sec:O2}, and observe that the fusion rules of our unconventionally-defined Wilson lines match the representation theory of $O(2)$. In Section~\ref{sec:higher_group} we discuss the higher-group symmetry of $O(2)$ gauge theory at the lattice level, and analyze its 't Hooft anomaly. We also use the higher-group to derive a charge relation for degrees of freedom on twist vortices, which we study in more detail in Section~\ref{sec:twists}. There we give explicit examples of worldvolume degrees of freedom which cancel anomaly inflow from the bulk, their consistency with the higher-group, and the possible junctions with Wilson and 't Hooft lines. In Section~\ref{sec:noninvertible} we construct the topological operators for the non-invertible electric and magnetic symmetries, analyze their fusion and action on charged operators, and discuss their selection rules. In Section~\ref{sec:phases} we give a possible phase diagram of $O(2)$ gauge theory on the lattice with dynamical magnetic monopoles and twist vortices. We conclude in Section~\ref{sec:conclusion} and give some directions for future work. 

\smallskip
Various technical details and ancillary results are collected in the Appendices. In Appendix~\ref{app:twisted} we give explicit formulae for twisted (or $\C$-covariant) exterior derivatives and cup products on the (hyper)cubic lattice, together with some useful identities. In Appendix~\ref{app:wilson_line} we prove the gauge-invariance of our unconventional definition of the $O(2)$ Wilson line. In Appendices~\ref{app:poc} and \ref{app:orbifold} we present Villain lattice discretizations of the orbifolded particle on a circle and compact boson theories, which feature as the worldvolume degrees of freedom of twist vortices in three and four dimensions, respectively. Finally, in Appendix~\ref{app:pin2} we compare the global symmetries of pure gauge theories with gauge group $O(2)$ and its double cover $Pin^-(2)$.

\section{Generalities: Charge Conjugation on the Lattice}
\label{sec:generalities} 

Throughout the paper we will be considering gauge theories defined on Euclidean spacetime lattices $\Lambda$ with periodic boundary conditions, i.e. with the topology of a torus. The lattice consists of sites (or $0$-cells, denoted by $s$), links ($1$-cells $\ell$), plaquettes ($2$-cells $p$), cubes ($3$-cells $c$), and hypercubes ($4$-cells $h$), etc. depending on the dimension. It will sometimes be useful to label an $i$-cell by a `root' site $s$ together with an ordered list of $i$ integers valued in $\{1,\ldots,d\}$ indicating the directions spanned by the $i$-cell. For instance we may denote a plaquette in the 13 plane with root site $s$ as $p = (s,13)$. A field living on an $i$-cell is referred to as an $i$-cochain, with $\calC^i(\Lambda,F)$ denoting the set of $F$-valued $i$-cochains (we will consider $F = \RR, \ZZ, \ZZ_2$ in this paper). We will also make use of the dual lattice $\tilde\Lambda$, and a notion of Hodge duality $\star$ exchanging $i$-cochains on $\Lambda$ with $(d-i)$-cochains on $\tilde\Lambda$. Finally, it will be useful to use cup products and their higher generalizations on the (hyper)cubic lattice~\cite{Chen:2021ppt}, which are discrete analogs of the continuum wedge product: the standard cup product joins together an $i$- and $j$-cochain to make a $(i+j)$-cochain. We refer the reader to Appendix~\ref{app:twisted} and Refs.~\cite{Sulejmanpasic:2019ytl,Jacobson:2023cmr} for more details regarding operations on lattice cochains.

\smallskip
Working at the lattice level allows us to study charge conjugation symmetry systematically by coupling the theory to discrete background gauge fields. For ordinary $0$-form symmetries, the coupling to background fields is achieved by replacing the finite difference operator $d$ acting on $0$-cochains with an appropriate covariant derivative. Since charge conjugation acts on generic $i$-cochains with $i \ge 0$, one has to first decide how an $i$-cochain transforms under \emph{local} background gauge transformations, and subsequently define an appropriate covariant derivative for generic $i$-cochains. This is handled using the technology of twisted differentials. 

\smallskip
To begin, let us describe the (for now background) charge conjugation gauge field, which is a $\ZZ_2$-valued $1$-cochain $\C \in \calC^1(\Lambda, \ZZ)$ equal to $\pm 1$ on each link. Alternatively, we can write $\C = e^{i \pi C}$ with $C \in \calC^1(\Lambda,\ZZ_2)$ equal to 0 or 1 on each link.\footnote{Throughout the paper we use sans-serif fonts to indicate cochains taking values in multiplicative presentation of the group, and ordinary fonts for the additive presentation.} Under a background gauge transformation, 
\begin{equation}
\C_{s,i} \to \G_s \, \C_{s,i} \,  \G_{s+\hat i}\,, \quad C \to C + dG \text{ mod } 2\,,
\end{equation} 
where $\G = e^{i \pi G} \in \calC^0(\Lambda, \ZZ_2)$ is the $\ZZ_2$-valued gauge parameter. In terms of cup products, we can write the transformation rule as $\C \to \G \cup \C \cup \G$. If we want the background field to be flat, then we restrict $\prod_{\ell\in\partial p} \C_\ell = 1$, or equivalently $dC = 0$ mod $2$. Then we have $\C \in Z^1(\Lambda, \ZZ_2)$. Modding out by gauge transformations, which are exact $1$-cochains (or coboundaries)  $dG \in B^1(\Lambda,\ZZ_2)$, $\C$ can be viewed as the representative of a class in $H^1(\Lambda, \ZZ_2) = Z^1(\Lambda,\ZZ_2)/B^1(\Lambda,\ZZ_2)$.  

\smallskip
Dynamical fields $X \in \calC^i(\Lambda,F)$ transform under charge conjugation as $X \to (-1)^{Q_\C} X$ with $Q_\C = 0,1$. We promote this to a local transformation via the rule
\begin{equation} \label{eq:local_action} 
X_{s,\, j_1 j_2 \ldots j_i} \to \G_s^{Q_\C}\, X_{s,\, j_1 j_2 \ldots j_i}\,, \text{ or } X \to \G^{Q_\C} \cup X\,. 
\end{equation}
Intuitively, a $\C$-odd $i$-cochain transforms under local $\C$-gauge transformations at the `root site' from which it emanates. Let $X$ be $\C$ odd, with $Q_\C = 1$. We define the (left) charge conjugation-covariant derivative, or twisted differential, to be
\begin{equation}\label{eq:twisted_diff} 
\dC X  \equiv dX + (\C-\1) \cup X = dX - 2 C \cup X\,,
\end{equation}
recalling that $C$ takes values 0 or 1. Here we have introduced the $i$-cochain $\1$ which is equal to $+1$ on each positively oriented $i$-cell (the degree of this cochain should be obvious from context). We refer the reader to Appendix~\ref{app:twisted} for an explicit formula for the above twisted differential, with relevant examples shown graphically in Fig.~\ref{fig:twisted_d}. Eq.~\eqref{eq:twisted_diff} indeed defines a covariant derivative --- under a gauge transformation (see Appendix~\ref{app:twisted} for the derivation)
\begin{equation}
\dC X \to d_{\G\cup \C \cup \G}(\G \cup X) = \G \cup \dC X\,. 
\end{equation}
A crucial point is that while the ordinary differential $d$ is nilpotent, i.e. $d^2 =0$, the twisted differential is in general not. Instead (again see Appendix~\ref{app:twisted}), 
\begin{equation}
\dC^2 X = \C \cup \C \cup X\,.
\end{equation}
Unpacking the cup product on the right hand side, 
\begin{equation}
(\C \cup \C)_{s,ij} = \C_{s,i}\C_{s+\hat i,j} (1-e^{i \pi (dC)_{s,ij}}) = -2 \C_{s,i}\C_{s+\hat i,j} \,\overline{(dC)}_{s,ij}\,,
\end{equation}
where $\overline{dC} \equiv dC \text{ mod } 2$.\footnote{Throughout the following, barred quantities refer to their reduction modulo $2$.} We learn that the twisted differential is nilpotent, $\dC^2 = 0$, \emph{only} if the $\C$ gauge field is flat. This is familiar from the continuum, where the exterior covariant derivative associated to a non-flat connection fails to be nilpotent. 

\begin{figure}[h!] 
   \centering
   \includegraphics[width=.7\textwidth]{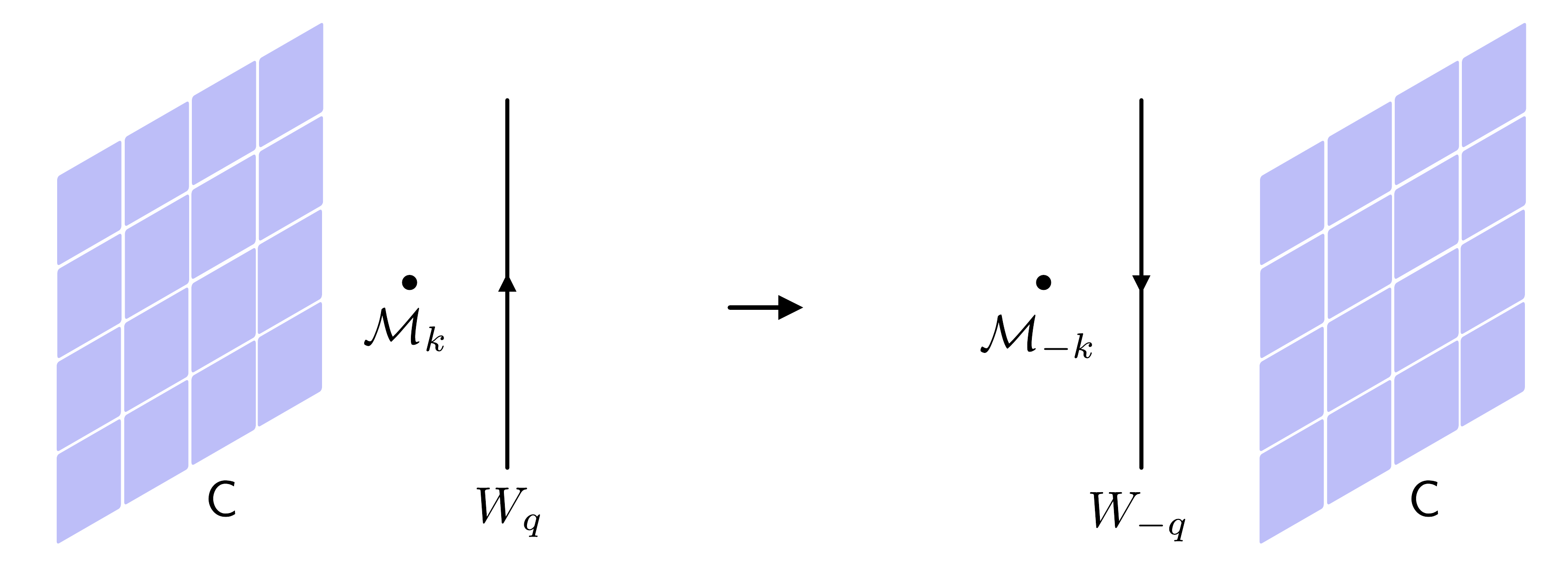} 
   \caption{The charge conjugation action is implemented by sweeping its associated codimension-1 symmetry defect past charged operators, such as the local monopole operator $\mathcal M_k$ or Wilson line $W_q$ in 3d $U(1)$ gauge theory. }
   \label{fig:caction}
\end{figure}

\smallskip
When $\C$ is a background field, we are free to consider only flat gauge field configurations. It is often useful to view a background $\ZZ_2$ gauge field configuration in terms of its Poincar\'e dual --- for each $\C$ we can write 
\begin{equation}
\C = e^{i \pi \star[\tilde \Omega_{d-1}]}\,, \quad C = \star[\tilde \Omega_{d-1}] \text{ mod } 2\,,
\end{equation}  
where $\tilde \Omega_{d-1}$ is a (not necessarily connected) codimension-1 surface on the dual lattice $\tilde\Lambda$, and $\star [\tilde \Omega_{d-1}] \in \calC^1(\Lambda, \ZZ_2)$ is a 1-cochain whose value on a link is the oriented number of times the link pierces $\tilde \Omega_{d-1}$. The flatness condition translates to the condition that $\partial\tilde \Omega_{d-1} = 0$ mod $2$, i.e. $\tilde \Omega_{d-1}$ cannot have a boundary unless a multiple of two codimension-1 surfaces end on it. Since $\tilde \Omega_{d-1}$ is only physical mod 2, its orientation is unimportant, so we can think of every \emph{flat} background $\C$ gauge field in terms of a collection of \emph{closed}, unoriented codimension-1 surfaces. Intuitively, moving such a surface past a charged operator implements the $\C$ action, as shown in Fig.~\ref{fig:caction}. 

\smallskip
To make the charge conjugation gauge field dynamical, we simply replace capital letters by lower case ones $\C \to \c$ and $C \to c$, which are summed over in the path integral with weight 
\begin{equation}
\sum_{\c \in \calC^1(\Lambda, \ZZ_2)} = \ \prod_{\ell\in\Lambda} \sum_{\c_\ell = \pm 1} = \  \prod_{\ell\in\Lambda} \sum_{c_\ell = 0,1}\,.
\end{equation}
This gives rise to a new Wilson loop observable, 
\begin{equation} \label{eq:det_line} 
\eta(\gamma) = \prod_{\ell \in \gamma} \c_\ell = \exp\left( i \pi \sum_{\ell\in\gamma} c_\ell \right)\,, 
\end{equation}
which is invariant under dynamical charge conjugation gauge transformations $\c \to \g \cup \c \cup \g$ (or $c \to c + dg$) provided $\gamma$ is a closed loop. In the context of $O(2)$ gauge theory, $\eta(\gamma)$ is simply the Wilson loop associated to the one-dimensional `determinant' representation which maps all group elements containing a reflection to $-1$.

\smallskip
Note that in the path integral, we sum over all $\ZZ_2$ 1-cochains $\c$ including field configurations which are not flat. We can eliminate such configurations from the path integral by introducing a Lagrange multiplier $v \in \calC^{d-2}(\Lambda,\ZZ_2)$, with coupling
\begin{equation} \label{eq:lagrange_multiplier} 
 i \pi \sum_{d\text{-cells}} dc \cup v\,.
\end{equation}
We will refer to the field $v$ as the `twist field.' It is a $(d-2)$-form $\ZZ_2$ gauge field with its own gauge redundancy $v \to v + dt$ for $t \in \calC^{d-3}(\Lambda, \ZZ_2)$. Provided the only term in the action involving $v$ is the one in Eq.~\eqref{eq:lagrange_multiplier}, the twist field sets $dc = 0$ mod $2$, and the Wilson line $\eta(\gamma)$ becomes a topological line. It also squares to 1, and so defines a $(d-2)$-form $\ZZ_2$ symmetry $\ZZ_{2,v}^{(d-2)}$ which shifts the twist field by a $\ZZ_2$ $(d-2)$-cocycle $\omega \in Z^{(d-2)}(\Lambda,\ZZ_2)$, 
\begin{equation}
\ZZ_{2,v}^{(d-2)}: v \to v + \omega\,,  \ d\omega = 0\,. 
\end{equation}
From the twist field we can construct codimension-$2$ operators
\begin{equation} \label{eq:twist_vortex} 
T(\Gamma_{d-2}) = \exp\left( i \pi \sum_{(d-2)\text{-cells}\in\Gamma} v \right)\,,
\end{equation}
whose effect is to force the charge conjugation gauge field to be non-flat precisely at the location of the insertion, 
\begin{equation}
dc = \star[\Gamma^\vee_{d-2}] \text{ mod } 2\,.
\end{equation}
In other words, we must sum over codimension-1 surfaces $\tilde\Omega_{d-1}$ subject to the constraint $\partial\tilde\Omega_{d-1} = \Gamma^\vee_{d-2}$. Here $\Gamma^\vee_{d-2}$ is a codimension-2 surface on the dual lattice obtained from $\Gamma_{d-2}$ by applying a shift in the $-\frac{\hat 1 + \hat 2 + \cdots + \hat d}{\sqrt{d}}$ direction, so that
\begin{equation}
[\Gamma^\vee_{d-2}]_{\tilde s,\, j_1,j_2,\ldots,j_{d-2}} = [\Gamma_{d-2}]_{\tilde s - \frac{1}{2}(\hat 1 + \hat 2 + \cdots + \hat d),\, j_1,j_2,\ldots,j_{d-2}}\,,
\end{equation}
and we define a similar shift in the $+\frac{\hat 1 + \hat 2 + \cdots + \hat d}{\sqrt{d}}$ direction so that $(\Gamma^\vee)^\wedge = (\Gamma^\wedge)^\vee = \Gamma$. We follow \cite{Heidenreich:2021xpr} and refer to the operator \eqref{eq:twist_vortex} as the `twist vortex.' It is the charged object under the dual, or `quantum,' symmetry generated by $\eta(\gamma)$. 

\smallskip
Up to this point, our discussion has been very general and can be applied to any lattice quantum field theory with a charge conjugation symmetry. Since we will be mostly focused on lattice gauge theories, we need to consider one further point, namely how dynamical gauge redundancies are modified in the presence of charge conjugation backgrounds. Simply put, a $\C$-odd $i$-form gauge field $Y \in \calC^i(\Lambda, F)$ which shifts by the exterior derivative $Y \to Y + dV$ of a $(i-1)$-form $V \in \calC^{i-1}(\Lambda, F)$ must now shift by the $\C$-covariant derivative, $Y \to Y + \dC V$. This ensures that the resulting gauge parameters compose according to the semi-direct product $F \rtimes \ZZ_2$, so that for instance doing a $\C$-gauge transformation after an $F$ gauge transformation flips the sign of the original $F$ gauge transformation, $(1,\G) \circ (V,1) = (\G V, \G)$. 

\begin{figure}[h!] 
   \centering
   \includegraphics[width=1\textwidth]{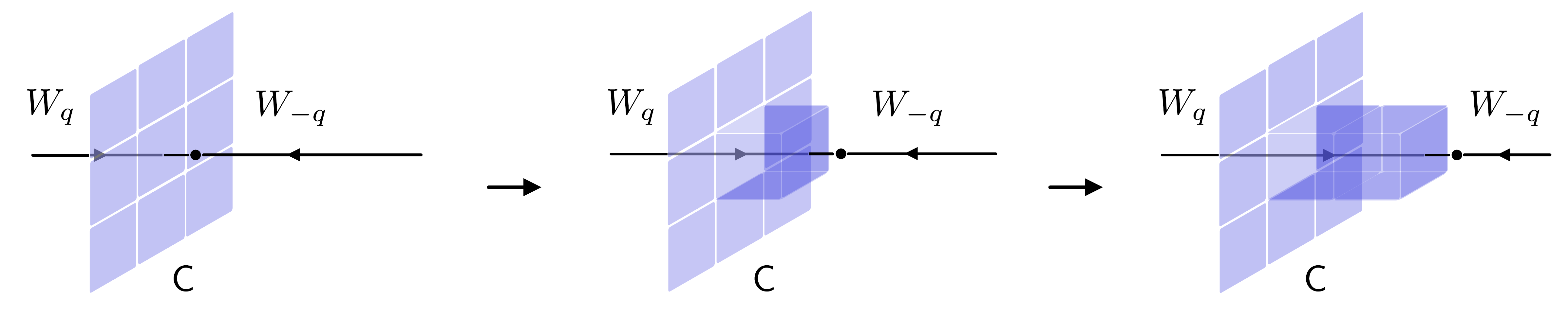} 
   \caption{The codimension-1 charge conjugation defect hosts a junction (represented by the black dot) between Wilson lines with charge $q$ and $-q$, which is gauge invariant thanks to the twisted gauge transformations \eqref{eq:twisted_gauge}. Background gauge transformations topologically deform the surface, `converting' $W_q$ into $W_{-q}$ in the process.}
   \label{fig:cjunction}
\end{figure}

\smallskip
As a concrete example, and the one which is most relevant for the remainder of the paper, consider a $U(1)$ gauge field in the Villain formulation. The ingredients are a real link field $a \in \calC^1(\Lambda, \RR)$ and an integer plaquette field $n \in \calC^2(\Lambda, \ZZ)$. Both fields are $\C$-odd, transforming as $a \to \G \cup a$ and $n \to \G \cup n$. In the presence of charge conjugation background fields, the small (parameterized by $\lambda \in \calC^0(\Lambda,\RR)$) and large (parameterized by $m \in \calC^1(\Lambda,\ZZ)$) gauge transformation rules become
\begin{equation}\label{eq:twisted_gauge}
a \to a + \dC\lambda + 2\pi m\,, \  n \to n + \dC m\,. 
\end{equation}
This twisted gauge transformation rule also has a pleasing interpretation in terms of the junctions between non-local operators which meet at a codimension-1 surface implementing the charge conjugation action, as depicted in Fig.~\ref{fig:cjunction}. In words, given a link variable which pierces the charge conjugation defect, the twist flips the sign of the gauge parameter on one side of the defect, allowing for a gauge-invariant junction between Wilson lines of charge $q$ and $-q$.

\smallskip
The final technical ingredient for our construction is a $\C$-twisted version of the cup product. Such twisted cup products were for instance discussed in Appendix A of Ref.~\cite{Benini:2018reh} on triangulations, and we give explicit formulae for the twisted cup products on (hyper)cubic lattices in Appendix~\ref{app:twisted} (in particular, see Fig.~\ref{fig:twisted_cup} for a figure with some relevant examples in 3d). Intuitively, the twisted cup product $X\cup_\C Y$ between $p$- and $q$-cochains $X$ and $Y$ with $\C$ charges $Q_\C^X$ and $Q_\C^Y$ is a $(p+q)$-cochain with $\C$ charge $Q_\C^X + Q_\C^Y$. The crucial feature of the twisted cup product is that it obeys a Leibniz rule with respect to the $\C$-covariant exterior derivative. In particular, when both $X$ and $Y$ are $\C$ odd, we have 
\begin{equation}
d (X \cup_\C Y) = \dC X \cup_\C Y + (-1)^p\, X \cup_\C \dC Y\,,
\end{equation} 
provided $\C$ is flat.\footnote{The twisted cup products are also non-associative in the presence of non-flat $\C$ gauge fields.} This allows us to `sum by parts' on closed manifolds.

\subsection{Constructing Gauge-Invariant Operators}
\label{sec:general_construction} 

The operators of a quantum field theory with gauged charge conjugation symmetry fall into distinct classes. One class consists of new operators which arise from gauging but which were not present before, such as the Wilson line $\eta$ and twist vortex $T$ we discussed above. The second class consists of operators which were present in the ungauged theory, and can be organized under their parity under charge conjugation --- roughly speaking, neutral operators survive in the gauge theory while odd operators are projected out. For local operators it is straightforward to form charge conjugation eigenstates by taking symmetric and antisymmetric linear combinations of operators exchanged by $\C$. The former give rise to gauge-invariant operators in the gauged theory, while the latter become twisted-sector operators, i.e. non-genuine local operators living at the endpoints of the topological line $\eta$. For instance, the local monopole operator $e^{i\sigma_s}$ in the 3d $U(1)$ gauge theory survives in the $O(2)$ theory as $\mathcal M(s) = e^{i \sigma_s} + e^{-i\sigma_s}$. On the other hand, for extended operators this procedure is more subtle, since the local action of charge conjugation in general only transforms a \emph{portion} of a non-local operator. As a result, the naive symmetric linear combination of e.g. a Wilson loop, $W_q(\gamma) + W_{-q}(\gamma)$, is not invariant under local charge conjugation gauge transformations. The purpose of this section is to make the intuitive notion of a linear combination of extended operators precise, at the nonperturbative lattice level.  

\smallskip
To illustrate the basic construction, we start with the quintessential extended operator, namely the Wilson line of $U(1)$ gauge theory. As we described above, the simple linear combination of lines $W_q + W_{-q}$ is not locally charge conjugation-invariant. As a naive first attempt to fix this issue, let us sum over the gauge orbits of each segment of the Wilson line \emph{before} taking the product over all links, and write  
\begin{equation} \label{eq:naive_W} 
W_q(\gamma) \stackrel{?}{=} \prod_{\ell\in \gamma} \sum_{ \hat\g_\ell = \pm 1}e^{i q\, \hat\g_\ell \,a_\ell} = \prod_{\ell\in \gamma} \cos(q\, a_\ell) \,. 
\end{equation}
While this is clearly $\c$ gauge invariant, it is not invariant under $U(1) \subset O(2)$ gauge transformations! Expanding the $2^{|\gamma|}$ terms in a given fixed $\c$ background, only two (or zero) terms from the above sum will be gauge invariant under $a \to a + \dc \lambda$. This is not detrimental, since the remaining gauge-non-invariant terms will be projected out in any correlation function with gauge-invariant operators. However, depending on the context (for instance a Monte Carlo simulation) it may be inconvenient or impractical to use such a non-gauge-invariant representation of the Wilson loop. What's more, we can do better.

\smallskip 
Before proceeding, let us note an amusing fact: \emph{none} of the terms in the expansion of Eq.~\eqref{eq:naive_W} are $U(1)$ gauge-invariant in a $\c$ background for which $\eta(\gamma) = -1$. Therefore the only field configurations which contribute to correlation functions of Wilson loops have $\eta(\gamma) = 1$, so we can write $W_q(\gamma) \,\eta(\gamma) = W_q(\gamma)$ as an operator identity. This says that the charge-$q$ Wilson line can `absorb' the $\c$ Wilson line, and suggests that a gauge-invariant definition of $W_q(\gamma)$ will involve a projection operator 
\begin{equation} \label{eq:projector} 
P(\gamma) = \frac{1+\eta(\gamma)}{2}\,.
\end{equation}

\smallskip
We now describe how to make $W_q(\gamma)$ fully gauge-invariant. Inspired by a related construction in Ref.~\cite{Alford:1992yx} (see also Ref.~\cite{Bais:2008xi} and more recently Ref.~\cite{Choi:2024rjm}), we begin by choosing a basepoint $*$ on $\gamma$, and for each link $\ell\in\gamma$, a path $\gamma_{\ell, *}$ lying in $\gamma$ which connects the basepoint to the site from which $\ell$ emanates. We refer to this set of paths $\{\gamma_{\ell,*}\}$ as the \emph{scaffolding} of the curve $\gamma$. Then we define\footnote{In the original construction of Ref.~\cite{Alford:1992yx} upon which the definition here is inspired, the projection factor $P(\gamma)$ is absent and replaced with the requirement that the union of all scaffolding paths contains no closed loops. Furthermore, the paths were not restricted to lie on $\gamma$, which introduces ambiguities in the choice of paths. } 
\begin{equation} \label{eq:gauge_invariant_wilson} 
W_q(\gamma) \equiv P(\gamma) \sum_{\hat\g_* = \pm 1} \exp\left(i q\, \hat\g_* \sum_{\ell \in \gamma} \eta(\gamma_{\ell,*})\, a_\ell \right)\,. 
\end{equation}
The operator is defined such that under $\c$ gauge transformations, each term in the exponent transforms covariantly \emph{at the basepoint}. In other words, for each $\ell\in\gamma$, we have that $\eta(\gamma_{\ell,*}) a_\ell \to \g_{*}\eta(\gamma_{\ell,*}) a_\ell$ under a general $\c$ gauge transformation. The operator is rendered $\c$-gauge invariant by summing over gauge transformations at the basepoint. Equivalently, we can view $\hat\g_*$ as a degree of freedom localized at the basepoint, and assign $\hat\g_* \to \g_* \hat\g_*$ under $\c$ gauge transformations. The above definition is independent of the choice of basepoint, since we can effectively shift its location by redefining $\hat\g_* \to \hat\g_*\, \eta(\gamma_{*,*'})$, where $\gamma_{*,*'}$ is a path in $\gamma$ connecting the old and new basepoints.

\begin{figure}[h!] 
   \centering
   \includegraphics[width=.95\textwidth]{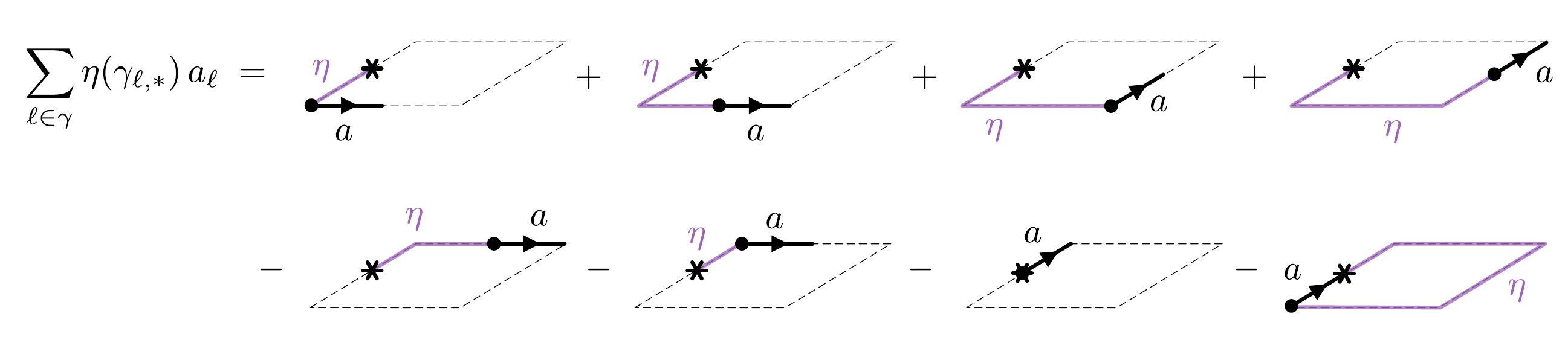} 
   \caption{An example of a curve $\gamma$ and the scaffolding paths $\gamma_{\ell,*}$ connecting the basepoint $*$ to the root site (indicated by the black circle) of each link. Under a $\c$ gauge transformation, each term transforms by multiplication by $\g_*$. }
   \label{fig:line_scaffolding}
\end{figure}

\smallskip
The operator appears to be quite non-local, depending on the scaffolding paths $\gamma_{\ell,*}$. On the other hand, we expect the dependence to be quite mild since $\eta$ is a topological operator. In fact, Eq.~\eqref{eq:gauge_invariant_wilson} is completely independent of the choice of paths $\gamma_{\ell,*}$. To see this, we note that the projection factor $P(\gamma)$ ensures that the $\c$-holonomy evaluated on the difference of any two paths from a given link to the basepoint will be trivial. When $\gamma$ is contractible this would be true without the projection factor (assuming the absence of twist vortices). However, when $\gamma$ is non-contractible, two paths from a given link to the basepoint may differ by $\gamma$ itself, leading to a sign ambiguity if $\eta(\gamma) = -1$. The factor $P(\gamma)$ enforces $\eta(\gamma) = +1$, thereby eliminating this ambiguity. We conclude that the Wilson loop does not depend on the choice of scaffolding used to define it in Eq.~\eqref{eq:gauge_invariant_wilson}.

\smallskip
Finally, it is not immediately obvious that we have achieved our original goal of defining a Wilson loop which is invariant under $U(1) \subset O(2)$ gauge transformations. The argument is slightly involved and is presented in Appendix~\ref{app:wilson_line}. 

\smallskip
If we restrict ourselves to configurations where $\c = 1$ near $\gamma$, the complicated expression in Eq.~\eqref{eq:gauge_invariant_wilson} reduces to the naive linear combination $W_q(\gamma) = W_q^{U(1)}(\gamma) + W_{-q}^{U(1)}(\gamma)$, while more generally, it reduces to the earlier definition in Eq.~\eqref{eq:naive_W} in any fixed field configuration of $\c$. Hence Eq.~\eqref{eq:gauge_invariant_wilson} provides a fully gauge-invariant alternative to the naive definition in Eq.~\eqref{eq:naive_W}.

\smallskip
The same construction can be generalized to an arbitrary operator in the pre-gauged theory defined on a (closed) submanifold $\Sigma_i$ which is built from the exponential of a $\c$-odd $i$-cochain $X$. First one chooses a scaffolding: a basepoint $*$ and a set of paths $\gamma_{i,*}$ lying in $\Sigma_i$ from the root site of each $i$-cell in $\Sigma_i$ to the basepoint. Then we can define the gauge invariant operator
\begin{equation} \label{eq:general_O} 
\calO_\alpha(\Sigma_i) = P(\Sigma_i) \sum_{\hat\g_* = \pm 1} \exp\left( i \alpha\, \hat\g_* \sum_{\text{$i$-cells}\in\Sigma_i} \eta(\gamma_{i,*})\, X_i \right)\,,
\end{equation}
where the projector is proportional to the sum over lines $\eta(\gamma)$ on the homology cycles of $\Sigma_i$,
\begin{equation} \label{eq:projector_2} 
P(\Sigma_i) = \frac{1}{|H_1(\Sigma_i,\ZZ_2)|}\sum_{\gamma \in H_1(\Sigma_i, \ZZ_2)} \eta(\gamma) \,.
\end{equation}
This ensures that the operator \eqref{eq:general_O} is independent of the choice of scaffolding. Up to the normalization factor, the above projector is an example of a condensation defect~\cite{Else:2017yqj,gaiotto2019condensations,Roumpedakis:2022aik,Choi:2022zal,Tantivasadakarn:2023zov} --- an operator defined by summing over (in other words condensing) higher codimension operators on its worldvolume. More specifically, this condensation defect is equivalent to higher-gauging the $\ZZ_{2,v}^{(d-2)}$ symmetry on $\Sigma_i$. Such condensation defects have been studied in detail from this perspective in Ref.~\cite{Roumpedakis:2022aik}. We can also write the condensation defect as a $\ZZ_2$ gauge theory on $\Sigma_i$ coupled to the charge conjugation gauge field,\footnote{The normalization factor is (see e.g. Appendix~A of Ref.~\cite{Choi:2021kmx})
\begin{equation}
\mathcal X = \frac{|H^{i-2}(\Sigma_i,\ZZ_2)||H^{i-4}(\Sigma_i,\ZZ_2)|\cdots}{|H^{i-1}(\Sigma_i,\ZZ_2)||H^{i-3}(\Sigma_i,\ZZ_2)|\cdots}\frac{2^{\# (i-3)\text{-cells}+\# (i-5)\text{-cells}+\cdots}}{2^{\# i\text{-cells}+\# (i-2)\text{-cells}+\cdots}}\,.
\end{equation}
}  
\begin{equation} \label{eq:condensation} 
P(\Sigma_i) = \mathcal X \sum_{\substack{u \in \calC^0(\Sigma_i, \ZZ_2),\\ b \in \calC^{i-1}(\Sigma_i, \ZZ_2)}} \exp\left( i \pi \sum_{\text{$i$-cells}\in \Sigma_i} c\cup b + u \cup db \right)\,.
\end{equation}
The 0-form field $u$ shifts by $u \to u + g$ under charge conjugation, and (for $i \ge 2$) we impose the gauge redundancy $b\to b+ dz$ with $z \in \calC^{i-2}(\Sigma_i,\ZZ_2)$. Notably, this is only respected provided $c$ is flat, which means that twist vortex insertions are not allowed to pierce $\Sigma_i$.\footnote{More specifically, we must assign $v \to v + z \cup \star[\Sigma^\wedge]$ in the presence of a condensation defect. This means that the intersection locus between $\Gamma$ and $\Sigma$ is not gauge-invariant under the worldvolume gauge redundancies. This can be cancelled by attaching a $b$ Wilson operator on the condensation defect, which generates the higher quantum symmetry~\cite{Roumpedakis:2022aik} acting as shifts of $u$. } This representation is equivalent to \eqref{eq:projector} since summing over $u$ sets $db = 0$ mod $2$, so that $b$ is Poincar\'e dual to closed curves $\gamma \in Z_1(\Sigma_i,\ZZ_2)$. The coupling $i \pi \sum c\cup b$ is equivalent to $i\pi \sum_\gamma c$, and summing over $b$ (equivalently $\gamma$) reproduces Eq.~\eqref{eq:projector}. This way of writing the projector also provides the following useful perspective: the projector \emph{ungauges} the charge conjugation symmetry at the location of the defect. To see this, we can instead sum over $b$ first, which sets $c = du$ mod $2$ along $\Sigma_i$, i.e. sets $c$ to be pure gauge. This is to be expected, since the condensation defect gauges the dual symmetry obtained from gauging $\c$ to begin with, and gauging the dual symmetry returns back the ungauged theory~\cite{Tachikawa:2017gyf}. 

\begin{figure}[h!] 
   \centering
   \includegraphics[width=.95\textwidth]{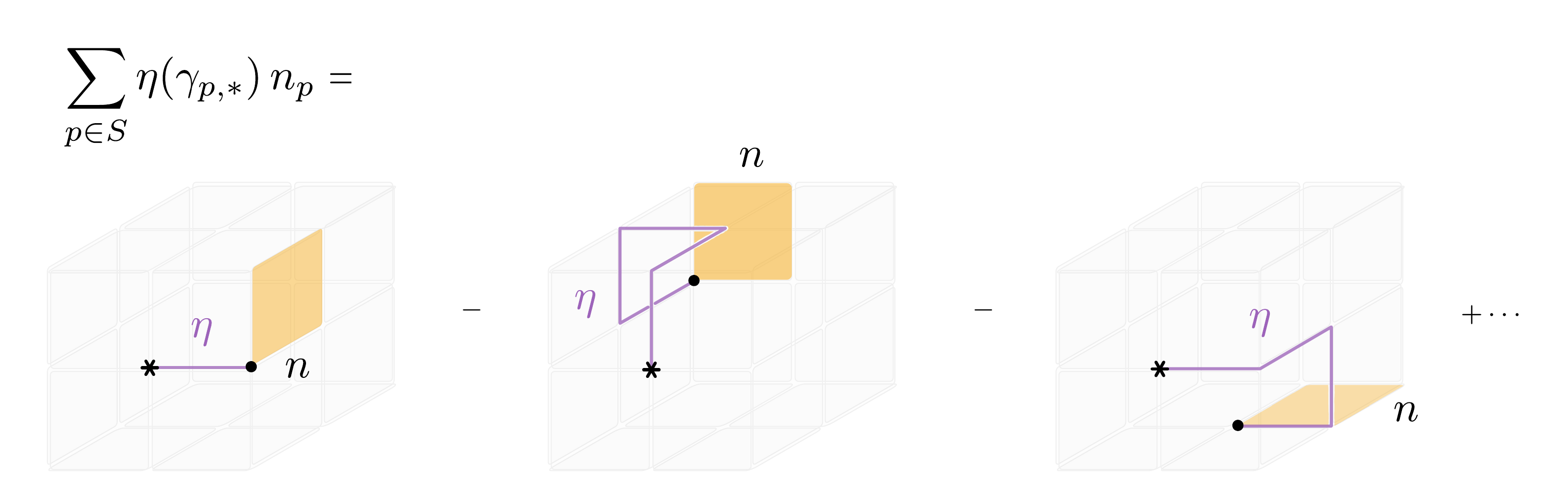} 
   \caption{An example of a closed surface $S$ on the lattice and some of the scaffolding paths $\gamma_{p,*}$ needed to define the surface operator $V_\alpha(S)$ in Eq.~\eqref{eq:Valpha}. }
   \label{fig:surface_scaffolding}
\end{figure}

\smallskip
As an example, let us apply the more general construction \eqref{eq:general_O} to the magnetic flux surfaces in $O(2)$ gauge theory. Specifically, we start with the magnetic flux operator in the $U(1)$ theory, 
\begin{equation}
V_\alpha^{U(1)}(S) = \exp\left(i \alpha \sum_{p \in S}  n_p \right)\,.
\end{equation}
We can construct the gauge-invariant image of this operator in the $O(2)$ theory by picking a scaffolding of $S$ (see Fig.~\ref{fig:surface_scaffolding} for an example) and writing 
\begin{equation} \label{eq:Valpha} 
V_\alpha(S) = P(S) \sum_{\hat\g_* = \pm 1} \exp\left(i \alpha\, \hat\g_* \sum_{p \in S} \eta(\gamma_{p,*})\, n_p \right)\,,
\end{equation}
where in this case the projector is 
\begin{equation}
P(S) = \frac{|H^0(S,\ZZ_2)|}{|H^1(S,\ZZ_2)|}\frac{1}{2^{\#\text{plaq.}+\# \text{sites}}} \sum_{\substack{u \in \calC^0(S, \ZZ_2),\\ b \in \calC^1(S, \ZZ_2)}} \exp\left( i \pi \sum_{p\in S} c\cup b + u \cup db \right)\,.
\end{equation}
One can use a generalization of the argument in Appendix~\ref{app:wilson_line} to show that the above operator is invariant under large $U(1)$ gauge transformations which shift $n \to n + d_\c m$. We will discuss the properties of $V_\alpha(S)$ in more detail later in Section~\ref{sec:noninvertible_magnetic} where we show that it generates a non-invertible magnetic symmetry. Finally, we note that the operator in $U(1)$ gauge theory that measures magnetic flux mod $2$ is charge conjugation-invariant, so if we set $\alpha = \pi$ we have the relation
\begin{equation}
V_\pi (S) = 2P(S) V_\pi^{U(1)} (S)\,. 
\end{equation}
In particular, both $V_\pi^{U(1)}$ and $V_\pi$ are valid operators in the $O(2)$ theory. 

\smallskip
The above construction can also be extended to manifolds $\Sigma_i$ with boundary --- one simply imposes Dirichlet boundary conditions for the gauge field $b$ and includes an additional projector for the boundary components $\partial \Sigma_i$. For instance, consider an open line $L_{x,y}$ consisting of $|L|$ sites, and with no self-intersections. To build an open Wilson line, we can start with a charge-$q$ matter field $\varphi$, transforming as $\varphi \to \varphi + q\lambda$, and write
\begin{equation}
W_q(L_{x,y}) = P(L_{x,y})P(\partial L_{x,y}) \sum_{\hat\g_* = \pm 1} \exp\left( i\hat\g_* \sum_{\ell\in L_{x,y}} \eta(\gamma_{\ell,*})(q a_\ell - (d_\c\varphi)_\ell) \right)\,. 
\end{equation}
The boundary projector is trivial and 
\begin{equation}
P(L_{x,y}) = \frac{1}{2^{|L|}} \sum_{\substack{u \in \calC^0(L_{x,y}, \ZZ_2),\\ b \in \calC^0(L_{x,y}, \ZZ_2), b_x = b_y = 0}} \exp\left( i \pi \sum_{\ell \in L_{x,y} } c\cup b + u \cup db \right) = 1\,,
\end{equation}
since the only $\ZZ_2$ $0$-cocycle on $L$ obeying the Dirichlet boundary conditions is $b = 0$. Intuitively, since a line segment has no non-trivial cycles, there are no ambiguities in the choices of scaffolding paths, so there is no need for a projector. Finally, if we note that
\begin{equation}
\sum_{\ell \in L_{x,y}} \eta(\gamma_{\ell,*})(d_\c \varphi)_\ell = \eta(\gamma_{y,*})\,  \varphi_y  -  \eta(\gamma_{x,*})\,  \varphi_x \,,
\end{equation}
we can write a more compact expression for the open Wilson line
\begin{equation} \label{eq:open_line} 
W_q(L_{x,y}) = \sum_{\hat\g_*} \exp\left(-i \hat\g_* \, \eta(\gamma_{x,*})\, \varphi_x\right)\exp\left(i q \, \hat\g_* \sum_{\ell \in L_{x,y}} \eta(\gamma_{\ell,*})\, a_\ell \right)\exp\left(i \hat\g_* \, \eta(\gamma_{y,*})\, \varphi_y  \right) \,.
\end{equation}

\smallskip
Finally, while we have been focusing on the case when the charge conjugation gauge field is flat, this is not strictly necessary. Relaxing the cocycle condition $dc = 0$ mod $2$ allows twist vortices as dynamical excitations, and breaks the dual $\ZZ_{2,v}^{(d-2)}$ symmetry. As a result, we can no longer interpret the projector \eqref{eq:condensation} as the condensation defect obtained by higher-gauging a symmetry (it is also no longer topological). Nonetheless, the projector still forces the charge conjugation gauge field to be pure gauge, making all $\c$-holonomies trivial on $\Sigma_i$. This is all that is needed for the operator \eqref{eq:general_O} to be gauge-invariant and independent of its scaffolding. 

\subsubsection{Twisted Sector Operators}
\label{sec:twisted_operators} 

One can also use the recipe described above to construct twisted sector operators, i.e. the images of $\c$-odd operators in the pre-gauged theory. Let us start with the more familiar case of a local operator, for instance the monopole operator in the twisted sector of 3d $O(2)$ gauge theory. Such an operator lives at the endpoint of the dual symmetry line $\eta$, 
\begin{equation}\label{eq:twisted_monopole} 
\widetilde{\mathcal{M}}_k(s)\, \eta(\gamma_{s,\infty})  = (e^{ik \sigma_s} - e^{-ik\sigma_s})\, \eta(\gamma_{s,\infty}) = \sum_{\hat\g_s = \pm 1} \hat \g_s\, \,\eta(\gamma_{s,\infty}) \, e^{i k \hat\g_s \sigma_s}\,,
\end{equation}
where $\gamma_{s,\infty}$ is a half-infinite line ending at the site $s$. The last presentation above generalizes to non-local operators in the twisted sector --- we simply insert the basepoint degree of freedom $\hat\g_*$ and multiply by a half-infinite line $\eta(\gamma_{*,\infty})$ ending at the basepoint. In other words, starting with a $\c$-odd $i$-cochain $X$, we have
\begin{equation} \label{eq:twisted_O} 
\widetilde\calO_\alpha(\Sigma_i;*)\, \eta(\gamma_{*,\infty}) = P(\Sigma_i) \sum_{\hat\g_* = \pm 1} \hat\g_*\,  \eta(\gamma_{*,\infty})\,  \exp\left( i \alpha\, \hat\g_* \sum_{\text{$i$-cells}\in\Sigma_i} \eta(\gamma_{i,*})\, X_i \right)\,.
\end{equation}
Note that while the two factors on the left-hand side individually depend on the choice of basepoint $*$, their product does not. Such non-local twisted-sector operators appear due to the action of non-invertible symmetries, as we discuss in Sec.~\ref{sec:noninvertible}.

\section{Villain Formulation of $O(2)$ Gauge Theory}
\label{sec:O2} 

We now turn to a specific application of the general ideas from the previous sections. Our starting point is the modified Villain action for $U(1)$ gauge theory in $d$ spacetime dimensions, which we take to be
\begin{equation}
S_{U(1)} = \frac{\beta}{2}\sum_{\text{plaq.}} (da - 2\pi n)^2 + i \sum_{d\text{-cells}} dn \cup \tilde a\,.
\end{equation}
This differs slightly from the action written down in Ref.~\cite{Sulejmanpasic:2019ytl} in that we place the Lagrange multiplier field $\tilde a \in \calC^{d-3}(\Lambda,\RR)$ on the lattice instead of the dual lattice. While this presentation obscures the lattice rotation symmetry, it allows a more convenient coupling to $\C$ gauge fields --- we will return to the fate of rotation symmetry at the end of this section. The gauge redundancies are 
\begin{equation}
\begin{split}
a &\to a + d\lambda + 2\pi m\,, \ n \to n + dm \,, \ \tilde a \to \tilde a + d\tilde\lambda + 2\pi \tilde m\,,\\
\lambda &\in \calC^0(\Lambda,\RR)\,, \ \tilde\lambda \in \calC^{d-4}(\Lambda,\RR)\,, \ m \in \calC^1(\Lambda,\ZZ)\,, \ \tilde m \in \calC^{d-3}(\Lambda,\ZZ)\,. 
\end{split}
\end{equation}
In four dimensions, $\tilde a$ is itself a 1-form $U(1)$ gauge field, namely the magnetic gauge field, and in three dimensions $\tilde a \equiv \sigma$ is a compact scalar, namely the dual photon. In two dimensions this field is absent, and instead we have the option to write a $2\pi$-periodic theta term $i \theta \sum n$ (we set the 4d lattice theta term~\cite{Sulejmanpasic:2019ytl,Anosova:2022cjm,Jacobson:2023cmr} to zero for simplicity).  

\smallskip 
To obtain the modified Villain action for $O(2)$ gauge theory, we simply covariantize the derivative and cup product and include the coupling from Eq.~\eqref{eq:lagrange_multiplier},\footnote{This amounts to gauging $\C$ with a particular choice of fractionalization class. We could in principle change this by shifting the background field for the electric 1-form symmetry (discussed below) by $B_e \to B_e + \frac{1}{2}dc$, and/or a similar shift for the magnetic symmetry in 4d. Furthermore, for simplicity we omit the 3d Dijkgraaf-Witten twist $\frac{i\pi}{2}c\cup dc$. We thank Z. Sun for discussions on these points.}
\begin{equation} \label{eq:O2_action} 
S_{O(2)} = \frac{\beta}{2}\sum_{\text{plaq.}} (\dc a - 2\pi n)^2 + i \sum_{d\text{-cells}} \dc n \cupc \tilde a + i \pi \sum_{d\text{-cells}} dc \cup v \,.
\end{equation}
This action is designed to be invariant under dynamical $\c$-gauge transformations parameterized by $\g = (-1)^g \in \calC^0(\Lambda,\ZZ_2)$, which act as
\begin{equation}
c \to c + dg \,, \ a \to \g \cup a \,,\ n \to \g \cup n \,,\ \tilde a \to \g \cup\tilde a \,. 
\end{equation}
For the moment we restrict ourselves to on-shell configurations where $dc = 0$ mod $2$, i.e. we do not consider insertions of twist vortices. In this case, the covariant derivative $\dc$ obeys the same Leibniz and integration-by-parts identities as $d$, and the action is invariant under
\begin{equation}
a \to a + \dc \lambda + 2\pi m\,,\ n \to n + \dc m\,,\ \tilde a \to \tilde a + \dc\tilde\lambda + 2\pi \tilde m\,,\ v \to v + dt\,.
\end{equation}
Continuing to assume that there are no twist vortex insertions, we can integrate by parts and dualize the above action by writing it as 
\begin{equation} \label{eq:auxiliary_action} 
\frac{1}{2\beta}\sum_{(d-2)\text{-cells}} z^2 + i \sum_{d\text{-cells}} (\dc a) \cupc  z - n \cupc  (\dc \tilde a + 2\pi z) + i \pi \sum_{d\text{-cells}}  dc \cup v \,,
\end{equation}
where $z \in \calC^{d-2}(\Lambda, \RR)$ is a $\c$ odd auxiliary field. Now the sum on $n$ sets $2\pi z + \dc \tilde a = 2\pi \tilde n$ for some $\c$ odd integer field $\tilde n \in \calC^{d-2}(\Lambda, \ZZ)$ which transforms as $\tilde n \to \tilde n + \dc \tilde m$. Integrating by parts again, we arrive at a dual representation,
\begin{equation} \label{eq:dual_action} 
\tilde S_{O(2)} = \frac{1}{2\beta(2\pi)^2}\sum_{(d-2)\text{-cells}} ( \dc \tilde a - 2\pi \tilde n )^2 + i \sum_{d\text{-cells}} a \cupc  \dc  \tilde n + i \pi \sum_{d\text{-cells}}  dc \cup v  \,. 
\end{equation}
In $d=4$ this is the S-dual $O(2)$ Lagrangian where the roles of the ($U(1)$ part of the) electric and magnetic gauge fields are reversed and the coupling $\beta \to \frac{1}{(2\pi)^2\beta}$, while in $d=3$ this is simply the modified Villain action for the 3d compact boson, with gauged charge conjugation symmetry. Hence the exact lattice dualities of the modified Villain $U(1)$ gauge theory commute with the operation of gauging charge conjugation.

\smallskip
Any attempt to perform numerical Monte Carlo simulations directly with the actions in either Eq.~\eqref{eq:O2_action} or Eq.~\eqref{eq:dual_action} would face a severe sign problem. However, the sign problem can in principle be avoided by simply integrating out the Lagrange multiplier fields. First, in the absence of twist vortex insertions we can sum over $v$ in favor of the constraint $dc = 0$ mod $2$. It is straightforward to propose Monte Carlo updates which satisfy this constraint, see for instance Refs.~\cite{Gattringer:2018dlw,Berkowitz:2023pnz} for examples. Second, integrating out the dual field $\tilde a$ in Eq.~\eqref{eq:O2_action} or the original gauge field $a$ in Eq.~\eqref{eq:dual_action} give the constraints
\begin{equation}
\dc n = 0 \ \Longrightarrow \ dn = 2c \cup n \,, \text{ or } \dc \tilde n = 0 \ \Longrightarrow \ d\tilde n = 2c \cup \tilde n\,. 
\end{equation}
One likely needs more sophisticated techniques, for instance a modified version of the surface worm algorithm~\cite{Mercado_2013}, to efficiently generate configurations satisfying such non-linear constraints. The point remains that at least in principle, it is possible to simulate $O(2)$ gauge theory using the non-abelian Villain formulation without a sign-problem. 

\smallskip
We also hasten to mention that the action Eq.~\eqref{eq:O2_action} does not respect the full cubic rotational symmetry of the lattice. This can be traced back to the fact that the local charge conjugation action \eqref{eq:local_action} itself breaks rotation symmetry by picking a preferred root site for each $i$-cochain. However, summing over flat charge conjugation gauge fields does not introduce any local dynamics, and should not affect local correlation functions sensitive to rotational symmetry. Therefore, at least for flat gauging of charge conjugation, we expect the full $SO(d)$ Euclidean rotational symmetry to emerge in the continuum limit. 

\subsection{Wilson Lines and Representations}
\label{sec:reps} 

We now describe how the spectrum and fusion of Wilson lines in our modified Villain theory match the representation theory of $O(2)$. In a more conventional lattice discretization, Wilson lines are defined as traces of path-ordered products of link matrices (i.e. characters in some representation), so that their fusion automatically coincides with tensor product decomposition. In our case this is less obvious, and matching the tensor product decomposition of $O(2)$ representations serves as a non-trivial check for our more unconventional Wilson loops. 

\smallskip 
The irreducible representations of $O(2)$ consist of two-dimensional irreps $\2_q$ (with $q >0$), a one-dimensional `det' representation $1_{\text{det}}$ (the adjoint), and a one-dimensional trivial representation $\1$, which satisfy 
\begin{equation}\label{eq:reps} 
\begin{split}
\1_{\det} &\otimes \1_{\det} = \1\,, \\
\2_q &\otimes \1_{\det} = \2_q \,, \\
\2_q &\otimes \2_{q'} = \2_{q+q'} \oplus \2_{|q-q'|}\,, \text{ for } q \not=q' \\
\2_q &\otimes \2_q = \2_{2q}  \oplus \1 \oplus \1_{\det} \,.
\end{split}
\end{equation}
The corresponding Wilson lines are identified as $\2_q \leftrightarrow W_q$ and $\1_{\det} \leftrightarrow \eta$. The first line above is matched by the fact that the $\eta$ line generates a $\ZZ_2$ symmetry, and the second line is matched by the projector in the definition of $W_q$ (this was already discussed near Eq.~\eqref{eq:projector} in Section~\ref{sec:general_construction}).   

\smallskip
Since we are at finite lattice spacing, we can compute the fusion of the remaining non-topological Wilson lines by simply placing them on the same contour $\gamma$ and re-expanding the result. Since the basepoints and scaffolding paths can be chosen arbitrarily, we choose them to be the same for the two lines being fused. Then, by rewriting the sum over gauge transformations at the basepoint, we find
\begin{multline} 
W_q(\gamma)W_{q'}(\gamma) = P(\gamma)P(\gamma)\sum_{\hat\g_* = \pm 1}\sum_{\hat\g_*'= \pm 1} \exp\left(i (q\hat\g_* + q'\hat\g_*') \sum_{\ell\in\gamma}\eta(\gamma_{\ell,*})a_\ell\right) \\
= P(\gamma)\sum_{\hat\g_* = \pm 1}\sum_{\hat\g_*'= \pm \hat\g_*} \exp\left(i \hat\g_*(q + q'\hat\g_*\hat\g_*') \sum_{\ell\in\gamma}\eta(\gamma_{\ell,*})a_\ell\right) \\
= W_{q+q'}(\gamma) + W_{q-q'}(\gamma)\,,
\end{multline}
where in the last line we performed the sum over $\hat\g_*'$. Note that $W_q = W_{-q}$ so this reproduces the third line of Eq.~\eqref{eq:reps}. We also reproduce the last line of Eq.~\eqref{eq:reps}, since setting $q' = q$ gives 
\begin{equation}
W_q(\gamma)W_q(\gamma) = W_{2q}(\gamma) + W_0(\gamma) = W_{2q}(\gamma) + 2P(\gamma) = W_{2q}(\gamma) + 1 + \eta(\gamma)\,.
\end{equation}
Finally, the same fusion rules apply to the 't Hooft lines in 4d, defined as Wilson lines of the magnetic gauge field $\tilde a$, 
\begin{equation}
H_{\tilde q}(\gamma) = P(\gamma) \sum_{\hat\g_* = \pm 1} \exp\left(i \tilde q\, \hat\g_*\sum_{\ell\in\gamma}\eta(\gamma_{\ell,*})\tilde a_\ell \right)\,.
\end{equation}

\section{Higher-Group Symmetry}
\label{sec:higher_group} 

In this section we study the group-like, invertible symmetries of the $O(2)$ gauge theory on the lattice. As discussed in Section~\ref{sec:generalities}, gauging the charge conjugation symmetry of the $U(1)$ theory gives rise to a dual $\ZZ_{2,v}^{(d-2)}$ symmetry which acts on the codimension-2 twist vortices. In addition, the gauging process breaks the $U(1)_e^{(1)}$ and $U(1)_m^{(d-3)}$ symmetries down to their $\ZZ_2$ subgroups which commute with the charge conjugation action. These three $\ZZ_2$ symmetries are however not independent, and combine into a non-trivial higher-group~\cite{Hsin:2019fhf}. 

\smallskip
We can uncover the higher-group structure by coupling to flat background fields for the 1-form electric and $(d-3)$-form magnetic symmetries. Specifically, we work with integer lifts of $\ZZ_2$ gauge fields, $B_e \in \calC^2(\Lambda, \ZZ)$, $B_m \in \calC^{d-2}(\Lambda,\ZZ)$, which are closed mod 2. The action minimally coupled to background fields is
\begin{multline} \label{eq:3d_minimal_backgrounds} 
S(B_e,B_m) = \frac{\beta}{2}\sum_{\text{plaq.}} (\dc a - \pi B_e - 2\pi n)^2 + i\sum_{d\text{-cells}} \left(\dc n + \frac{1}{2}\dc B_e \right) \cupc  \tilde a \\
+ i \pi \sum_{d\text{-cells}} \left(n+\frac{1}{2}B_e\right)\cup B_m + i \pi \sum_{d\text{-cells}} c \cup dv \,. 
\end{multline}
Before checking the invariance under background gauge transformations, let us make sure that the dynamical gauge redundancies are respected. Since the background fields are classical sources, they cannot transform under dynamical gauge redundancies. As a result, the presence of $B_e$ in the gauge field kinetic term spoils the usual $\c$ gauge transformations, and we must augment the transformation rule of the Villain field $n$ by 
\begin{equation}
n \to \g \cup n + \frac{1}{2}(\g - \1)\cup B_e \,,
\end{equation}
such that the combination $n + \frac{1}{2}B_e$ transforms covariantly. Even with this modification, the second line of Eq.~\eqref{eq:3d_minimal_backgrounds} is not $\c$ gauge-invariant, but shifts by 
\begin{equation} \label{eq:g_variation} 
\begin{split}
\Delta S = i\pi \sum_{d\text{-cells}} (\g -\1)\cup (n+\frac{1}{2}B_e) \cup B_m  = i \pi \sum_{d\text{-cells}} g \cup B_e \cup B_m \text{ mod } 2\pi i \,.
\end{split}
\end{equation}
This gauge variation is simply reflecting the mixed anomaly between $\ZZ_{2,e}^{(1)}\times\ZZ_{2,m}^{(d-3)}$ and $\C$ in the pure $U(1)$ gauge theory. As described in Ref.~\cite{Tachikawa:2017gyf}, gauging a symmetry which participates linearly in such a discrete anomaly gives rise to some kind of extension of the other symmetries participating in the anomaly. In this case, gauging $\C$ gives rise to a higher-group. The anomalous variation \eqref{eq:g_variation} simply states that the point-like intersection of a codimension-2 $\ZZ_{2,e}^{(1)}$ symmetry operator (surface in 4d and line in 3d) and a $\ZZ_{2,m}^{(d-3)}$ surface is $\C$-odd. When we gauge $\C$, a Wilson line $\eta$ must terminate at the intersection point to render it gauge invariant. Since $\eta$ is the generator of $\ZZ_{2,v}^{(d-2)}$, this means that turning on general backgrounds for $\ZZ_{2,e}^{(1)}\times\ZZ_{2,m}^{(d-3)}$ requires us to turn on (non-flat) backgrounds for $\ZZ_{2,v}^{(d-2)}$. 

\begin{figure}[h!] 
   \centering
   \includegraphics[width=.35\textwidth]{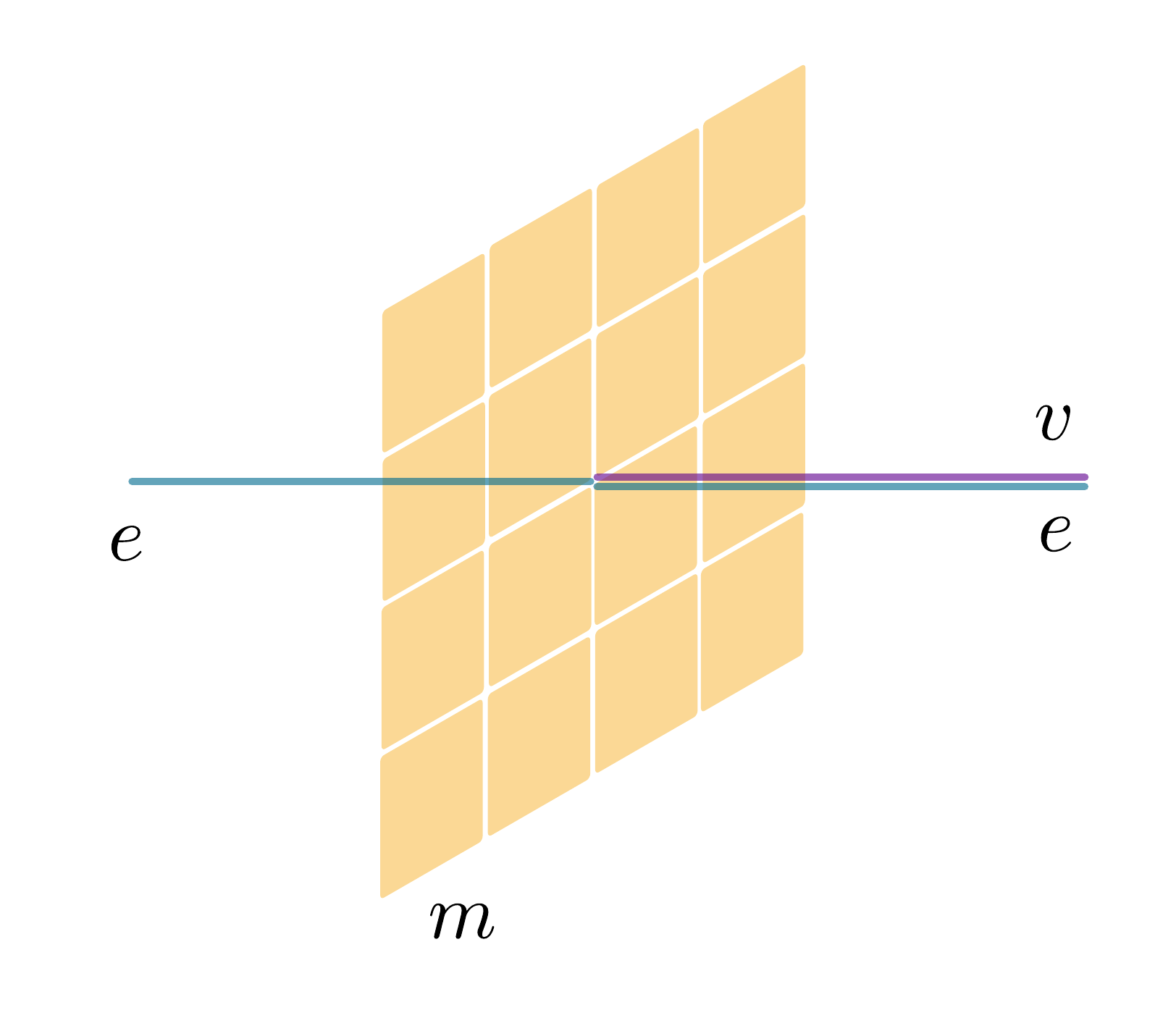} 
   \caption{The modified cocycle condition characterizing the 2-group in three dimensions implies that when a $\ZZ_{2,e}^{(1)}$ line pierces a codimension-1 $\ZZ_{2,m}^{(0)}$ surface, it emerges as the line for $\ZZ_2 \subset \ZZ_{2,e}^{(1)} \times \ZZ_{2,v}^{(1)}$. In other words, the magnetic symmetry acts non-trivially on the 1-form symmetry by exchanging the $e$ and $ev$ lines. }
   \label{fig:rho}
\end{figure}

\smallskip
Indeed, let us turn on a background field $B_v \in \calC^{d-1}(\Lambda, \ZZ)$ for the $(d-2)$-form twist vortex symmetry, 
\begin{equation} \label{eq:Z2v_coupling} 
i \pi \sum_{d\text{-cells}} c \cup dv \ \to \  i \pi \sum_{d\text{-cells}} c \cup (dv-B_v)
\end{equation}
which itself shifts by 
\begin{equation}
i \pi \sum_{d\text{-cells}} g \cup dB_v
\end{equation}
under $\c$ gauge transformations. This variation would vanish for an ordinary flat, $(d-1)$-form $\ZZ_2$ gauge field satisfying $dB_v = 0$ mod $2$, but we can cancel Eq.~\eqref{eq:g_variation} by twisting the flatness condition, 
\begin{equation} \label{eq:higher_group} 
dB_v = B_e \cup B_m \text{ mod } 2\,,
\end{equation}
leading to a higher-group structure. This equation captures the fact that a $\ZZ_{2,v}^{(d-2)}$ symmetry line must emanate from the point-like intersection of $\ZZ_{2,e}^{(1)}$ and $\ZZ_{2,m}^{(d-3)}$ generators, as we described above. In three dimensions, we can equivalently say that when a $\ZZ_{2,e}^{(1)}$-generating line pierces a codimension-1 $\ZZ_{2,m}^{(0)}$ surface, it emerges as the line that generates the diagonal $\ZZ_2 \subset \ZZ_{2,e}^{(1)} \times \ZZ_{2,v}^{(1)}$, i.e. the 0-form magnetic symmetry permutes the 1-form symmetry lines $e$ and $ev$. This is illustrated in Fig.~\ref{fig:rho}.

\begin{figure}[h!] 
   \centering
   \includegraphics[width=\textwidth]{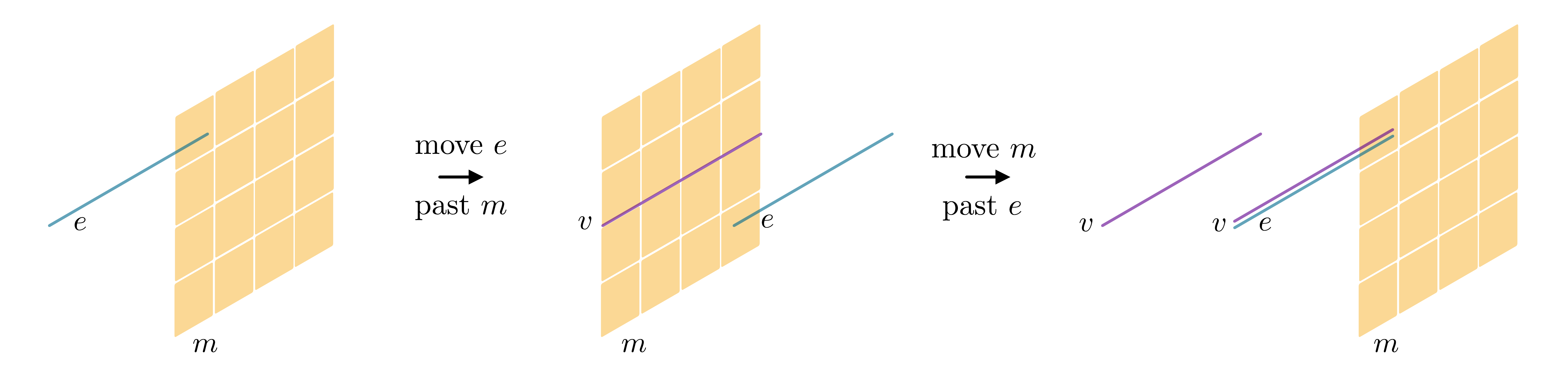} 
   \caption{The higher-group gauge transformations \eqref{eq:2group_gt} imply that a $\ZZ_{2,v}^{(d-2)}$ line gets nucleated whenever a $\ZZ_{2,e}^{(1)}$ generator and a $\ZZ_{2,m}^{(d-3)}$ surface cross. In 4d, the $\ZZ_{2,e}^{(1)}$ generator above is a surface spanning the suppressed fourth dimension while the $\ZZ_{2,m}^{(d-3)}$ and $\ZZ_{2,v}^{(d-2)}$ generators are localized. }
   \label{fig:2group}
\end{figure}

\smallskip
In order for this twisted cocycle condition to be background gauge invariant (i.e. for the aforementioned junction to be topological), the background field for the twist vortex symmetry must transform under gauge transformations for the electric and magnetic symmetries, 
\begin{equation} \label{eq:2group_gt} 
\begin{split}
B_e &\to B_e + dV_e + 2K_e\,, \\ 
B_m &\to B_m + dV_m + 2K_m\,, \\
B_v &\to B_v + dV_v + 2K_v +  V_e \cup B_m + B_e \cup V_m + dV_e \cup V_m\,,
\end{split}
\end{equation}
where $V_e \in \calC^1(\Lambda, \ZZ), V_m \in \calC^{d-3}(\Lambda, \ZZ), V_v \in \calC^{d-2}(\Lambda, \ZZ)$ parameterize `small' $\ZZ_2$ gauge transformations and $K_e \in \calC^2(\Lambda, \ZZ), K_m \in \calC^{d-2}(\Lambda, \ZZ), K_v \in \calC^{d-1}(\Lambda, \ZZ)$ parameterize changes in the choices of integer lifts. This modified gauge transformation rule for $B_v$ has a nice geometric interpretation in terms of symmetry defects --- it says that whenever we pass a $\ZZ_{2,e}^{(1)}$ generator through a $\ZZ_{2,m}^{(d-3)}$ surface (or visa versa), we generate a $\ZZ_{2,v}^{(d-2)}$ line. This is depicted in Fig.~\ref{fig:2group}. The higher-group shift \eqref{eq:2group_gt} is also consistent with `gauge transformations of gauge transformations' of the sort described in Refs.~\cite{Gukov:2013zka,Kapustin:2013uxa,Cordova:2018cvg} whereby e.g. we simultaneously take $V_e \to V_e + 2 X_e, K_e \to K_e - dX_e$. Such a shift represents a trivial gauge transformation and can be absorbed by taking $K_v \to K_v - X_e \cup B_m$. 

\subsection{Anomalies} 
\label{eq:higher_group_anomalies} 

\smallskip
Now we can check the background gauge transformation properties of the action coupled to the higher-group background fields. The dynamical fields shift by 
\begin{equation} \label{eq:2group_dyn} 
\begin{split}
\ZZ_{2,m}^{(d-3)}:& \  \tilde a \to \tilde a + \pi V_m\,, \\ 
\ZZ_{2,e}^{(1)}:& \  a \to a + \pi V_e\,, \ n \to  n + \frac{1}{2}(\c-\1)\cup V_e - K_e\,, \\ 
\ZZ_{2,v}^{(d-2)}:& \  v \to v + V_v\,. 
\end{split} 
\end{equation}
The unconventional shift of $n$ ensures that the combination $\dc a - 2\pi n$ shifts by $\pi dV_e$ and not $\pi \dc V_e$. Under a background gauge transformation the action (Eq.~\eqref{eq:3d_minimal_backgrounds} plus Eq.~\eqref{eq:Z2v_coupling}) is not invariant, but transforms by terms involving the background fields and their gauge parameters (assuming no twist vortex insertions)
\begin{equation} \label{eq:putative_anomaly} 
\Delta S  =  i\pi \sum_{d\text{-cells}} \left[ (B_e + dV_e) \cup K_m + \frac{1}{2}V_e \cup dB_m \right]  \in i \pi \ZZ\, . 
\end{equation}
While at face value this seems to indicate that the mixed anomaly between $\ZZ_{2,e}^{(1)}$ and $\ZZ_{2,m}^{(d-3)}$ in the $U(1)$ theory survives in $O(2)$, since these symmetries are involved in a higher-group there are more counter-terms available and it is possible for this putative anomaly to be trivialized. This turns out to be the case in three dimensions.\footnote{This is entirely analogous to the fact that the $D_8$ symmetry of the $\ZZ_2$ orbifold of the compact boson CFT is anomaly free~\cite{Thorngren:2021yso}.} To proceed, we note that the above phase can be cancelled by a bulk inflow term
\begin{equation} \label{eq:inflow} 
\mathcal A = \frac{i\pi}{2}\sum_{(d+1)-\text{cells} } B_e \cup dB_m\,,
\end{equation}
assuming we can extend $B_e$ and $B_m$ to some bulk $(d+1)$-dimensional lattice while maintaining their flatness. In three dimensions where $B_m$ is a 1-cochain we can make use of the following identity (see Eq.~\eqref{eq:useful_2} in Appendix~\ref{app:twisted}),
\begin{equation} \label{eq:applied_useful} 
d\overline{B}_m = 2 B_m \cup B_m \text{ mod } 4\,,
\end{equation}
which holds provided $dB_m = 0$ mod $2$, to write the inflow action as 
\begin{multline}
\mathcal A = \frac{i\pi}{2}\sum_{\text{hyper-cu.} } B_e \cup dB_m - B_e \cup d\overline{B}_m + 2 B_e \cup B_m \cup B_m \\
=\frac{i\pi}{2}\sum_{\text{hyper-cu.} } B_e \cup d(B_m- \overline{B}_m) + 2 dB_v \cup B_m \\
= \frac{i\pi}{2} \sum_{\text{hyper-cu.} } d\left[ B_e \cup (B_m- \overline{B}_m)+2B_v \cup B_m \right]\,.
\end{multline}
In the second line we used the twisted cocycle condition Eq.~\eqref{eq:higher_group}, and in the third line we used mod 2 flatness of $B_e$ and $B_m$ and the fact that $B_m - \overline{B}_m \in 2\ZZ$. We see that the anomaly inflow action is the total derivative of a local 3d counter-term whose variation cancels Eq.~\eqref{eq:putative_anomaly}. The correct coupling to 2-group background fields in three dimensions is then
\begin{multline} 
S(B_e,B_m,B_v) = \frac{\beta}{2}\sum_{\text{plaq.}} (\dc a - \pi B_e - 2\pi n)^2 + i\sum_{\text{cubes}} \left(\dc n + \frac{1}{2}\dc B_e \right) \cupc  \sigma \\
+ i \pi \sum_{\text{cubes}} \left(n+\frac{1}{2}B_e- B_v\right)\cup \overline{B}_m + i \pi \sum_{\text{cubes}} c \cup (dv-B_v) \,, 
\end{multline}
which is indeed invariant under the 2-group gauge transformations Eqs.~\eqref{eq:2group_gt}, \eqref{eq:2group_dyn}. As a result, there is no obstruction to gauging the entire 2-group global symmetry in three dimensions. In four dimensions the inflow action in Eq.~\eqref{eq:inflow} is non-trivial, so the mixed anomaly between $\ZZ_{2,e}^{(1)}$ and $\ZZ_{2,m}^{(1)}$ from the $U(1)$ theory survives as a non-trivial 3-group anomaly in the $O(2)$ theory. 

\subsection{Action on Twist Vortices}

\begin{figure}[h!] 
   \centering
   \includegraphics[width=.95\textwidth]{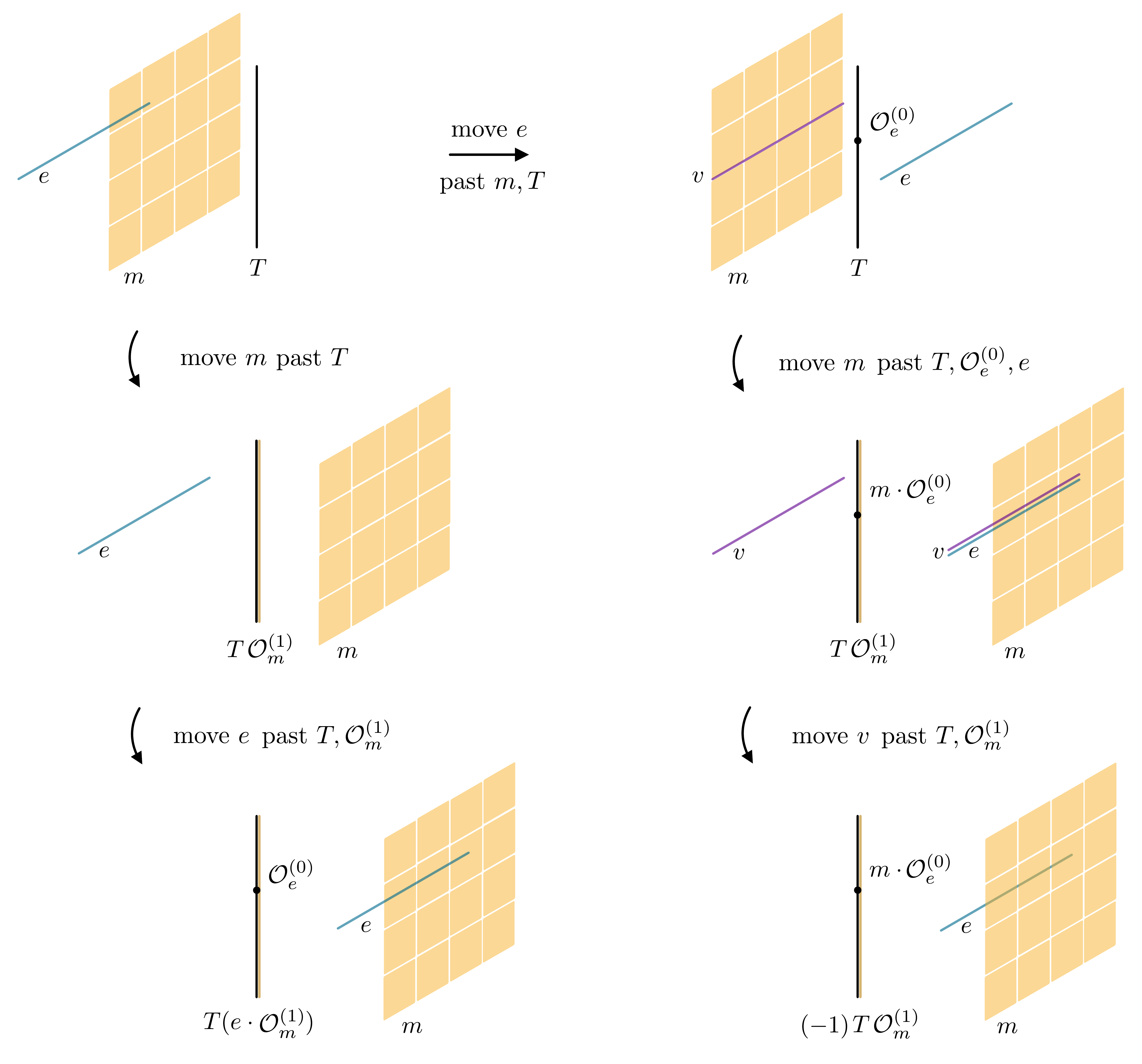} 
   \caption{In three dimensions, the electric symmetry line $e$ leaves behind a local operator $\calO_e^{(0)}$ on the twist vortex $T$ while the magnetic symmetry surface $m$ leaves behind a line operator $\calO_m^{(1)}$. Passing the symmetry operators across the twist vortex in two different ways yields a relation \eqref{eq:projectivity} between the charges of these operators.  }
   \label{fig:projective_twist_3d}
\end{figure}

\begin{figure}[h!] 
   \centering
   \includegraphics[width=.85\textwidth]{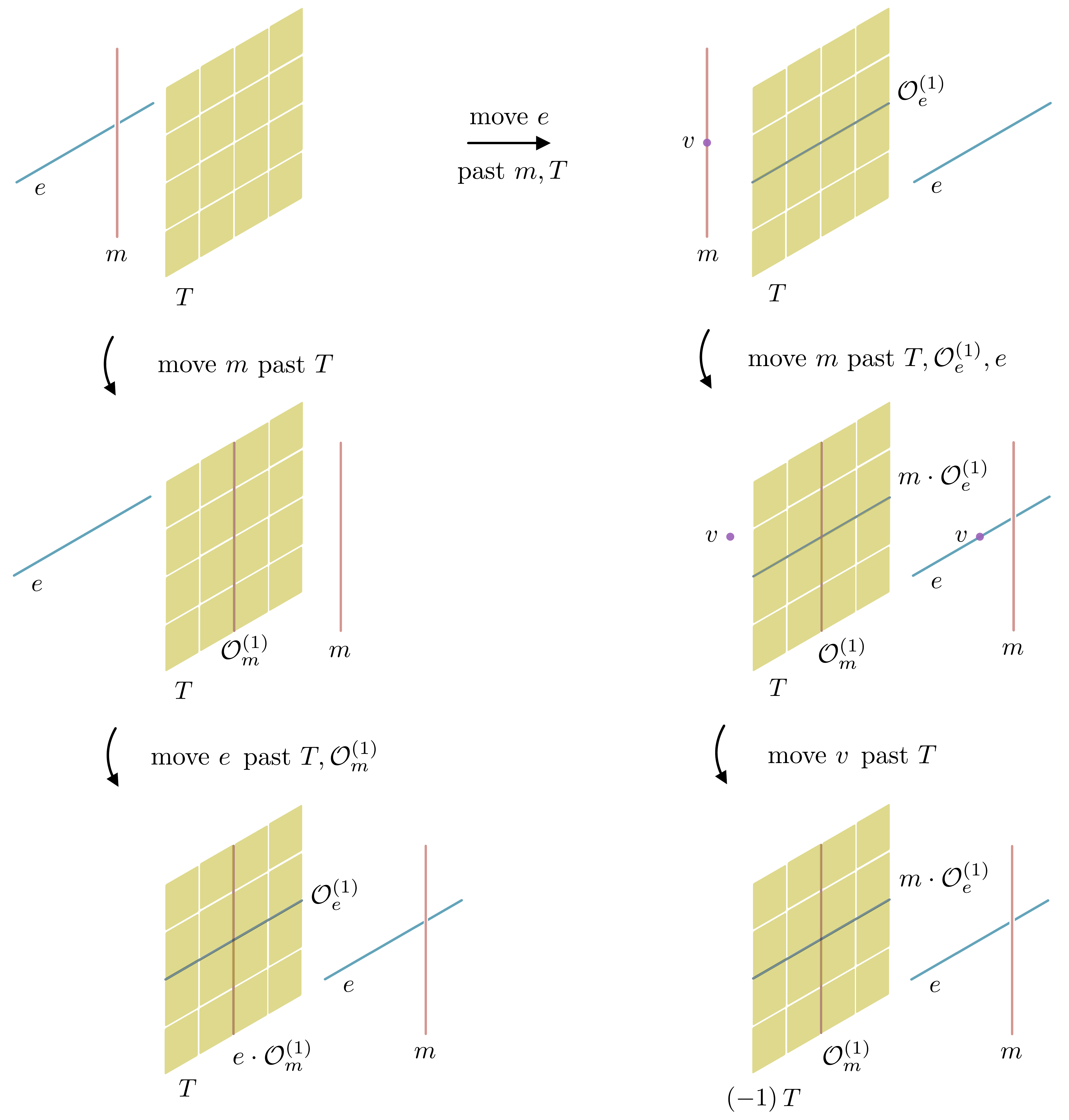} 
   \caption{In four dimensions, the electric and magnetic symmetry generators leave behind line operators $\calO_m^{(1)}$ and $\calO_e^{(1)}$ on the twist vortex. Passing the symmetry operators across the twist vortex in two different ways yields a relation \eqref{eq:projectivity} between the 1-form charges of these operators. The topological operators in this figure span the fourth dimension, while the twist vortex does not. }
   \label{fig:projective_twist_4d}
\end{figure}

\smallskip
One important implication of the higher-group structure \eqref{eq:higher_group} is that the worldvolumes of extended operators charged under the highest form symmetry (here $\ZZ_{2,v}^{(d-2)}$) must carry degrees of freedom which break the lower form symmetries ($\ZZ_{2,e}^{(1)}$ or $\ZZ_{2,m}^{(d-3)}$). This is a consequence of the fact that $\ZZ_{2,e}^{(1)}$ and $\ZZ_{2,m}^{(d-3)}$ fail to commute when acting on the operators charged under $\ZZ_{2,v}^{(d-2)}$, namely the twist vortices. To see this, we consider a general twist vortex operator $T$ and pass a magnetic symmetry surface $m$ and a codimension-2 electric symmetry generator $e$ across it in two different (but ultimately equivalent) ways, as in Figs.~\ref{fig:projective_twist_3d} and \ref{fig:projective_twist_4d} (see Ref.~\cite{Barkeshli:2022wuz} for a similar analysis in topological lattice models described by discrete gauge theories).\footnote{Alternatively we can study what happens in a local neighborhood when we move the point-like higher-group junction around the twist vortex. } 

\smallskip
Naively, the twist vortex is neutral under both $\ZZ_{2,m}^{(d-3)}$ and $\ZZ_{2,e}^{(1)}$. We might expect that the associated symmetry operators can move freely past the twist vortex, but this would be inconsistent --- passing the symmetry operators past the twist vortex in two different ways would give results which differ by a sign. To get around this, we have to allow a more general action of $\ZZ_{2,m}^{(d-3)}$ and $\ZZ_{2,e}^{(1)}$. In general, when the $\ZZ_{2,m}^{(d-3)}$ surface crosses $T(\Gamma_{d-2})$ it may deposit a line operator $\calO_m^{(1)}$ on $\Gamma_{d-2}$. Consistency with the bulk topological $\ZZ_{2,e}^{(d-3)}$ operators and their fusion implies that $\calO_m^{(1)}$ must itself be topological on $\Gamma_{d-2}$ and that $(\calO_m^{(1)})^2 = 1$. Similarly, the codimension-2 $\ZZ_{2,e}^{(1)}$ generator may act on $T(\Gamma_{d-2})$ by depositing a $(d-3)$-dimensional operator $\calO_e^{(d-3)}$ on $\Gamma$, and consistency with the bulk $\ZZ_{2,e}^{(1)}$ operators implies that $\calO_e^{(d-3)}$ must be topological on $\Gamma_{d-2}$ and $(\calO_e^{(d-3)})^2=1$.\footnote{In three dimensions, since $\calO_m^{(1)}$ is a codimension-0 operator on $\Gamma$, the fact that the bulk $\ZZ_{2,m}^{(0)}$ surface is topological implies that it realizes a topological junction between $T$ and $T\calO_m^{(1)}$.} From the perspective of the $\tilde d =(d-2)$-dimensional twist vortex worldvolume $\Gamma$, $\calO_m^{(1)}$ generates a $\ZZ_2$ $(\tilde d-2)$-form symmetry while $\calO_e^{(\tilde d-1)}$ generates a 0-form symmetry.\footnote{These symmetries are not required to act faithfully on genuine operators in the worldline theory. For instance, they could be completely trivial, or only act on the junctions of twist vortices with bulk operators. }

\smallskip
The worldline operators $\calO_m^{(1)}$ and $\calO_e^{(d-3)}$ have the right dimensionalities to carry charge under the bulk electric and magnetic symmetries, respectively. Let us denote the $\ZZ_{2,e}^{(1)}$ charge of $\calO_m^{(1)}$ by $Q_{\calO_m}^e$ and the $\ZZ_{2,m}^{(d-3)}$ charge of $\calO_e^{(d-3)}$ by $Q_{\calO_m}^e$. Passing electric and magnetic symmetry operators through the twist vortex in the two ways shown in Figs.~\ref{fig:projective_twist_3d} and \ref{fig:projective_twist_4d} gives the following relation between these charges:  
\begin{equation} \label{eq:projectivity} 
Q_{\calO_e}^m + Q_{\calO_m}^e =  1 \text{ mod } 2\,. 
\end{equation}
This implies that the electric and magnetic symmetries fail to commute in their action on the twist vortex. The featureless twist vortex with $\calO_e^{(d-3)} =1$ and $\calO_m^{(1)} =1$ fails to satisfy Eq.~\eqref{eq:projectivity}, so there must exist non-trivial worldvolume degrees of freedom on $\Gamma_{d-2}$.\footnote{The featureless twist vortex is anyway not gauge-invariant, and requires charged degrees of freedom in order to cancel this worldvolume anomaly. This is true even in the absence of the higher-group. Matching the relation in Eq.~\eqref{eq:projectivity} is an additional requirement which constrains the possible worldvolume degrees of freedom.} In the next section we will give explicit examples of twist vortex degrees of freedom which match the above consistency relation. 

\smallskip
The fact that the twist vortex carries non-trivial degrees of freedom charged under $\ZZ_{2,e}^{(1)}$ or $\ZZ_{2,m}^{(d-3)}$ (or both) gives an intuitive explanation of the fact that $\ZZ_{2,e}^{(1)} \times \ZZ_{2,m}^{(d-3)}$ does not form a `good subgroup' of the higher-group. From the background field perspective, we cannot turn on generic backgrounds for these symmetries without turning on backgrounds for $\ZZ_{2,v}^{(d-2)}$, so if $\ZZ_{2,v}^{(d-2)}$ is explicitly broken then at least one of $\ZZ_{2,e}^{(1)}$ or $\ZZ_{2,m}^{(d-3)}$ must be broken as well. This hierarchical breaking pattern is reflected in the worldvolume degrees of freedom on the twist vortex --- if we explicitly break $\ZZ_{2,v}^{(d-2)}$ by allowing dynamical twist vortices then we necessarily introduce dynamical objects which are either electrically or magnetically charged and which break $\ZZ_{2,e}^{(1)}$ or $\ZZ_{2,m}^{(d-3)}$ (or both). This has important implications for the phase diagram of the theory with dynamical twist vortices, which we explore in Section~\ref{sec:phases}. 

\section{Twist Vortices and Junctions}
\label{sec:twists} 

We now return to a more detailed analysis of the twist vortices, or Alice strings, introduced in Section~\ref{sec:generalities}.\footnote{The results in this section apply equally well to twist \emph{defects} for charge conjugation symmetry in $U(1)$ gauge theory, i.e. codimension-1 charge conjugation defects with (non-topological) boundaries. } In the current context of $O(2)$ gauge theory, inserting a twist vortex is equivalent to constraining the holonomy of the gauge field around a codimension-2 surface to lie in the conjugacy class of reflections. In other words, we pick up a $\c$ gauge transformation when encircling the twist vortex. So far, we have ignored the effects of twist vortices when analyzing the gauge transformation and symmetry properties of the action Eq.~\eqref{eq:O2_action}. A basic problem is that the naive twist vortex defined in Eq.~\eqref{eq:twist_vortex}, 
\begin{equation}
T(\Gamma_{d-2}) \stackrel{?}{=} \exp\left( i \pi \sum_{(d-2)\text{-cells}\in\Gamma} v \right)\,,
\end{equation}
is in fact not gauge-invariant. The effect of this insertion is simply to set $dc = 1$ mod $2$ for plaquettes pierced by $\Gamma^\vee$. In such a background, the action in Eq.~\eqref{eq:O2_action} remains invariant under $\c$ gauge transformations but transforms under $U(1) \subset O(2)$ gauge transformations by 
\begin{multline} \label{eq:gauge_variation} 
 i \sum_{d\text{-cells}} \dc^2\lambda \cup_\c z + \dc n \cupc \dc \tilde\lambda + \dc^2 m \cup_\c (\tilde a+\dc\tilde\lambda) \\
 = i \sum_{d\text{-cells}} (\c \cup \c \cup\lambda) \cup_\c z+ \dc n \cupc \dc \tilde\lambda + (\c\cup\c\cup m) \cup_\c (\tilde a + \dc\tilde\lambda)\,,
\end{multline}
where we have made use of the auxiliary variable $z$ from Eq.~\eqref{eq:auxiliary_action}. Note that both terms above are nonzero precisely where $\c \cup \c$, which is proportional to $dc$ mod $2$, is nonzero, i.e. at the location of the twist vortex insertion. Our strategy will be to introduce dynamical degrees of freedom on the twist vortex which can cancel the above gauge-non-invariance, effectively implementing a lattice version of Callan-Harvey anomaly inflow~\cite{Callan:1984sa}.\footnote{A similar gauge variation localized on twist vortices was found in the context of $\ZZ_N$ gauge theory in Ref.~\cite{Kaidi:2022cpf}, where the authors simply restricted gauge transformations to vanish at the appropriate loci to render the twist vortex gauge invariant. Our approach is more general, and is well-suited for non-topological theories.} The choice of these degrees of freedom is not unique, and we will present minimal examples that do the job. We will also describe how these worldline degrees of freedom give rise to junctions between the twist vortices and Wilson and (in 4d) 't Hooft lines, and how such junctions are consistent with the electric and magnetic symmetries. 

\smallskip
Notably, the electric- and magnetically-charged worldvolume degrees of freedom in our lattice setup exactly parallel the zero modes found on Alice strings in the continuum which arise from Higgsing some UV gauge group down to a non-abelian subgroup~\cite{Alford:1990ur,Alford:1990mk}. Moreover, we can dial parameters on the twist vortex worldvolume to make either the electric or magnetic particles light and, in particular limits, condense. In these extreme limits we can view the twist vortex itself as a (non-topological) condensation defect where the bulk (electric or magnetic) gauge field is Higgsed on a codimension-2 surface in spacetime. Such condensation defects have been studied in the context of topological phases of condensed matter systems (see for instance Refs.~\cite{Else:2017yqj,Barkeshli:2022wuz,Tantivasadakarn:2023zov}) where they are also dubbed `Cheshire strings.'

\subsection{Three Dimensions}
\label{sec:3d_tv} 

In three dimensions the twist vortex-localized gauge variation we need to cancel is (renaming $\tilde a \equiv \sigma$ to be the dual photon)
\begin{equation} \label{eq:3d_gauge_variation} 
 i \sum_{\text{cubes}} (\c \cup \c \cup\lambda) \cup_\c z + (\c\cup\c\cup m) \cup_\c \sigma \,.
\end{equation}
The specific examples we will consider involve a compact scalar on the twist vortex worldline, in other words a particle on a circle (but with charge conjugation symmetry gauged). This system admits a Villain lattice description described in Appendix~\ref{app:poc}, whose ingredients are a real scalar $\varphi \in \calC^0(\Gamma, \RR)$ and associated integer Villain field $w \in \calC^1(\Gamma, \ZZ)$, with gauge redundancy $\varphi \to \varphi + 2\pi r,  w \to w + \dc r$, where $r \in \calC^0(\Gamma,\ZZ)$. Both fields are odd under charge conjugation, $\varphi \to \g\cup\varphi$ and $w \to \g\cup w$. Before gauging charge conjugation, this quantum mechanical system has a $U(1)$ global shift symmetry $\varphi \to \varphi + \text{constant}$ and a theta angle $\theta$ with periodicity $2\pi$ (formally a $(-1)$-form symmetry). An important fact about this system is that there is a generalized mixed anomaly between the $U(1)$ global symmetry and the periodicity of the $\theta$ parameter~\cite{Gaiotto:2017yup,Kikuchi:2017pcp,Cordova:2019jnf} which is preserved in the Villain lattice discretization, as reviewed in Appendix~\ref{app:poc}. This anomaly is precisely what we need in order to cancel the gauge variation \eqref{eq:gauge_variation}, in other words the anomaly inflow, from the bulk. Recalling that $z = \tilde n - \frac{1}{2\pi} \dc\sigma$ on shell, the terms in Eq.~\eqref{eq:3d_gauge_variation} match exactly the anomalous terms in Eq.~\eqref{eq:poc_anomaly}. 

\smallskip
We begin by coupling the shift symmetry of $\varphi$ to the bulk gauge field with charge $q_\varphi$, replacing the usual kinetic term with
\begin{equation}
\frac{\kappa_v}{2} \sum_{\text{links}\,\in \Gamma} (\dc\varphi - q_\varphi a - 2\pi w)^2\,. 
\end{equation}
The fields on the twist vortex worldline transform under bulk gauge transformations as $\varphi \to \varphi + q_\varphi \lambda$ and $w \to w - q_\varphi m$. The first term in Eq.~\eqref{eq:gauge_variation}, i.e. the non-invariance under small $U(1)$ gauge transformations, can be written as
\begin{equation}
-2i \sum_{\text{cubes}} \overline{dc} \cup \lambda \cup z\,,
\end{equation}
where $\overline{dc} \equiv dc$ mod $2$, and can be cancelled by including the coupling
\begin{equation} \label{eq:phi} 
\frac{2i}{q_\varphi} \sum_{\text{links}\,\in \Gamma}  \varphi \cup z\,.
\end{equation}
To find the effect of this term in the usual presentation of the action, we simply integrate out $z$ to find that the gauge field kinetic term gets replaced by 
\begin{equation} \label{eq:modified_kinetic} 
\frac{\beta}{2}(\dc a - 2\pi n)_p^2 \ \to \  \frac{\beta}{2}( \dc a -\frac{1}{q_\varphi}\dc^2 \varphi - 2\pi n)_p^2 = \frac{\beta}{2}( \dc a -\frac{1}{q_\varphi}\c\cup\c\cup \varphi - 2\pi n)_p^2\,.
\end{equation}
Note that this is a non-trivial change only for the plaquettes $p \in \star[\Gamma^\vee]$. For this modification to preserve the compactness of $\varphi$, we must assign $n \to n - \frac{1}{q_\varphi}\c \cup \c \cup r$ under the gauge redundancy \eqref{eq:compactness}. Recalling that $\c \cup \c$ is even, this shift is well-defined only if $|q_\varphi| = 1,2$. The original monopole-suppression term in the action is not invariant under this additional twist vortex-localized transformation of $n$, but shifts by  
\begin{equation} \label{eq:variation_2} 
-\frac{i}{q_\varphi} \sum_{\text{cubes}} \dc (\c\cup\c\cup r) \cup_\c\sigma = -\frac{i}{q_\varphi} \sum_{\text{cubes}}(\c \cup \c \cup  \dc r) \cup_\c\sigma\,,
\end{equation}
where we made repeated use of Eq.~\eqref{eq:derivative}. 

\smallskip
Conveniently, the transformation properties of $w$ are such that we can cancel both the second term in Eq.~\eqref{eq:gauge_variation}, as well as Eq.~\eqref{eq:variation_2}, by adding to the action the coupling\footnote{This can be written as a sum over links on $\Gamma$, but it takes a complicated form since the twisted cup product is not associative. To save space we write this term as if it were a bulk term, with the knowledge that it localizes to terms involving $w$ and $\sigma$ on $\Gamma$. For instance if we consider a straight twist vortex, we could write this as
\begin{equation}
-\frac{2i}{q_\varphi} \sum_{\text{links}\,\in \Gamma} w \cup_\c \sigma \,. 
\end{equation}
}
\begin{equation}
 \frac{i}{q_\varphi} \sum_{\text{cubes}} (\c\cup\c\cup w) \cup_\c\sigma\,.
\end{equation}
Note that since $|q_\varphi| = 1,2$ and $\c\cup\c$ is even this preserves the periodicity of $\sigma$. This couples the bulk monopole operator to the topological charge density of the particle on a circle, and can be thought of as gauging its $(-1)$-form symmetry, i.e. promoting the theta parameter to a dynamical axion. 

\smallskip
In summary, we have the improved, gauge invariant twist vortex operators\footnote{\label{fn:gaugefix} The `sum' over $\varphi \in \calC^0(\Gamma,\RR)$ is schematic. Since the worldvolume theory has a $\ZZ$ gauge redundancy, one must gauge-fix in order to get a finite result for the expectation value of the twist vortex. One possibility is to choose the `Villain gauge' where $\varphi$ is restricted to lie in the range $(-\pi,\pi]$, and the gauge variation $q_\varphi \lambda$ is also projected to this interval. }
\begin{align} \label{eq:type_12_3d} 
T_{q_\varphi}(\Gamma) =\sum_{\substack{\varphi \in \calC^0(\Gamma, \RR), \\ w \in \calC^1(\Gamma, \ZZ)}} \exp\Bigg[i \sum_{\text{links} \in\Gamma} \left( \pi  v - \frac{2}{q_\varphi}\varphi \cup z\right) - \frac{\kappa_v}{2} \sum_{\text{links}\in\Gamma} (&\dc\varphi - q_\varphi a - 2\pi w)^2 \\
- \frac{i}{q_\varphi} \sum_{\text{cubes}} &(\c \cup \c \cup w) \cup_\c \sigma \Bigg] \,, \quad |q_\varphi| = 1,2\,,  \nonumber 
\end{align}
which are invariant under the combined gauge redundancies
\begin{equation}
\begin{split}
\varphi &\to \varphi + 2\pi r + q_\varphi\lambda\,, \ w \to w + \dc r - q_\varphi m \,, \\
a &\to a + \dc\lambda + 2\pi m \,, \ n \to n + \dc m  - \frac{1}{q_\varphi} \c \cup \c  \cup r\,, \ \sigma \to \sigma + 2\pi \tilde m\,. 
\end{split}
\end{equation}
The twist vortices fall into two classes, depending on whether $|q_\varphi| = 1,2$, which we will refer to as type-$e$ and type-$m$, respectively. Roughly, the type-$e$ vortex breaks the $\ZZ_{2,e}^{(1)}$ electric symmetry of the bulk and preserves the $\ZZ_{2,m}^{(0)}$ magnetic symmetry, while the type-$m$ vortex breaks the magnetic symmetry and preserves the electric symmetry. From the perspective of the bulk, the particle on the circle serves as a charge-$q_\varphi$ Higgs field localized on $\Gamma$. If we take the limit $\kappa_v \to \infty$ we pin $a = \frac{1}{q_\varphi}(\dc\varphi- 2\pi w)$, which restricts the bulk gauge field to be a $\ZZ_{|q_\varphi|}$ gauge field on the twist vortex worldline. In this strict limit, the twist vortex can be thought of as a kind of condensation defect, where charge-$q_\varphi$ excitations proliferate on its worldvolume.

\subsubsection{Consistency with the 2-Group}

In Section~\ref{sec:higher_group} we derived a consistency relation between the charges of operators deposited by electric and magnetic symmetry generators on the twist vortex worldvolume. Here we describe how the specific twist vortices in Eq.~\eqref{eq:type_12_3d} satisfy these conditions. 
\begin{itemize}
\item \underline{Type-$e$}: If we take $q_\varphi = 1$, the twist vortex is invariant under shifts $\sigma \to \sigma + \pi V_m$ and therefore respects the $\ZZ_{2,m}^{(0)}$ symmetry. In other words, nothing gets deposited on $\Gamma$ as we pass a magnetic symmetry surface past it, so $\calO_m^{(1)} = 1$ and $Q_{\calO_m}^e = 0$. On the other hand, performing a 1-form symmetry transformation does not leave the twist vortex invariant, but leaves behind an insertion of 
\begin{equation} \label{eq:3d_calOe} 
e^{-\frac{\kappa_v}{2}\sum_{\ell \in \Gamma} (\dc\varphi - a - \pi V_e - 2\pi w)_\ell^2 -(\dc\varphi - a - 2\pi w)_\ell^2} \, e^{-\frac{i}{2} \sum_{\text{cubes}} (\c \cup \c \cup V_e) \cupc \sigma}\,.
\end{equation}
Since $V_e = \star[\tilde D]$, with $\tilde D$ a surface on the dual lattice, this effectively inserts a local operator $\calO_e^{(0)}$ at the location where $\tilde D$ pierces $\Gamma$. In the presence of this operator, we must modify the $\c$ gauge transformation of $w$ to $w \to \g \cup w + \frac{1}{2}(\g-1)\cup V_e$. To see that this worldline operator is topological, we can perform a shift $\varphi \to \varphi - \pi U_e$ together with $w\to w+\frac{1}{2}(\c-1)\cup U_e$ and $n \to n - \frac{1}{2}\c\cup\c \cup U_e$, which leaves the modified gauge field kinetic term \eqref{eq:modified_kinetic} invariant but alters the insertion to 
\begin{equation}
e^{-\frac{\kappa_v}{2}\sum_{\ell \in \Gamma} (\dc\varphi - a - \pi (V_e+dU_e) - 2\pi w)_\ell^2 -(\dc\varphi - a - 2\pi w)_\ell^2} \,  e^{-\frac{i}{2} \sum_{\text{cubes}} (\c \cup \c  \cup (V_e+dU_e)) \cupc \sigma}\,.
\end{equation}
This effectively moves the location of $\calO_e^{(0)}$ along the twist vortex, making it topological. One can also verify that the operator $\calO_e^{(0)}$ has $\ZZ_2$ fusion rules consistent with the bulk $\ZZ_{2,e}^{(1)}$ line. The last factor in Eq.~\eqref{eq:3d_calOe} is charged under the magnetic symmetry so that $Q_{\calO_e}^m = 1$. Hence and $Q_{\calO_e}^m + Q_{\calO_m}^e = 1+0= 1$, and Eq.~\eqref{eq:projectivity} is satisfied.

\item \underline{Type-$m$}: If we take $q_\varphi = 2$, we can preserve $\ZZ_{2,e}^{(1)}$ by assigning $w \to w - V_e$, so in this case $\calO_e^{(0)} = 1$. On the other hand, performing a magnetic symmetry transformation $\sigma \to \sigma + \pi V_m$ inserts a line operator $\calO_m^{(1)}$ 
\begin{equation}
e^{i \pi \sum_{\ell \in \Gamma} w \cup V_m } 
\end{equation}
along the twist vortex. Since $V_m = \star[\tilde Y]$ for some 3-volume $\tilde Y$, this operator is localized along the line segment where $\Gamma$ lies in $\tilde Y$. Since $w \to w - V_e$ under the 1-form symmetry, we now have that $Q_{\calO_e}^m = 0$ and $Q_{\calO_m}^e = 1$, and Eq.~\eqref{eq:projectivity} is again satisfied. 

\end{itemize}


\subsubsection{Junctions}

Both types of twist vortices host electrically charged matter on their worldlines which we can use to terminate bulk Wilson lines. Specifically, we can construct open Wilson lines as described in Eq.~\eqref{eq:open_line} which end (roughly speaking) on the twist vortex-localized operator $e^{i\varphi}$. This operator has charge 1 on a type-$e$ twist vortex, so all Wilson lines $W_q$ with $q \ge 1$ can end on it. In the type-$m$ case, this operator has charge 2, so that only even charge Wilson lines can end (alternatively, $e^{i\varphi}$ can serve as a junction between $W_1$ in the bulk and line operators on the twist vortex such as $W_1$ or $\calO_m^{(1)}$). 

\begin{figure}[h!] 
   \centering
   \includegraphics[width=.8\textwidth]{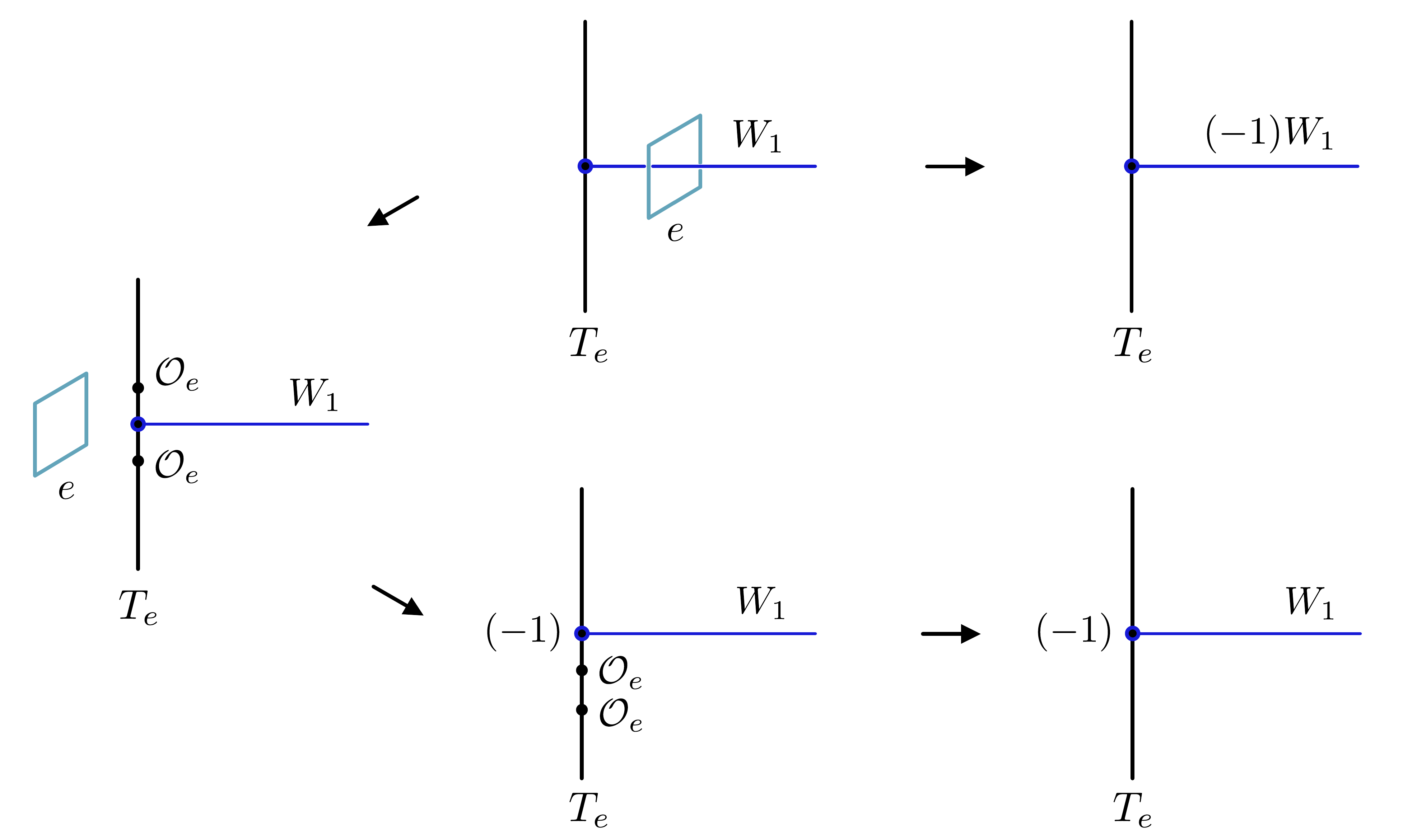} 
   \caption{Junction between a minimally charged Wilson line $W_1$ and a type-$e$ twist vortex $T_e$. When a $\ZZ_{2,e}^{(1)}$ symmetry line crosses $T_e$, it deposits the topological local operator $\calO_e^{(0)}$ which acts on the point-like junction with the Wilson line. As a result, the junction is consistent with the $\ZZ_{2,e}^{(1)}$ symmetry. }
   \label{fig:3djunction}
\end{figure}

\smallskip
At first glance, the fact that the minimal Wilson line can end on the type-$e$ vortex seems to violate the $\ZZ_{2,e}^{(1)}$ symmetry. In fact, the junction is perfectly consistent with the electric 1-form symmetry --- the local operators $\calO_e^{(0)}$ left behind by the $\ZZ_{2,e}^{(1)}$ line act on the junction operator $e^{i\varphi}$, as discussed above. This is depicted in Fig.~\ref{fig:3djunction}.  

\begin{figure}[h!] 
   \centering
   \includegraphics[width=.5\textwidth]{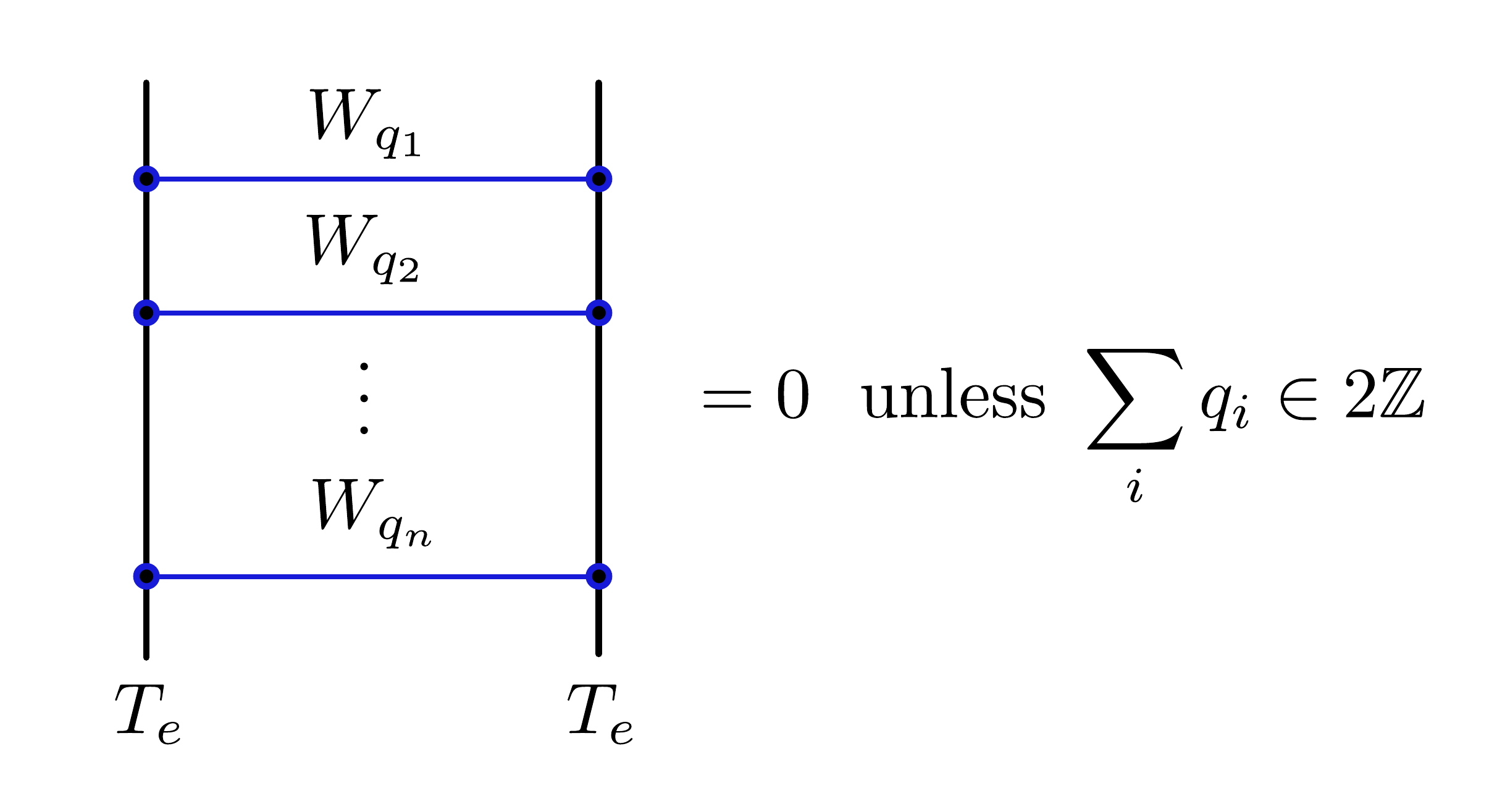} 
   \caption{An example of a selection rule for Wilson loop junctions arising due to $\ZZ_{2,e}^{(1)}$ symmetry.}
   \label{fig:ladder}
\end{figure}

\smallskip
Finally, the $\ZZ_{2,e}^{(1)}$ electric 1-form symmetry gives selection rules for the allowed junction configurations. In particular, the net charge of a `ladder' of Wilson loops stretched between two twist vortices has to be even, otherwise the configuration vanishes (see Fig.~\ref{fig:ladder}). It is straightforward to find configurations in the strong coupling expansion that contribute to e.g. a charge-2 Wilson line stretched between two non-contractible twist vortices.\footnote{The invertible symmetries do not give rise to constraints on the same configuration but where the twist vortices are both contractible. However, such a configuration is constrained by the non-invertible electric symmetry discussed in Section~\ref{sec:noninvertible_electric}.  } We discuss such configurations in more detail in Section~\ref{sec:selection_rules}.

\subsection{Four Dimensions}
\label{sec:4d_tv}

The four dimensional uplift of the story involves the (orbifolded) compact boson on the twist vortex worldsheet $\Sigma$. We review the Villain discretization of this system and compute the 't Hooft anomaly for its global $U(1)_m^{(0)} \times U(1)_w^{(0)}$ symmetry in Appendix~\ref{app:orbifold}. The ingredients are a compact scalar and its dual $\varphi, \tilde\varphi \in \calC^0(\Sigma, \RR)$, and an integer Villain field $w \in\calC^1(\Sigma,\ZZ)$ subject to the gauge redundancies
\begin{equation} 
\varphi \to \varphi + 2\pi r\,, \  w \to w +\dc r\,,\ \tilde\varphi \to \tilde\varphi + 2\pi \tilde r\, .
\end{equation}
The naive twist vortex induces a gauge variation of the action 
\begin{equation} \label{eq:4d_gauge_variation} 
 i \sum_{d\text{-cells}} -2\overline{dc}\cup \lambda\cup  z+ \dc n \cupc \dc \tilde\lambda + (\c\cup\c\cup m) \cup_\c (\tilde a + \dc\tilde\lambda)\,.
\end{equation}
Note that despite appearances the $\dc n \cupc \dc \tilde\lambda$ term is localized on the twist vortex --- the non-flatness of $c$ is what prevents this term from being a total derivative. Since $z =\tilde n - \frac{1}{2\pi} \dc \tilde a$ on-shell, the terms above are in 1-1 correspondence with the anomalous terms in Eq.~\eqref{eq:cb_anomaly}. Hence, paralleling the 3d discussion, we can cancel the above gauge variation by coupling the momentum and winding symmetries of the compact boson to the bulk electric and magnetic gauge fields. 

\smallskip
Cutting to the chase, we again have a choice of the charge $q_\varphi$ (with $|q_\varphi| = 1,2$) of $\varphi$ with respect to the electric gauge field $a$. The magnetic charge of $\tilde\varphi$ with respect to the dual gauge field $\tilde a$ is then fixed to be $2/q_\varphi$. The gauge-invariant twist vortex is 
\begin{multline} \label{eq:type_12_4d} 
T_{q_\varphi}(\Sigma) =\sum_{\substack{\varphi \in \calC^0(\Sigma, \RR), \tilde\varphi \in \calC^0(\Sigma,\RR), \\ w \in \calC^1(\Sigma, \ZZ)  }} \exp\Bigg[i \sum_{\text{plaq.} \in\Sigma} \left( \pi  v + \frac{2}{q_\varphi} \varphi \cup z\right) + \frac{\kappa_v}{2} \sum_{\text{links}\in\Sigma} (\dc\varphi - q_\varphi a - 2\pi w)^2 \\
 + i\sum_{\text{hyper-cu.}}\frac{q_\varphi}{2} \dc n \cupc \dc\tilde\varphi - \frac{1}{2}(\c \cup \c \cup w) \cupc \left(\frac{2}{q_\varphi}\tilde a - \dc\tilde\varphi \right) \Bigg]\,.
\end{multline}
Again, the measure is meant to be schematic, see the remarks in Footnote~\ref{fn:gaugefix}. It is straightforward to check that the insertion of such a twist vortex operator is invariant under the following gauge redundancies: 
\begin{equation}
\begin{split}
\varphi &\to \varphi + 2\pi r + q_\varphi\lambda\,, \ w \to w + \dc r - q_\varphi m \,,\  \tilde\varphi \to \tilde\varphi +2\pi \tilde r + \frac{2}{q_\varphi} \tilde\lambda\,, \\
a &\to a + \dc\lambda + 2\pi m \,, \ n \to n + \dc m  - \frac{1}{q_\varphi} \c \cup \c  \cup r\,, \ \tilde a \to \tilde a +\dc\tilde\lambda +q_\varphi \pi \dc \tilde r + 2\pi \tilde m\,. 
\end{split}
\end{equation}
The twist vortices again fall into two classes which we call type-$e$ (with $|q_\varphi| = 1$, i.e. minimally charged electric matter and non-minimally charged magnetic matter) and type-$m$ (with $|q_\varphi| = 2$, i.e. non-minimally charged electric matter and minimally charged magnetic matter). These two choices are clearly exchanged by the bulk electric-magnetic duality which involves T-duality on the twist vortex worldsheet. 

\smallskip
We can also consider an extreme limit where we send $\kappa_v \to \infty$ so as to pin $a = \frac{1}{q_\varphi}(\dc\varphi - 2\pi w)$. Since there is no kinetic term for the dual field $\tilde\varphi$, integrating it out further sets $\dc w = - q_\varphi n = 0$ mod $q_\varphi$.  In other words in this limit the electric gauge field is Higgsed to a \emph{flat} $\ZZ_{|q_\varphi|}$ gauge field on $\Sigma$, and we can think of the twist vortex as a genuine condensation defect. If on the other hand we take $\kappa_v = 0$, we can sum over $w$ to set $\tilde a = \frac{q_\varphi}{2}(\dc\tilde\varphi - 2\pi \tilde w)$ for some $\tilde w \in \calC^1(\Sigma,\ZZ)$. Integrating out $\varphi$ then sets $\dc\tilde w = \frac{2}{q_\varphi} \tilde n$. We can therefore view this limit as Higgsing the \emph{magnetic} gauge field to $\ZZ_{|2/q_\varphi|}$. T-duality of the orbifolded compact boson, which takes $\kappa_v \to \frac{1}{(2\pi)^2\kappa_v}$, exchanges these two limits.

\subsubsection{Consistency with the 3-Group}

Here we describe how the specific twist vortices in Eq.~\eqref{eq:type_12_4d} satisfy the relation Eq.~\eqref{eq:projectivity} between the 1-form charges of the lines $\calO_m^{(1)}$ and $\calO_e^{(1)}$ deposited by the bulk $\ZZ_{2,e}^{(1)}$ and $\ZZ_{2,m}^{(1)}$ symmetries.
\begin{itemize}
\item \underline{Type-$e$}: When $q_\varphi = 1$ the twist vortex is transparent to the magnetic symmetry, so that $\calO_m^{(1)} = 1$, while an electric 1-form transformation generates an insertion of 
\begin{equation} \label{eq:4d_calOe} 
e^{-\frac{\kappa_v}{2}\sum_{\ell \in \Sigma} (\dc\varphi - a - \pi V_e - 2\pi w)_\ell^2 -(\dc\varphi - a - 2\pi w)_\ell^2} \, e^{-\frac{i}{2} \sum_{\text{hyper-cu.}} (\c \cup \c \cup V_e) \cupc (\tilde a - \frac{1}{2}\dc\tilde\varphi)}\,.
\end{equation}
Here $V_e = \star[\tilde Y]$ is dual to a three-volume $Y$ whose intersection with $\Sigma$ is a line. We identify the above insertion as $\calO_e^{(1)}$. The location of the insertion can be shifted by redefining $\varphi \to \varphi + \pi dU_e$, $w \to w+\frac{1}{2}(\c-\1)\cup U_e$ and $n \to n - \frac{1}{2}\c \cup \c \cup U_e$ as in the 3d case. Because the last factor involves the magnetic gauge field, this operator is charged under the magnetic symmetry and $Q_{\calO_e}^m =1$. Eq.~\eqref{eq:projectivity} is satisfied because $Q_{\calO_e}^m + Q_{\calO_m}^e = 1+ 0 = 1$. 

\item \underline{Type-$m$}: If $q_\varphi =2$, we can make the twist vortex transparent to the electric 1-form symmetry by taking $w \to w - V_e$, so that $\calO_e^{(1)} = 1$. If we perform a magnetic symmetry transformation $\tilde a \to \tilde a + \pi V_m$, the result is to insert
\begin{equation}
e^{i \pi \sum_{p \in \Sigma} w \cup V_m} \,,
\end{equation}
which will be some line operator lying on $\Sigma$, which we identify as $\calO_m^{(1)}$. We can move its location by shifting $\tilde\varphi \to \tilde\varphi + \pi dU_m$, which does not affect the first term in the second line of Eq.~\eqref{eq:type_12_4d}, but changes the insertion to  
\begin{equation}
e^{i \pi \sum_{p \in \Sigma} w \cup (V_m + dU_m)} \,,
\end{equation}
which topologically deforms the line on which the operator is supported. Because $w$ transforms under the electric 1-form symmetry, we have $Q_{\calO_m}^e = 1$, while $Q_{\calO_e}^m = 0$ so again Eq.~\eqref{eq:projectivity} is satisfied.
\end{itemize}

\subsubsection{Junctions}

\begin{figure}[h!] 
   \centering
   \includegraphics[width=.95\textwidth]{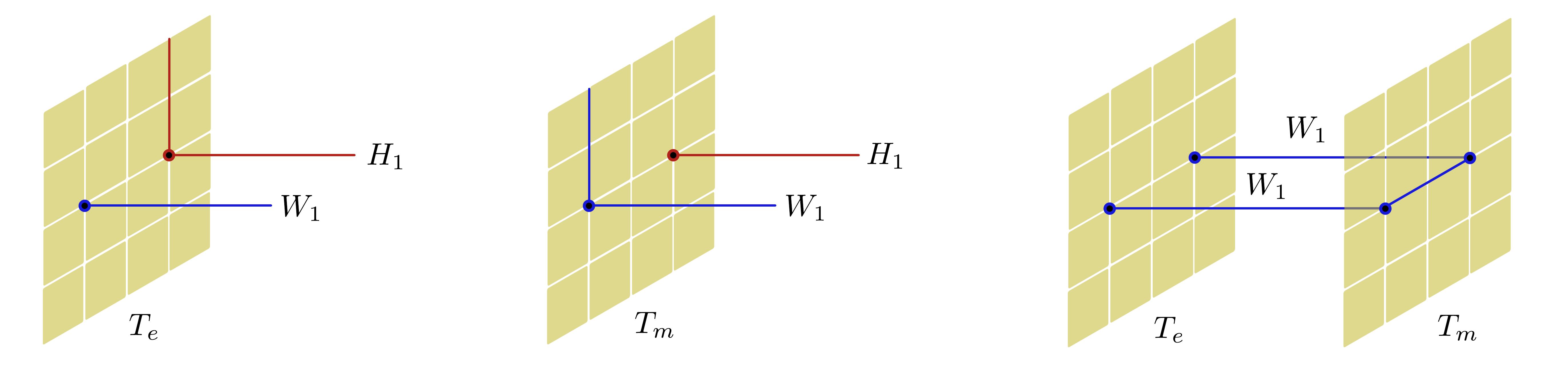} 
   \caption{Consistent junctions on type-$e$ and type-$m$ twist vortices $T_e$ and $T_m$, which carry charge-1 (resp. charge-2) electric matter and charge-2 (resp. charge-1) magnetic matter. On the right we show a configuration which survives selection rules due to $\ZZ_{2,v}^{(2)}$ and $\ZZ_{2,e}^{(1)}$ symmetries. The $T_e$ and $T_m$ twist vortices are exchanged under electric-magnetic duality, which is realized as T-duality on the twist vortex worldsheet.}
   \label{fig:4djunction}
\end{figure}

As the names suggest, a minimal Wilson line can terminate on the electric Higgs field $e^{i\varphi}$ on the type-$e$ twist vortex while a minimal 't Hooft line can terminate on the magnetic Higgs field $e^{i\tilde\varphi}$ on a type-$m$ twist vortex. These junctions are shown in Fig.~\ref{fig:4djunction}, and are clearly exchanged by electric-magnetic duality. As in the 3d case, the non-trivial degrees of freedom on the twist vortex ensure that these junctions are consistent with the bulk electric and magnetic 1-form symmetries. Specifically, when an electric (magnetic) symmetry surface crosses a type-$e$ (type-$m$) twist vortex, it deposits the topological line operator $\calO_e^{(1)}$ ($\calO_m^{(1)}$) which can act on the point-like junction with the Wilson ('t Hooft) line. The invertible symmetries constrain the allowed junction configurations --- for instance, any collection of junctions which involves Wilson or 't Hooft lines with odd net charge vanishes. 

\section{Non-Invertible Symmetries}
\label{sec:noninvertible} 

When we gauge the charge conjugation symmetry of $U(1)$ gauge theory we naively expect to break the $U(1)_e^{(1)}$ and $U(1)_m^{(d-3)}$ symmetries down to their $\ZZ_2$ subgroups which commute with the charge conjugation action. This turns out to be just part of a much larger categorical symmetry structure in the $O(2)$ theory that also includes \emph{non-invertible} analogs of $U(1)_e^{(1)}$ and $U(1)_m^{(d-3)}$. These types of non-invertible symmetries arise after gauging 0-form symmetries that act as outer automorphisms of group-like symmetries, and have been studied previously in a number of works, see for instance Refs.~\cite{Teo:2015xla,Nguyen:2021yld,Heidenreich:2021xpr,Thorngren:2021yso,Sharpe:2021srf,Bhardwaj:2022yxj,Arias-Tamargo:2022nlf,Bhardwaj:2022scy,Antinucci:2022eat,Schafer-Nameki:2023jdn,Damia:2023gtc,Gutperle:2024vyp,Hsin:2024aqb,Hofman:2024oze}.  Roughly, the idea is that the invertible symmetry operator $U_\theta$ of e.g. the $U(1)_e^{(1)}$ symmetry in the $U(1)$ theory gets replaced by the operator $U_\theta + U_{-\theta}$ in the $O(2)$ theory, which does not have an inverse. We can make this recipe precise by applying the general construction from Section~\ref{sec:general_construction} to the electric and magnetic symmetry operators of the $U(1)$ theory. This gives the $(O(2)/\ZZ_2)^{(1)}_e$ and $(O(2)/\ZZ_2)^{(d-3)}_m$ coset symmetry operators, whose non-invertibility is made obvious from the fact that projection operators are part of their definition. In this Section we describe our concrete realization of the symmetry defects and study their fusion rules, action on operators, and associated selection rules. 

\subsection{Electric Symmetry}
\label{sec:noninvertible_electric}

We start with the symmetry operator for the $U(1)_e^{(1)}$ electric symmetry of $U(1)$ gauge theory, namely 
\begin{equation}
U_\theta^{U(1)}(\tilde\Gamma) = \prod_{p\in\star[\tilde\Gamma]}  \exp\left[-\frac{\beta}{2}(\dc a - \theta \star[\tilde\Gamma] -2\pi n)_p^2 + \frac{\beta}{2}(\dc a -2\pi n)_p^2\right] \,.
\end{equation}
This operator effectively replaces $n \to n + \frac{\theta}{2\pi}\star[\tilde\Gamma]$ in the gauge field kinetic term, which is a non-trivial shift for all plaquettes pierced by the codimension-2 surface $\tilde\Gamma$ on the dual lattice. These operators satisfy the $U(1)$ fusion rule $U_\theta^{U(1)} U_{\theta'}^{U(1)} = U_{\theta+\theta'\text{ mod } 2\pi}^{U(1)}$. Obviously the above operator is not $\c$ invariant, and applying (a slightly modified version of) the recipe from Section~\ref{sec:general_construction}, we define
\begin{equation}
U_\theta(\tilde\Gamma) = P(\tilde\Gamma^\vee) \sum_{\hat\g_* = \pm 1} \prod_{p \in \star[\tilde\Gamma]} \exp\left[-\frac{\beta}{2}(\dc a - \hat\g_*\, \eta(\gamma_{p,*})\,\theta \star[\tilde\Gamma] -2\pi n)_p^2 + \frac{\beta}{2}(\dc a -2\pi n)_p^2\right]\,,
\end{equation}
where $\tilde\Gamma^\vee$ is $\tilde\Gamma$ shifted down to the original lattice where the scaffolding paths $\gamma_{p,*}$ and basepoint $*$ live. This is the Gukov-Witten operator associated with the $O(2)$ conjugacy class of rotations through an angle $\theta \sim - \theta \sim \theta + 2\pi$. 

\smallskip
Roughly speaking, we can topologically deform $U_\theta(\tilde\Gamma)$ by performing field redefinitions of the form $a_\ell \to a_\ell + \hat\g_*\,\eta(\gamma_{\ell,*})\, \theta$ (the projector $P(\tilde\Gamma^\vee)$ is topological on its own, and comes for the ride). As an explicit example, let us take $\tilde\Gamma = \partial \tilde\Sigma$ where $\tilde\Sigma$ is dual to a single link $(s,i)$. In 3d this corresponds to a plaquette-sized contractible loop, so that $\star[\tilde\Gamma]$ is non-zero on the four plaquettes touching the link $(s,i)$, i.e. the coboundary of $(s,i)$. Taking the expectation value of this contractible operator computes the quantum dimension of $U_\theta(\tilde\Gamma)$. To evaluate the expectation value we simply redefine $a_{(s,i)} \to a_{(s,i)} + \hat\g_* \, \theta$, effectively removing the insertion. What is left over is the projector, which is equal to 1 for a contractible curve, and the sum over $\hat\g_*$, which gives a factor of two. Hence we find the quantum dimension 
\begin{equation}
\langle U_\theta(\partial\tilde\Sigma) \rangle = 2.
\end{equation} 

\smallskip
As another example, we compute the parallel fusion of two symmetry operators $U_\theta(\tilde\Gamma)$ and $U_{\theta'}(\tilde\Gamma')$. Since the location of the basepoints of the two operators is arbitrary, we choose them to lie on the same line in the direction of fusion. Performing field redefinitions to join the lines, and using $P(\tilde\Gamma^\vee)^2 = P(\tilde\Gamma^\vee)$, we find
\begin{multline} \label{eq:electric_fusion} 
U_\theta(\tilde\Gamma)U_{\theta'}(\tilde\Gamma) \\
= P(\tilde\Gamma^\vee) \sum_{\hat\g_*, \hat\g_*'= \pm 1}\, \prod_{p \in \star[\tilde\Gamma]} \exp\Bigg[-\frac{\beta}{2}(\dc a - \eta(\gamma_{p,*}) (\hat\g_* \theta + \hat\g_*'\theta') \star[\tilde\Gamma] -2\pi n)_p^2 + \frac{\beta}{2}(\dc a -2\pi n)_p^2 \Bigg] \\
= U_{\theta+\theta'}(\tilde\Gamma) + U_{\theta-\theta'}(\tilde\Gamma)\,.
\end{multline}
In the special case that $\theta = \theta'$, this reduces to 
\begin{equation}
U_\theta(\tilde\Gamma)U_\theta(\tilde\Gamma) =  U_{2\theta}(\tilde\Gamma) +2P(\tilde\Gamma^\vee) \,.
\end{equation}
This matches the continuum results in Refs.~\cite{Heidenreich:2021xpr,Bhardwaj:2022yxj,Antinucci:2022eat}. 

\subsubsection{Action on Wilson Lines}

Let us now examine the action on Wilson lines. We start with a configuration where $\tilde\Gamma$ is contractible and links a Wilson line, and denote the basepoints of $W$ and $U$ as $*$ and $*'$ which respectively host $\ZZ_2$ degrees of freedom $\hat\g_*$ and $\hat\g_{*'}$. Performing $\hat\g_{*'}$-dependent shifts of $a$ to remove the operator $U$ gives 
\begin{equation}\label{eq:contractible_line_action}
\begin{split}
U_\theta(\tilde\Gamma) W_q(\gamma) &= P(\gamma)\sum_{\hat\g_*, \hat\g_{*'} = \pm 1}\exp\left(i q \theta \, \hat\g_{*'} \,\eta(\gamma_{*',*}) \hat\g_{*}\right)   \exp\left(i q\, \hat\g_* \sum_{\ell \in \gamma} \eta(\gamma_{\ell,*})\, a_\ell \right) \\
&= 2\cos(q\theta)\, W_q(\gamma)\,,
\end{split}
\end{equation}
where $\gamma_{*',*}$ is a path in $\gamma$ from $*'$ to $*$. 

\begin{figure}[h!] 
   \centering
   \includegraphics[width=\textwidth]{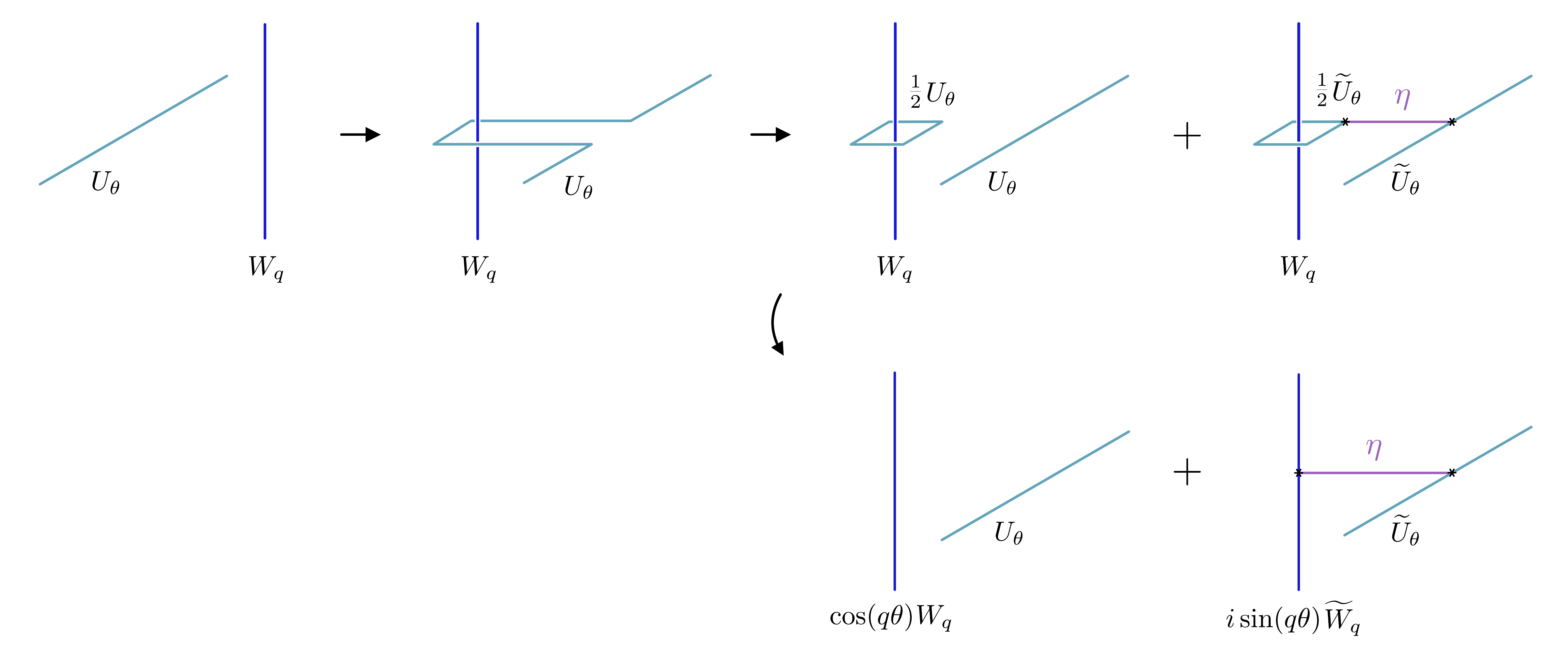} 
   \caption{The action of the non-invertible electric 1-form symmetry on a Wilson line. The result is a sum of two terms, the latter involving the twisted-sector Wilson line $\widetilde{W}_q$ and Gukov-Witten operator $\widetilde{U}_\theta$ connected by the dual symmetry line $\eta$. }
   \label{fig:1-form_action}
\end{figure}

\smallskip
More generally, we can study what happens when $\tilde\Gamma$ locally crosses $\gamma$. In that case, we can locally wrap part of the Gukov-Witten operator on the Wilson line. Let us write $W_q(\gamma) = \sum_{\hat\g_*} W_q(\gamma; \hat\g_*)$ and $U_\theta(\tilde\Gamma) = \sum_{\hat\g_{*'}} U_\theta(\tilde\Gamma; \hat\g_{*'})$. The result is 
\begin{equation} \label{eq:line_action} 
\begin{split}
U_\theta(\tilde\Gamma) W_q(\gamma) &= \sum_{\hat\g_*, \hat\g_{*'} = \pm 1}\exp\left(i q \theta \, \hat\g_{*'} \eta(\gamma_{*',*}) \hat\g_{*}\right)U_\theta(\tilde\Gamma';\hat\g_{*'})W_q(\gamma; \hat\g_*) \\
&= \sum_{\hat\g_*, \hat\g_{*'} = \pm 1}\left[ \cos(q\theta) + i\sin(q\theta)\, \hat\g_{*'}\eta(\gamma_{*',*})\hat\g_{*} \right] U_\theta(\tilde\Gamma';\hat\g_{*'})W_q(\gamma; \hat\g_*) \\
&= \cos(q\theta)U_\theta(\tilde\Gamma')W_q(\gamma) + i \sin(q\theta)\widetilde{U}_\theta(\tilde\Gamma';*')\eta(\gamma_{*',*})\widetilde{W}_q(\gamma;*)\,.
\end{split}
\end{equation}
Here $\tilde\Gamma'$ is the final configuration of the Gukov-Witten operator and $\gamma_{*',*}$ is a path connecting the basepoint $*'$ of the Gukov-Witten operator to the basepoint $*$ of the Wilson loop. The final expression consists of two terms --- the first reproduces the action in Eq.~\eqref{eq:contractible_line_action} while the second involves twisted-sector operators 
\begin{equation}
\widetilde{U}_\theta(\tilde\Gamma';*') = \sum_{\hat\g_{*'}=\pm1} \hat\g_{*'}\, U_\theta(\tilde\Gamma'; \hat\g_{*'}) \,, \ \widetilde{W}_q(\gamma;*) = \sum_{\hat\g_*=\pm1}\hat\g_*\,  W_q(\gamma; \hat\g_*)
\end{equation}
connected by a dual symmetry line $\eta$. This is shown in Fig.~\ref{fig:1-form_action}.\footnote{Notably, the fact that locally pinching $U_\theta$ produces $1+\eta$ but not $U_{2\theta}$ as one might have expected from the fusion rule \eqref{eq:electric_fusion} is consistent with the vanishing tadpole condition discussed in e.g. Ref.~\cite{Chang:2018iay}.}

\subsubsection{Three-Loop Braiding}
\label{sec:3_loop} 

\begin{figure}[h!] 
   \centering
   \includegraphics[width=.85\textwidth]{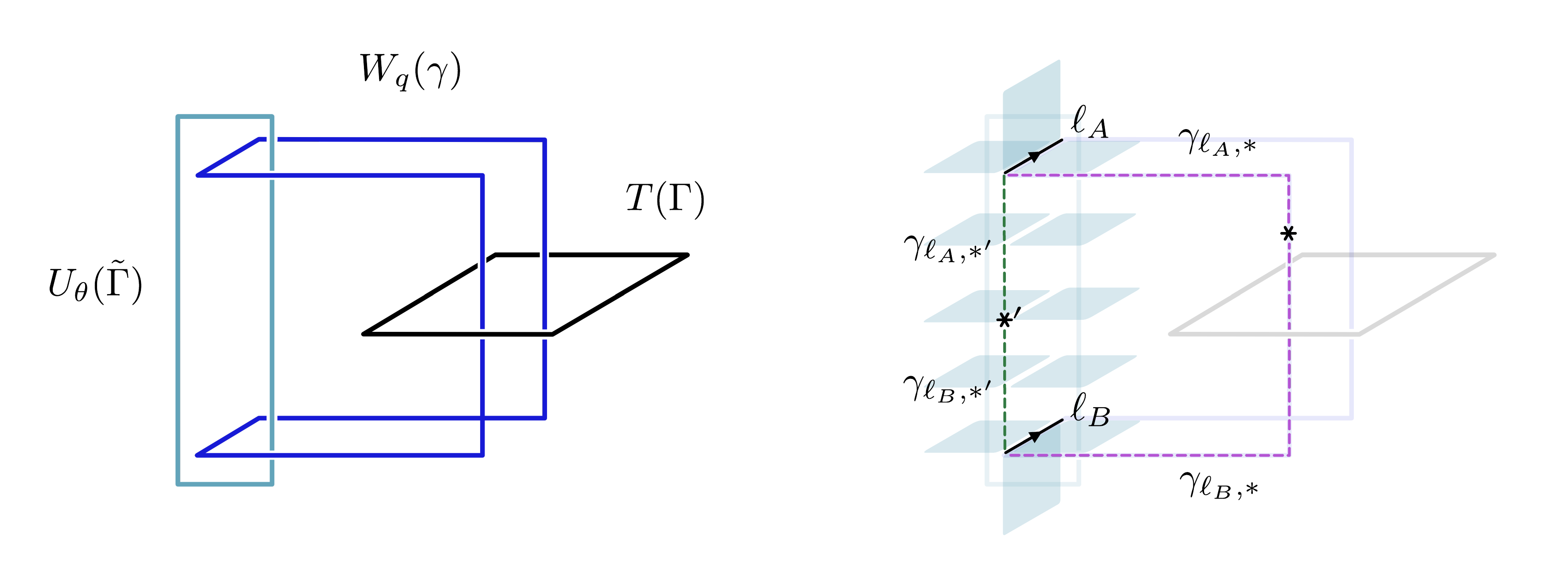} 
   \caption{A Borromean ring configuration involving a Wilson loop $W_q$, a twist vortex $T$, and a Gukov-Witten operator $U_\theta$. On the right, we show the scaffolding paths from the basepoints of the Wilson loop ($*$) and Gukov-Witten operator ($*'$) to the links $\ell_A, \ell_B$. The union of these paths forms a closed loop which links non-trivially with the twist vortex.  }
   \label{fig:3_loops}
\end{figure}

As an application, we consider the three-loop braiding discussed in Refs.~\cite{Bucher:1991bc,Alford:1992yx}, limiting ourselves to 3d for simplicity. We consider the configuration shown in Fig.~\ref{fig:3_loops} consisting of a Wilson loop $W_q(\gamma)$, a Gukov-Witten operator $U_\theta(\tilde\Gamma)$ for the conjugacy class $\theta$, and a twist vortex $T(\Gamma)$ (i.e. a Gukov-Witten operator for the conjugacy class of reflections). The loops form a Borromean ring configuration where no two loops are linked with each other but no loop can be topologically shrunk to a point without crossings. 

\smallskip
We can remove $U_\theta(\tilde \Gamma)$ by shifting the link variables in $\star[\tilde\Sigma]$ where $\partial\tilde\Sigma = \tilde\Gamma$. In particular, we must perform field redefinitions of links $\ell_A$ and $\ell_B$ which lie along the Wilson loop, as indicated in Fig.~\ref{fig:3_loops}. Specifically, we shift $a_{\ell_A} \to a_{\ell_A} + \hat\g_{*'}\eta(\gamma_{\ell_A,*'})\theta$ and $a_{\ell_B} \to a_{\ell_B} + \hat\g_{*'}\eta(\gamma_{\ell_B,*'})\theta$, where $\hat\g_{*'}$ is the basepoint degree of freedom associated to $U_\theta$. Performing these shifts results in
\begin{equation}
\begin{split}
 U_\theta(\tilde\Gamma)T(\Gamma)W_q(\gamma)  &=  T(\Gamma) \sum_{\hat\g_*,\hat\g_{*'} = \pm 1} W_q(\gamma;\hat\g_*)\, e^{ i q \hat \g_* \eta(\gamma_{\ell_A,*})\eta(\gamma_{\ell_A,*'}) \hat\g_{*'} \theta }e^{ -i q \hat \g_* \eta(\gamma_{\ell_B,*})\eta(\gamma_{\ell_B,*'}) \hat\g_{*'} \theta } \\
&=  T(\Gamma) \sum_{\hat\g_*,\hat\g_{*'} = \pm 1} W_q(\gamma;\hat\g_*)\, e^{ i q \hat \g_* \left[\eta(\gamma_{*,*'}^A)- \eta(\gamma_{*,*'}^B)\right] \hat\g_{*'} \theta } \,,
\end{split}
\end{equation}
where $\gamma_{*,*'}^{A/B} = \gamma_{\ell_{A/B},*} \circ \gamma_{\ell_{A/B},*'}$ are two paths from $*$ to $*'$. Crucially, the difference between these two paths links non-trivially with $\Gamma^\vee$, so that $\eta(\gamma^A_{*,*'}) = - \eta(\gamma^B_{*,*'})$. As a result, when we sum over $\hat\g_{*'}$ we get a non-trivial factor that depends on $q$ and $\theta$,
\begin{equation}
\langle U_\theta(\tilde\Gamma)T(\Gamma)W_q(\gamma)\rangle = 2\cos(2q\theta)\, \langle T(\Gamma) W_q(\gamma) \rangle = \cos(2q\theta) \langle U_\theta(\tilde\Gamma)\rangle\langle T(\Gamma) W_q(\gamma) \rangle  . 
\end{equation}

\subsubsection{Selection Rules}
\label{sec:selection_rules}

Because of its peculiar action, the selection rules arising from the non-invertible electric symmetry are not as strong as those arising from its invertible counterpart. Specifically, selection rules from non-invertible symmetries typically arise only in infinite volume~\cite{Cherman:2022eml}. In the invertible case, we can derive selection rules by wrapping a non-contractible Wilson loop with a codimension-2 surface and removing it in two ways --- shrinking the surface on the loop gives a phase, while expanding the surface to remove it does not. In the non-invertible case, the second step of expanding the surface can only be performed when some spacetime directions are infinite. 

\begin{figure}[h!] 
   \centering
   \includegraphics[width=.7\textwidth]{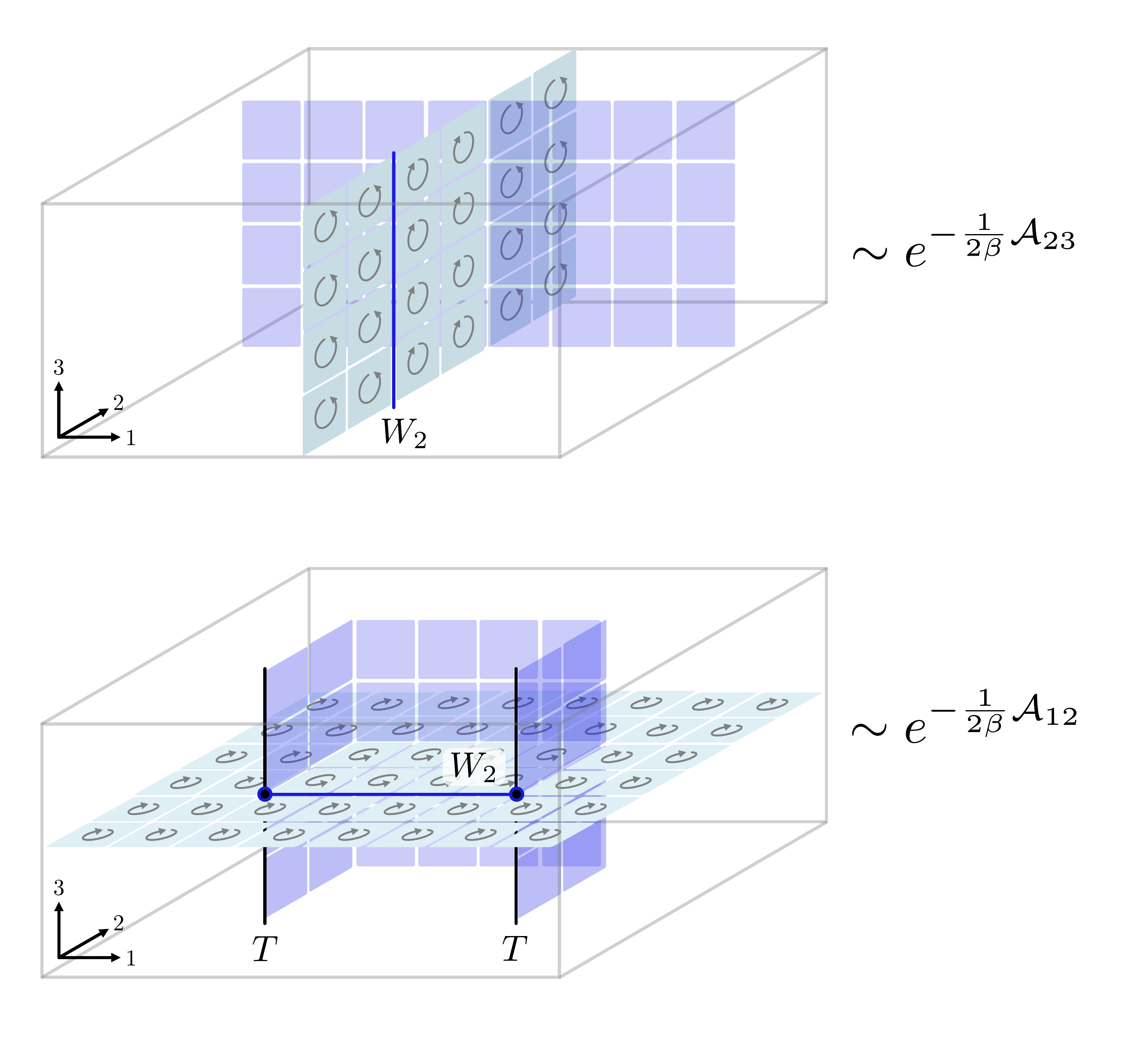} 
   \caption{Leading-order contributions in the strong coupling expansion for the expectation values of a charge-2 Polyakov loop and a charge-2 Wilson line ending on twist vortices. Both results are non-zero but vanish exponentially quickly with the volume. The transparent blue plaquettes indicate $\tilde\Omega^\vee$ while the light teal plaquettes represent the electric flux whose sign is indicated by the direction of the arrows. The electric flux flips sign when it crosses the $\c$ defect and changes by 2 units when it crosses the Wilson loop. Here $\mathcal A_{ij}$ refers to the area of the $ij$ cycle of the torus. }
   \label{fig:strongcoupling}
\end{figure}

\smallskip
To illustrate the idea, we consider two examples. First, take a charge-2 Wilson loop wrapped on a non-contractible cycle, i.e. a charge-2 Polyakov loop. This operator is neutral under $\ZZ_{2,e}^{(1)}$, so based on invertible symmetries alone we might expect it to have an order one expectation value. On the other hand, it is charged under the non-invertible electric symmetry. To give a concrete comparison to the case of a charge-1 Polyakov loop, we work in the strong coupling expansion. Namely, we drop the Lagrange multiplier suppressing monopoles, and apply the dualization procedure from Section~\ref{sec:O2}. Summing over $n$ constrains $z = \tilde n \in \calC^{d-2}(\Lambda,\ZZ)$, which is interpreted as the integer electric flux (more specifically, $\tilde n = \star[ \Sigma^\wedge]$ where $\Sigma$ is a surface of electric flux). We begin by considering a charge-1 Polyakov loop extending in the $\hat 3$ direction --- we denote the spacetime cycle by $\gamma_3$. It is easy to see that on a finite torus there are simply \emph{no} strong coupling diagrams one can write down that contribute to its expectation value, so that
\begin{equation}
\langle W_1(\gamma_3) \rangle = 0\,,
\end{equation}
consistent with unbroken $\ZZ_{2,e}^{(1)}$ symmetry. In contrast, one \emph{can} find contributions to the expectation value of a charge-2 Polyakov loop on the torus, see Fig.~\ref{fig:strongcoupling}. The non-vanishing contributions require activating a charge conjugation defect on a non-contractible cycle. The leading term consists of a sheet of flux spanning the 2-cycle with the smallest area which is parallel to the Polyakov loop --- in this case the $23$ plane with area $\mathcal A_{23} = 4\times 6 = 24$. As a result, we find
\begin{equation}
\langle W_2(\gamma_3) \rangle = 2\, e^{-\frac{1}{2\beta}\mathcal A_{23}} + \mathcal O(e^{-\frac{2}{\beta}})\,.  
\end{equation}
Notably, the result is nonzero, but vanishes exponentially quickly in the spacetime volume. This is consistent with the general remarks we made above. 

\smallskip
As another example, consider a charge-2 Wilson loop ending on a pair of non-contractible twist vortices. This configuration is unconstrained by the invertible 1-form symmetry but transforms non-trivially under the non-invertible symmetry. Again we find that in the strong coupling expansion, the leading contributions decay exponentially quickly in the area $\mathcal A_{12}$ of one of the spacetime 2-cycles, see Fig.~\ref{fig:strongcoupling}.  

\begin{figure}[h!] 
   \centering
   \includegraphics[width=.8\textwidth]{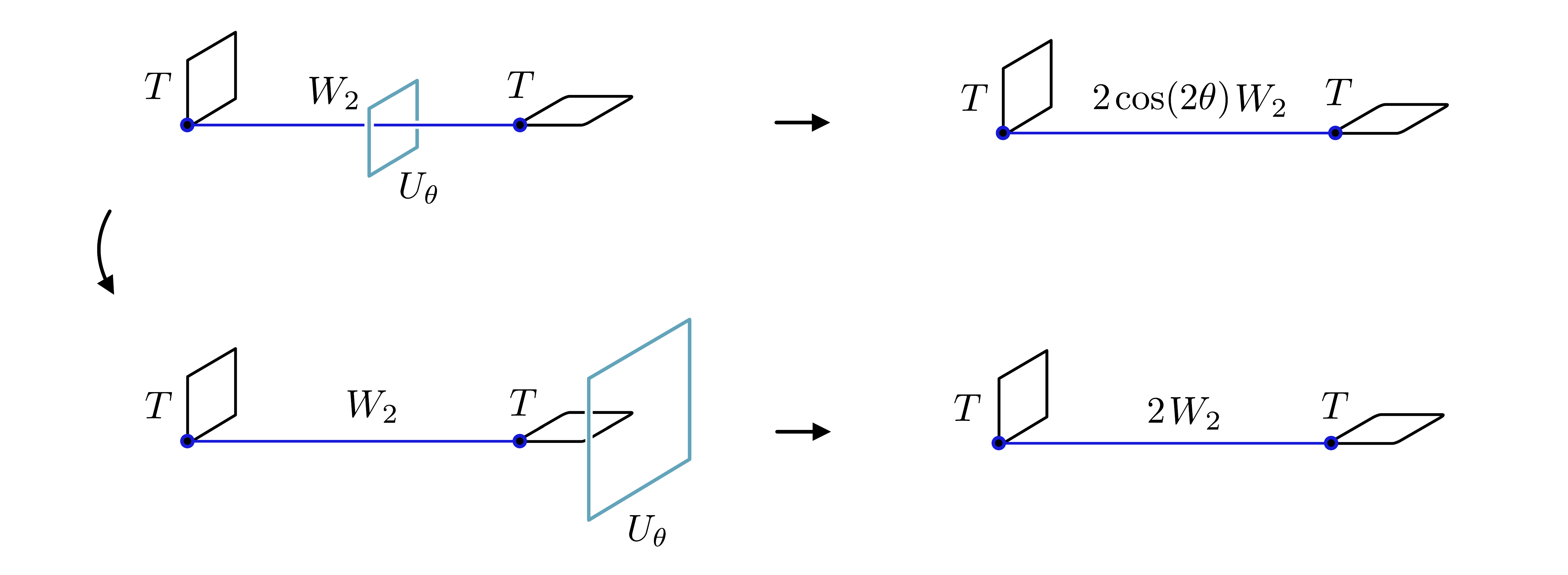} 
   \caption{Using the non-invertible electric symmetry to derive a selection rule for open charge-2 Wilson lines ending on contractible twist vortices.   }
   \label{fig:open_W2}
\end{figure}

\smallskip
Finally, there are examples of selection rules arising from non-invertible symmetries that lead to exact zeros even in finite spacetimes. For instance, while the expectation value of a charge-2 Wilson loop stretched between two \emph{non}-contractible twist vortices only vanishes in the infinite volume limit, a charge-2 Wilson loop stretched between two contractible twist vortices is effectively an open charge-2 Wilson line, and has exactly vanishing correlation functions due to the non-invertible symmetry. The argument is shown in Fig.~\ref{fig:open_W2} --- shrinking a non-invertible electric symmetry operator in two ways leads to the relation
\begin{equation}
\langle U_\theta \, T \, W_2 \, T \rangle = \langle U_\theta\rangle \langle T\, W_2\, T \rangle = \cos(2\theta)\langle U_\theta \rangle \langle T\, W_2\, T \rangle\,,
\end{equation}
and since $\langle U_\theta \rangle = 2$ we have that $\langle T \, W_2 \, T \rangle = 0$. Moreover, since this argument can be repeated in the presence of arbitrary additional operator insertions, all correlation functions of the above operator vanish. This is consistent with the `endability' criteria discussed in Refs.~\cite{Rudelius:2020orz,Cherman:2022eml}, which states that a line charged under a 1-form symmetry, invertible or not, cannot end on local operators. In this case, the putative open line operator simply vanishes.

\subsection{Magnetic Symmetry}
\label{sec:noninvertible_magnetic}

The properties of the non-invertible magnetic symmetry are analogous to the electric symmetry, so we will just briefly outline the main points. The symmetry is generated by the operator we constructed in Section~\ref{sec:generalities},
\begin{equation}  
V_\alpha(S) = P(S) \sum_{\hat\g_* = \pm 1} \exp\left(i \alpha\, \hat\g_* \sum_{p \in S} \eta(\gamma_{p,*})\, n_p \right)\,, 
\end{equation}
where the projector is
\begin{equation}
\begin{split}
P(S) &= \frac{1}{|H^1(S,\ZZ_2)|} \sum_{\gamma\in H^1(S,\ZZ_2)} \eta(\gamma) \\
&= \frac{|H^0(S,\ZZ_2)|}{|H^1(S,\ZZ_2)|}\frac{1}{2^{\#\text{plaq.}+\# \text{sites}}} \sum_{\substack{u \in \calC^0(S, \ZZ_2),\\ b \in \calC^1(S, \ZZ_2)}} \exp\left( i \pi \sum_{p\in S} c\cup b + u \cup db \right)\,.
\end{split}
\end{equation}
Like in the electric case, these symmetry operators have quantum dimension 2: the expectation value of a $V_\alpha(S)$ with $S = \partial \Omega$ equal to the boundary of a ball reduces to the sum over the basepoint degree of freedom, so that 
\begin{equation}
\langle V_\alpha(\partial\Omega) \rangle = 2\, .
\end{equation}
The parallel fusion works as in the electric case, and we find 
\begin{equation}
\begin{split}
V_\alpha(S)V_\beta(S) &= V_{\alpha+\beta}(S) + V_{\alpha-\beta}(S)\, \text{ for } \alpha \not=\beta\,,\\
V_\alpha(S)V_\alpha(S) &= V_{2\alpha}(S) + 2P(S). 
\end{split}
\end{equation}

\subsubsection{Action on Monopole Operators}

In 4d, the action of the magnetic symmetry on 't Hooft lines is completely analogous to the action of the electric symmetry on Wilson lines. The only subtlety is that in our setup both the magnetic symmetry and the 't Hooft line live on the original lattice, so that the correct notion of linking is that the 't Hooft line supported on $\gamma$ links with $S^\wedge$, where $S^\wedge$ is obtained from $S$ by a positive half-lattice-unit translation in each direction. In 3d, the magnetic symmetry is a 0-form symmetry acting on monopole operators $\mathcal M_k(s) = \sum_{\hat\g_s= \pm 1} e^{i k \hat\g_s \sigma_s}$. If we wrap such a local operator with a contractible surface, it is easy to see that
\begin{equation} \label{eq:local_0form_action} 
V_\alpha(S) \mathcal M_k(s)  = P(\gamma)\sum_{\hat\g_s, \hat\g_* = \pm 1}\exp\left(i k \alpha \, \hat\g_* \,\eta(\gamma_{*,s}) \hat\g_s\right)   \exp\left(i k\, \hat\g_s \sigma_s \right) = 2\cos(k\alpha)\, \mathcal M_k(s)\,.
\end{equation}

\begin{figure}[h!] 
   \centering
   \includegraphics[width=\textwidth]{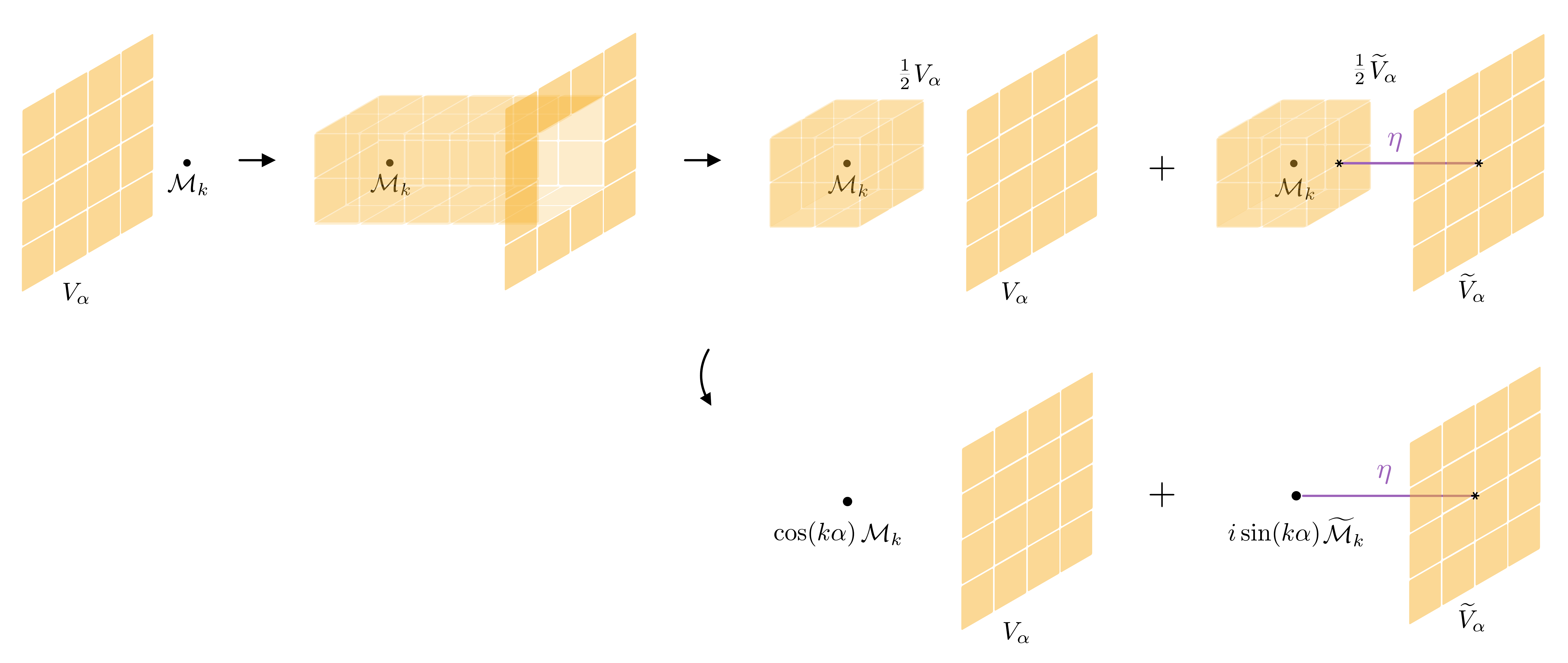} 
   \caption{Action of the non-invertible magnetic 0-form symmetry on a local monopole operator in 3d. }
   \label{fig:0-form_action}
\end{figure}

\smallskip
To study the more general action, we write $V_\alpha(S) = \sum_{\hat\g_*} V_\alpha(S; \hat\g_*)$. Then when a magnetic surface is swept from $S \to S'$ past a local monopole operator, the action is (see Fig.~\ref{fig:0-form_action})
\begin{equation} \label{eq:magnetic_action} 
\begin{split}
V_\alpha(S) \mathcal M_k(s) &= \sum_{\hat\g_s, \hat\g_* = \pm 1}\exp\left(i k \alpha \, \hat\g_* \eta(\gamma_{*,s}) \hat\g_s\right) V_\alpha(S';\hat\g_*)\exp\left(i k\, \hat\g_s \sigma_s \right) \\
&= \sum_{\hat\g_s, \hat\g_* = \pm 1}\left[ \cos(k\alpha) + i \sin(k\alpha)\, \hat\g_* \eta(\gamma_{*,s})\hat\g_s\right] V_\alpha(S';\hat\g_*)\exp\left(i k\, \hat\g_s \sigma_s \right)  \\
&= \cos(k\alpha)V_\alpha(S')\mathcal{M}_k(s) + i \sin(k\alpha) \widetilde{V}_\alpha(S';*)\eta(\gamma_{*,s})\widetilde{\mathcal{M}}_k(s)\,,
\end{split}
\end{equation}
where $\gamma_{*,s}$ is a path from the basepoint of the $V_\alpha$ operator to $s$. As in the electric case, the result consists of two terms, the second of which involves the twisted-sector operators 
\begin{equation}
\widetilde{V}_\alpha(S';*) = \sum_{\hat\g_*=\pm1} \hat\g_*\, V_\alpha(S'; \hat\g_*)\,, \ \widetilde{\mathcal{M}}_k(s) = \sum_{\hat\g_s=\pm1} \hat\g_s\, e^{i k\hat\g_s\sigma_s}\,.  
\end{equation}
To see that the first term in the last line of Eq.~\eqref{eq:magnetic_action} reproduces the local action in Eq.~\eqref{eq:local_0form_action}, we simply note that if $S$ (and therefore $S'$) are contractible, shrinking $S'$ to a point gives $V_\alpha(S') = 2$ while $\widetilde{V}_\alpha(S';*) = 0$.

\section{Phase Diagram with Monopoles and Vortices}
\label{sec:phases} 

So far we have focused on $O(2)$ gauge theory with monopole ($dn \not =0$) and twist vortex ($dc \not = 0 \text{ mod }2$) configurations suppressed. We now consider deformations away from this limit. Relaxing the no-monopole constraint is straightforward: we simply drop the Lagrange multiplier term involving the magnetic gauge field.\footnote{Alternatively, we can introduce magnetic Higgs fields in 4d or a suitable potential for the dual photon in 3d and achieve the same effect. } Introducing dynamical twist vortices is more subtle --- even in the absence of higher-group symmetry we have to include degrees of freedom on the twist vortices in order to render them gauge-invariant. We emphasize that at the lattice level we are free to choose what these degrees of freedom are, and the choice is not unique. Here we consider a minimal scenario consisting of a single Villain scalar $\varphi$ with charge $q_\varphi$, such that the full action is
\begin{multline} 
\label{eq:magnetic_matter} 
S = \frac{\beta}{2} \sum_{\text{plaq.} \in \Lambda} (\dc(a - \frac{1}{q_\varphi}\dc\varphi) -2\pi n)^2 + \frac{\kappa}{2}\sum_{\text{plaq.} \in \Lambda} (dc - 2h)^2 \\
+ \frac{\kappa_v}{2}\sum_{\text{links} \in \partial\Omega} (\dc\varphi- q_\varphi a - 2\pi w)^2 \,, 
\end{multline}
where $h \in \calC^2(\Lambda, \ZZ)$. The $\kappa$ term is a Villain-type kinetic term for a non-topological $\ZZ_2$ gauge theory and depends explicitly on the value of $dc$ --- unlike the action \eqref{eq:O2_action} which only depends on $\overline{dc}$. Now it becomes important that the reflection gauge transformations act as $c \to c + \overline{dg}$, so that $dc \to dc + d\,\overline{dg}$. Since $d\,\overline{dg}$ vanishes modulo 2, we can make the action gauge invariant by assigning $h \to h + \frac{1}{2}d\,\overline{dg}$.\footnote{Alternatively, we work with an integer lift $c \in \calC^1(\Lambda,\ZZ)$ with gauge redundancies $c \to c + dg + 2l$ with $g \in \calC^0(\Lambda,\ZZ)$ and $l\in \calC^1(\Lambda,\ZZ)$, and assign $h \to h + dl$. } The remaining gauge redundancies are
\begin{equation}
\begin{split}
a &\to a + \dc\lambda + 2\pi m \,, \ n \to n + \dc m  - \frac{1}{q_\varphi} \c \cup \c  \cup r\,, \\
\varphi &\to \varphi + 2\pi r + q_\varphi\lambda\,, \ w \to w + \dc r - q_\varphi m \,. 
\end{split}
\end{equation}
Recall that for the above shift of $n$ to be well-defined, we must take $|q_\varphi|= 1$ or $2$. The value of $q_\varphi$ is a choice which is made once and for all, and is part of the definition of our theory with dynamical twist vortices. Note that the action is manifestly real and positive definite and would be an interesting target for Monte Carlo simulations. 

\smallskip
For any finite value of $\kappa$ the $\ZZ_{2,v}^{(2)}$ symmetry is explicitly broken, which means that the non-invertible $O(2)/\ZZ_2$ electric 1-form symmetry is also explicitly broken to its invertible $\ZZ_{|q_\varphi|}$ subgroup. Since we have dynamical unit-charge magnetic monopoles, there is also no magnetic 1-form symmetry. Therefore, the only exact symmetry of the model is $\ZZ_{|q_\varphi|}$, which is non-trivial only if we choose $q_\varphi = \pm 2$. The explicitly broken symmetries can however re-emerge in various limiting regions in parameter space. There are two extreme limits where the model simplifies and its behavior can be straightforwardly analyzed:

\begin{figure}[h!] 
   \centering
   \includegraphics[width=.65\textwidth]{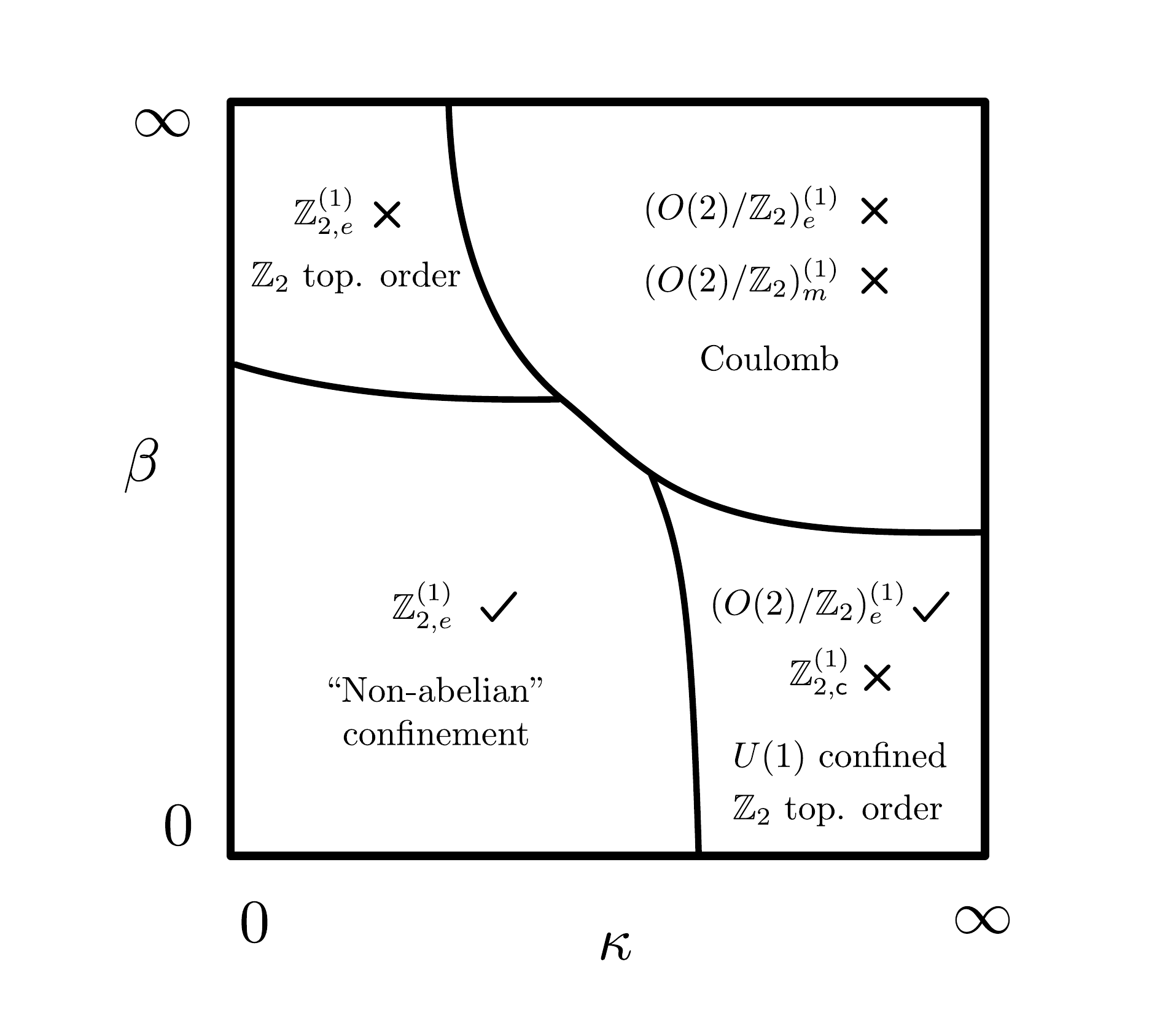} 
   \caption{Possible phase diagram for $O(2)$ gauge theory in four dimensions with dynamical monopoles and twist vortices. We have chosen the dynamical twist vortices to carry charge-2 electric degrees of freedom so that $\ZZ_{2,e}^{(1)}$ is an exact symmetry everywhere in the phase diagram. As $\beta \to 0$ there is an emergent $\ZZ_{2,c}^{(1)}$ 1-form symmetry which is spontaneously broken for large $\kappa$. As $\kappa \to \infty$ there is an emergent $\ZZ_{2,v}^{(2)}$ due to the absence of twist vortices, which in turn enhances $\ZZ_{2,e}^{(1)}$ to the non-invertible $(O(2)/\ZZ_2)^{(1)}_e$ electric symmetry. When both $\kappa$ and $\beta$ are large there is a gapless Coulomb phase with an emergent, and spontaneously broken, $(O(2)/\ZZ_2)^{(1)}_m$ magnetic symmetry. We have assumed that the twist vortex-localized hopping parameter $\kappa_v$ is sufficiently large such that proliferating twist vortices enables the condensation of charge-2 electric particles in the Higgs phase in the upper left hand corner.   }
   \label{fig:phases}
\end{figure}

\begin{itemize}

\item \underline{Weak $\c$ coupling limit ($\kappa = \infty$)}: This limit corresponds to the flat gauging of charge conjugation with $dc = 0$ mod $2$ where twist vortices are non-dynamical (so that there is an exact $\ZZ_{2,v}^{(2)}$ dual symmetry), and the $\c$-gauging does not introduce any local dynamics. Therefore the phase diagram in this limit is the same as that of the standard Villain $U(1)$ gauge theory, with a first order confinement-to-Coulomb transition as we increase $\beta$~\cite{PhysRevD.21.2291,Arnold:2002jk}. In the $O(2)$ context we should view the Coulomb phase at large $\beta, \kappa$ as a phase where the non-invertible electric and (emergent) magnetic symmetries are spontaneously broken. When $\kappa$ is large but finite, the $\ZZ_{2,v}^{(2)}$ symmetry is approximate and emerges at long distances.\footnote{This assumes that $\ZZ_{2,v}^{(2)}$ is spontaneously broken at $\kappa =\infty$, so that twist vortices are not confined, see e.g. Ref.~\cite{Cherman:2023xok} for a discussion of related issues. This is however expected because in the ungauged theory, charge conjugation symmetry is \emph{not} spontaneously broken in the 4d $U(1)$ lattice gauge theory. }

\item \underline{Strong $U(1)$ coupling limit ($\beta = 0$)}: In this limit we drop the first term in Eq.~\eqref{eq:magnetic_matter} and integrate out the $U(1)$ part of the gauge field --- this removes the last term in Eq.~\eqref{eq:magnetic_matter} in favor of an overall $\kappa_v$-dependent constant. Hence the model reduces to a pure, non-topological $\ZZ_2$ gauge theory with exact $\ZZ_{2,c}^{(1)}$ 1-form symmetry acting on the $\c$ gauge field. In four dimensions pure $\ZZ_2$ gauge theory on the lattice has a first-order confinement-to-deconfinement transition as we increase $\kappa$ from zero~\cite{PhysRevLett.42.1390,PhysRevD.20.1915}, where $\ZZ_{2,c}^{(1)}$ is spontaneously broken in the large $\kappa$, topologically ordered phase. From the perspective of the charge conjugation gauge field, turning on $\beta$ corresponds to adding fundamental matter, which does not immediately destroy the transition at $\beta = 0$~\cite{Fradkin:1978dv}. 

\end{itemize}

Finally, there is a third limit where the dynamics should simplify, but where our assumptions about the behavior of the model are more speculative:\footnote{In particular, the effects of broken lattice rotation symmetry are potentially significant in this regime.}

\begin{itemize}

\item \underline{Strong $\ZZ_{2,c}$ coupling limit ($\kappa = 0$)}: In this limit there is no per-unit-area action penalty for twist vortex configurations, and we expect them to proliferate in the vacuum. This in turn liberates the charged worldsheet degrees of freedom which become genuine bulk excitations. Whether or not these charged degrees of freedom themselves condense depends on $\kappa_v$ --- if $\kappa_v$ is small, the charged particles are heavy, while if $\kappa_v$ is large, they are light. If $\kappa_v$ is sufficiently large we expect charged particle worldlines to proliferate in the full 4d spacetime, Higgsing the bulk gauge field to $\ZZ_{|q_\varphi|}$. If $|q_\varphi|= 2$, then as we increase $\beta$ we will again find a first-order confinement-deconfinement transition, while if $|q_\varphi| =1$ the dependence on $\beta$ will be smooth. 

\end{itemize}

In Fig.~\ref{fig:phases} we show a possible phase diagram which is consistent with the expected behavior in the limiting cases discussed above.

\section{Summary and Discussion}
\label{sec:conclusion} 

In this work we described how to gauge charge conjugation symmetry in generic Euclidean lattice field theories, and applied the construction to $O(2)$ gauge theory in three and four spacetime dimensions. The result is a non-abelian version of the modified Villain formulation. The benefit of this approach is that in contrast to a more conventional Wilson formulation, our lattice discretization preserves the higher-group and non-invertible symmetries of the continuum $O(2)$ theory. We presented a recipe for how to define gauge-invariant extended operators in the $O(2)$ theory starting from operators in the $U(1)$ theory --- the construction is inspired by related proposals from the '90s~\cite{Alford:1992yx} and employs the more recent technology of condensation defects~\cite{Else:2017yqj,gaiotto2019condensations,Roumpedakis:2022aik}. We used this method to explicitly construct Wilson and 't Hooft lines as well as the non-invertible symmetry operators under which they are charged. We studied various implications of the generalized symmetries at the lattice level, including their exact and approximate selection rules, constraints on worldvolume degrees of freedom of extended operators, and impact on the phase diagram upon including dynamical magnetic monopoles and twist vortices.

\smallskip
There are a number of worthwhile extensions of our construction. The methods used in this paper can be extended to construct other disconnected, non-abelian lattice gauge theories such as $U(1)^{N-1}\rtimes S_N$~\cite{Nguyen:2021yld,Antinucci:2022eat}, and to the gauging of non-normal subgroups of global symmetries in lattice systems more generally. By adding matter fields with electric charge $N$ we can Higgs our $O(2)$ gauge theory down to discrete, dihedral $D_{2N}$ gauge theories. The resulting phase diagram with electric and magnetic matter will be quite rich, with phases labelled by the realization of non-invertible $\text{Rep}(D_{2N})$ and emergent $O(2)/\ZZ_2$ symmetries. The lattice theories we constructed can in principle be simulated without a sign problem, and it would be particularly interesting to numerically study dynamical twist vortices and the effects of their worldvolume degrees of freedom. 

\smallskip 
Another natural next step would be to generalize existing lattice discretizations of $U(1)$ Chern-Simons theories~\cite{Chen:2019mjw,Jacobson:2023cmr,Jacobson:2024hov} to the non-abelian $O(2)$ case~\cite{Barkeshli:2009fu,Cordova:2017vab,Hsin:2024aqb}. In that context, the electric Wilson lines become topological and the braiding relations between non-abelian anyons will involve the three-loop braiding discussed in Section~\ref{sec:3_loop}. This in turn is related to studying the 't Hooft anomalies of non-invertible symmetries, see e.g. Refs.~\cite{Chang:2018iay,Thorngren:2019iar,Kaidi:2023maf,Zhang:2023wlu,Choi:2023xjw,Cordova:2023bja,Antinucci:2023ezl,Cordova:2023jip,Hsin:2024aqb}. Since the modified Villain formulation of $U(1)$ gauge theory preserves the mixed 't Hooft anomaly between the invertible electric and magnetic symmetries, it is natural to expect that one could study the putative mixed anomaly between their non-invertible counterparts in $O(2)$ gauge theory. Our lattice construction may also help clarify the notion of a background gauge field for a non-invertible symmetry. We also note that we studied the action of the non-invertible coset symmetries on the most natural genuine local and line operators, but there is a zoo of correlation functions that would be interesting to explore further, such as the action on twist vortices and their junctions with Wilson and 't Hooft lines, and on non-genuine operators in the twisted sector. Similarly, it would be interesting to study the impact of Dijkgraaf-Witten (in 3d) and theta (in 4d) terms on the physics of extended operators --- for instance, the twist vortex will support dyonic degrees of freedom in the presence of a theta term~\cite{Bucher:1991qhl}.

\smallskip
Finally, while we presented a minimal non-abelian generalization of the Villain formulation, the techniques used in this paper are limited to disconnected gauge groups which can be obtained by gauging outer automorphisms of connected gauge groups. The question of how to construct a Villain-type formulation of generic non-abelian gauge theories remains an open and interesting problem (however see Ref.~\cite{Chen:2024ddr} for a recent proposal involving ideas from higher category theory). 

\vspace{1cm}
\noindent\textbf{Acknowledgments} \\

\noindent The author is grateful to M. Bullimore, N. Iqbal, M. Nguyen, T. Rudelius, T. Sulejmanpasic, Z. Sun, and R. Thorngren for useful suggestions and discussions, and to A. Cherman for comments on a draft. T.J. is supported by a Julian Schwinger Fellowship from the Mani L. Bhaumik Institute for Theoretical Physics at UCLA, and is grateful to Durham University and the Simons Center for Geometry and Physics for hospitality during the completion of this work.

\appendix

\section{Twisted Differentials and Cup Products}
\label{app:twisted} 

\subsection{Twisted Differentials} 

It is convenient to label an $i$-chain using a root site $s$ and $i$ coordinates given in increasing order. 

\begin{figure}[h!] 
   \centering
   \includegraphics[width=.95\textwidth]{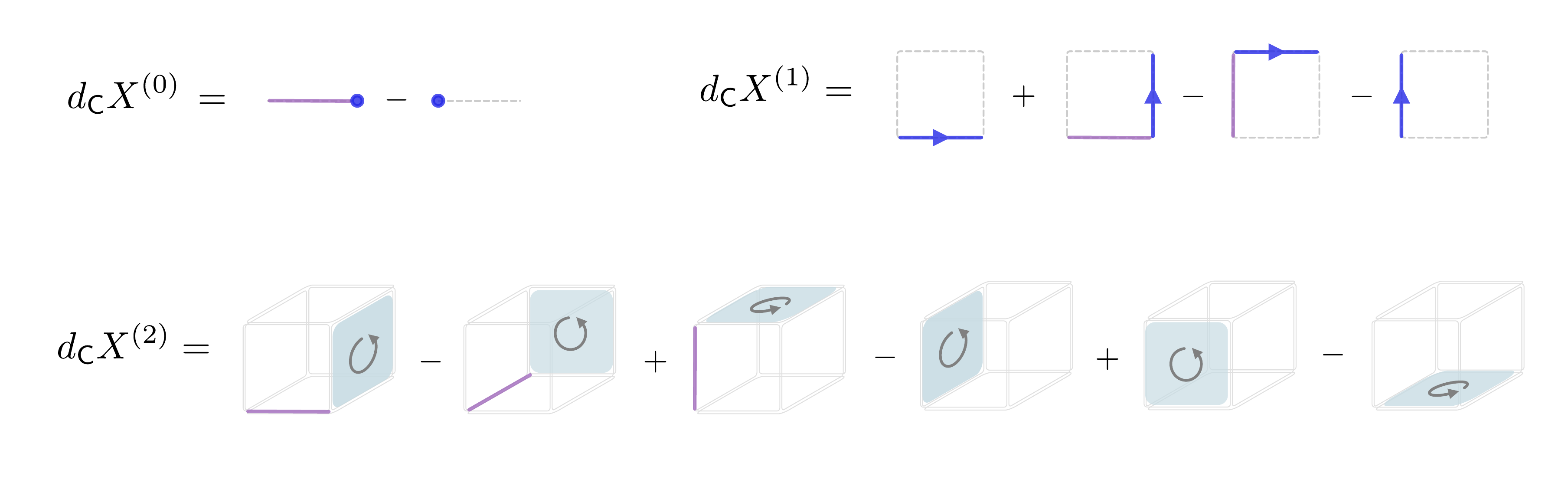} 
   \caption{Examples of twisted, or $\C$-covariant, exterior derivatives on the lattice. The purple lines represent the charge conjugation link field $\C$ --- it should be visually clear that each term transforms covariantly. }
   \label{fig:twisted_d}
\end{figure}

The ordinary differential acting on an $i$-cochain $X$ is defined as
\begin{equation}
(dX)_{s,\, j_1 \ldots j_{i+1}}  = (-1)^{i}\, \sum_{\substack{k_l \in \{j_1,\ldots,j_{i+1}\}, \\ k_1<\cdots< k_i}} \epsilon_{k_1\ldots k_{i+1}} (X_{s+\hat k_{i+1}, k_1\ldots k_i} - X_{s, k_1\ldots k_i})\, . 
\end{equation}
If $X$ is $\C$ odd, the two sets of terms on the right hand side transform under local charge conjugation gauge transformations at different points. The $\C$-covariant, or $\C$-twisted differential is defined by 
\begin{equation} 
(\dC X)_{s,\, j_1 \ldots j_{i+1}}  = (-1)^{i}\,  \sum_{\substack{k_l \in \{j_1,\ldots,j_{i+1}\}, \\ k_1<\cdots< k_i}} \epsilon_{k_1\ldots k_{i+1}} (\C_{s,k_{i+1}} X_{s+\hat k_{i+1}, k_1\ldots k_i} - X_{s, k_1\ldots k_i})\, . 
\end{equation}
It is easy to see from this definition that every term on the right hand side transforms at the same point $s$, i.e. covariantly. See Fig.~\ref{fig:twisted_d} for spelled out examples in three dimensions. 

\smallskip
The twisted differential has a compact expression in terms of cup products, 
\begin{equation} \label{eq:twisted_differential_cup} 
\dC X = dX + (\C-\1)\cup X\,, 
\end{equation}
where here $\1$ is the 1-cochain equal to $+1$ for every positively oriented link. More generally we will define $\1$ to be the $i$-cochain with this same property, with the degree of the cochain implicit. Note that $\1$ is flat, $d\1 = 0$, and the 1-cochain $\1$ has the property that 
\begin{equation}\label{eq:derivative}
\1 \cup X -(-1)^i\, X\cup \1 = dX
\end{equation}
for any $i$-cochain $X$. We can then verify that Eq.~\eqref{eq:twisted_differential_cup} defines a suitable covariant derivative by examining the gauge variation 
\begin{equation}
\begin{split}
\dC X \to d_{\G\cup \C \cup \G}(\G \cup X) &= d(\G \cup X) + (\G\cup\C\cup \G-\1)\cup \G\cup X \\
&= d\G \cup X + \G \cup dX + \G \cup \C \cup X - \1 \cup \G \cup X \\
&= \G \cup dX + \G \cup \C \cup X - \G \cup \1 \cup X = \G \cup \dC X \,,
\end{split}
\end{equation}
where we used associativity of the cup product, the fact that $\G \cup \G = \1$, and Eq.~\eqref{eq:derivative}. We note that the following formula,
\begin{equation}
(\C-\1)\cup(\C-\1) = \C \cup \C - (\1 \cup \C + \C \cup \1) + \1 \cup \1 = \C \cup \C -d\C\,,
\end{equation}
which was established using \eqref{eq:derivative} and the fact that $\1\cup\1 = 0$, can be used to show that the twisted differential satisfies
\begin{equation}
\begin{split}
\dC^2X = d^2X + d(\C-\1) \cup X- (\C-\1) \cup dX + (\C-\1) \cup dX + (\C-\1)\cup (\C-\1)\cup X \\
= (\C \cup \C) \cup  X \,.
\end{split}
\end{equation}

\subsection{Twisted Cup Products} 

\begin{figure}[h!] 
   \centering
   \includegraphics[width=.95\textwidth]{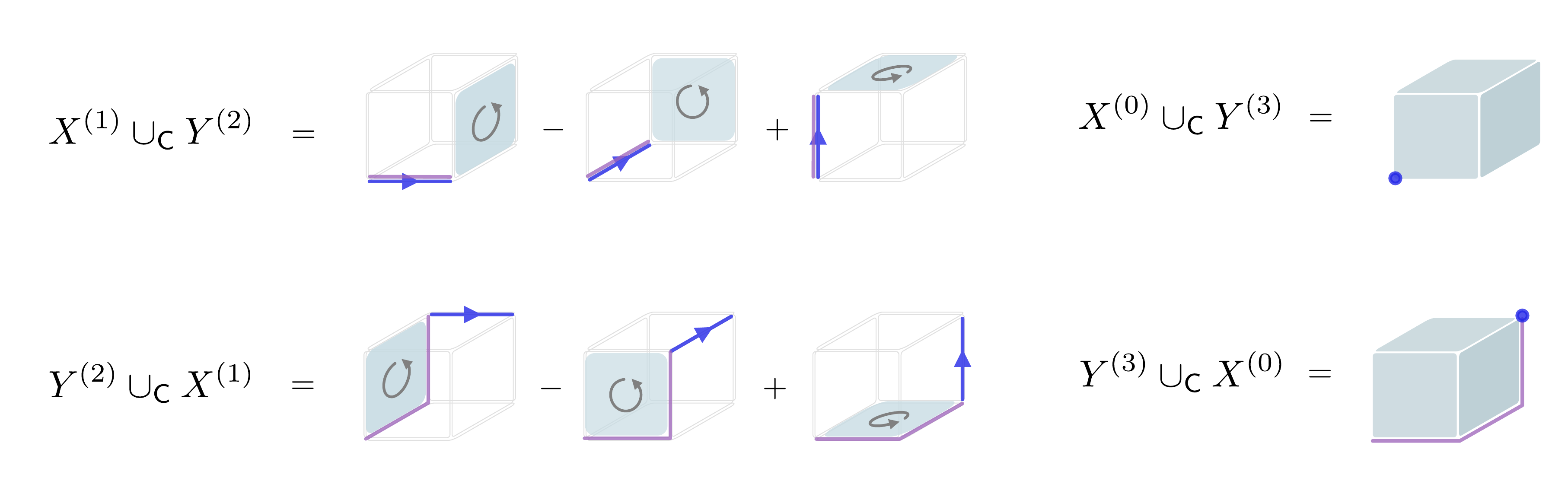} 
   \caption{Examples of twisted, cup products on the cubic lattice between two $\C$-odd cochains. The purple lines represent the charge conjugation link field $\C$ --- it should be visually clear that each term is $\C$-invariant. }
   \label{fig:twisted_cup}
\end{figure}

The ordinary cup product between an $p$-cochain $X$ and $q$-cochain $Y$ is 
\begin{equation}
(X \cup Y)_{s,j_1 \ldots j_{p+q}} =  \sum_{\substack{k_l \in \{j_1, \ldots, j_{p+q}\}, \\ k_1< \cdots < k_p, \\ k_{p+1} < \cdots < k_{p+q}}}\epsilon_{k_1 \ldots k_{p+q}}\,  X_{s,k_1\ldots k_p} \, Y_{s+\hat k_1 + \cdots +\hat k_p, k_{p+1}\ldots k_{p+q}}\,.
\end{equation}
If both $X$ and $Y$ are $\C$ odd, this combination is not $\C$-invariant since $X$ and $Y$ on the right hand side transform at different sites. To remedy this, we connect the root site of $Y$ to the root site of $X$ with a $\C$ Wilson line, 
\begin{align}
(X \cup_\C Y)_{s,j_1 \ldots j_{p+q}} = \sum_{\substack{k_l \in \{j_1, \ldots, j_{p+q}\}, \\ k_1< \cdots < k_p, \\ k_{p+1} < \cdots < k_{p+q}}}  \epsilon_{k_1 \ldots k_{p+q}}\, & X_{s,k_1\ldots k_p} \,\C_{s,k_1}\C_{s+\hat k_1,k_2}\cdots \\
& \C_{s+\hat k_1 +\cdots +\hat k_{p-1},k_p}\,  Y_{s+\hat k_1 + \cdots +\hat k_p, k_{p+1}\ldots k_{p+q}}\,. \nonumber
\end{align}
In order to write down an action for $O(2)$ gauge theory in 3d and 4d, we in particular need the following twisted cup product between a 3-cochain and 0-cochain, 
\begin{equation}
(\dC n \cup_\C \sigma)_{x,123} = (\dC n)_{x,123}\, \C_{x,1}\C_{x+\hat 1,2}\C_{x+\hat 1+\hat 2,3}\, \sigma_{x+\hat 1+ \hat 2 + \hat 3},
\end{equation}
and the twisted cup product between a 3-cochain and 1-cochain,
\begin{multline}
(\dC n \cup_\C \tilde a)_{x,1234} = (\dC n)_{x,123} \, \C_{x,1}\C_{x+\hat 1,2}\C_{x+\hat 1 + \hat 2,3} \, \tilde a_{x+\hat 1 +\hat 2 +\hat 3,4} \\
- (\dC n)_{x,124} \, \C_{x,1}\C_{x+\hat 1,2}\C_{x+\hat 1 + \hat 2,4} \, \tilde a_{x+\hat 1 +\hat 2 +\hat 4,3} \\
-(\dC n)_{x,234} \, \C_{x,2}\C_{x+\hat 2,3}\C_{x+\hat 2 + \hat 3,4} \, \tilde a_{x+\hat 2 +\hat 3 +\hat 4,1}\\
+(\dC n)_{x,134} \, \C_{x,1}\C_{x+\hat 1,3}\C_{x+\hat 1 + \hat 3,4} \, \tilde a_{x+\hat 1 +\hat 3 +\hat 4,1}\,.
\end{multline}
See Fig.~\ref{fig:twisted_cup} for more examples in 3d. 

\subsection{Cup Product Identities}

The (untwisted) cup product obeys 
\begin{equation}
X \cup Y- (-1)^{pq} Y \cup X = (-1)^{p+q+1}\left[ d(X \cup_1 Y) - dX \cup_1 Y - (-1)^p \, X \cup_1 dY \right]\,. 
\end{equation}
Higher cup products on the hypercubic lattice were defined in Refs.~\cite{Chen:2021ppt,Jacobson:2023cmr}. They obey a generalization of the above equation,
\begin{equation}
X \cup_i Y- (-1)^{pq+i} Y \cup_i X = (-1)^{p+q+1+i}\left[ d(X \cup_{i+1} Y) - dX \cup_{i+1} Y - (-1)^p \, X \cup_{i+1} dY \right]\,. 
\end{equation}
The generalization to twisted higher cup products is straightforward, although we do not need to use them in the current paper. 

\smallskip
The cup product identities give rise to a useful formula involving integer cochains. Let $X \in \calC^{i}(\Lambda,\ZZ)$. Then 
\begin{equation}
\begin{split}
2 X \cup_{i-1} X &= (-1)^i \left[ d(X \cup_i X) - dX \cup_i X - (-1)^i X \cup_i dX \right]\,,\\
 dX \cup_{i+1} dX &= dX \cup_i X - (-1)^i X \cup_i dX\,,
\end{split}
\end{equation}
hence 
\begin{equation}
 d(X\cup_i X)  = 2(-1)^i (X \cup_{i-1} X+ X \cup_i dX) + dX \cup_{i+1}dX \,.
\end{equation}
Now suppose $dX = 0$ mod $2$, so that 
\begin{equation} \label{eq:useful_1} 
d(X\cup_i X) = 2 X \cup_{i-1}X \text{ mod } 4\,.
\end{equation}
We can simplify this equation further by noting that $(X\cup_i X)_{s,j_1\cdots j_i} = (X_{s,j_1\cdots j_i})^2$. Using the fact that $a^2 \text{ mod } 4 = a \text{ mod } 2$ for any integer $a$, we have
\begin{equation}
X \cup_i X \text{ mod } 4 = X \text{ mod } 2. 
\end{equation}
Then we find 
\begin{equation} \label{eq:useful_2}
d\overline{X} = 2 X \cup_{i-1}X \text{ mod } 4, 
\end{equation}
which we used in Eq.~\eqref{eq:applied_useful} in the main text. 

\section{Gauge-Invariance of the $O(2)$ Wilson Line}
\label{app:wilson_line} 

Here we show that the $O(2)$ Wilson loop defined in Eq.~\eqref{eq:gauge_invariant_wilson} and reproduced here for convenience, 
\begin{equation} 
W_q(\gamma) \equiv P(\gamma) \sum_{\hat\g_* = \pm 1} \exp\left(i q\, \hat\g_* \sum_{\ell \in \gamma} \eta(\gamma_{\ell,*})\, a_\ell \right)\,. 
\end{equation}
is gauge-invariant. First, as we described in the main text, it is clearly invariant under $\ZZ_2 \subset O(2)$ transformations. It is less obvious that it is invariant under $U(1) \subset O(2)$ gauge transformations of the form $a \to a + \dc \lambda$. In the main text we showed that the above operator does not depend on the specific choice of scaffolding paths $\gamma_{\ell,*}$. Therefore, we are free to make a particular choice which makes our analysis simpler. To do so we first label a curve $\gamma$ in terms of a basepoint $*$ and an ordered set $(i_1, \ldots,i_{L_{\gamma}})$ where $i_j \in \{\pm 1,\ldots,\pm d\}$ and $L_\gamma$ is the length of $\gamma$. Since $\gamma$ is closed, $\sum_j i_j = 0$ modulo the number of sites $L_1,L_2,\ldots,L_d$ of the torus. A change in basepoint can be absorbed into a cyclic permutation of the indices $i_j$. In terms of this data, an ordinary Wilson loop in $U(1)$ gauge theory can be expressed as the exponential of
\begin{equation}
\sum_{\ell\in\gamma} a_\ell = \sum_{j=1}^L a_{* + \sum_{t=1}^{j-1} \hat i_t, i_j}. 
\end{equation}
In the current context, we choose paths $\gamma_{\ell,*}$ to lie along $\gamma$ such that 
\begin{multline}
\sum_{\ell \in \gamma} \eta(\gamma_{\ell,*})\, a_\ell =  \sum_{j=1}^{L} \left( \prod_{k=1}^{j-1} \c_{*+\sum_{r=1}^{k-1}\hat i_r, i_k}\right)a_{*+\sum_{t=1}^{j-1}\hat i_t, i_j} \\
= a_{*, i_1} + \c_{*,\hat i_1} a_{*+\hat i_1, i_2} + \c_{*,\hat i_1}\c_{*+\hat i_1,i_2} a_{*+\hat i_1+\hat i_2, i_3} +  \\
\cdots+ \c_{*,\hat i_1}\c_{*+\hat i_1,i_2}\cdots\c_{*+\sum_{r=1}^{L-2}\hat i_r,i_{L-1}}a_{* - \hat i_L,i_{L}}\,. 
\end{multline}
The variation of the above expression under $a \to a + \dc\lambda$ is 
\begin{multline}
\sum_{j=1}^{L} \left( \prod_{k=1}^{j-1} \c_{*+\sum_{r=1}^{k-1}\hat i_r, i_k}\right)\left(\c_{*+\sum_{t=1}^{j-1}\hat i_t, i_j} \lambda_{*+\sum_{t=1}^j\hat i_t} - \lambda_{*+\sum_{t=1}^{j-1}\hat i_t}\right) \\
=\sum_{j=1}^{L} \left[ \left( \prod_{k=1}^{j} \c_{*+\sum_{r=1}^{k-1}\hat i_r, i_k}\right)\lambda_{*+\sum_{t=1}^j\hat i_t} - \left( \prod_{k=1}^{j-1} \c_{*+\sum_{r=1}^{k-1}\hat i_r, i_k}\right) \lambda_{*+\sum_{t=1}^{j-1}\hat i_t}\right] \,.
\end{multline}
Now we observe that all terms cancel in the sum over $j$ except two terms, which reduce to
\begin{equation}
\left(\prod_{k=1}^L \c_{*+\sum_{r=1}^{k-1}\hat i_r, i_k}\right) \lambda_{* + \sum_{t=1}^L \hat i_t} - \lambda_{*} = (\eta(\gamma)-1)\lambda_{*} \,. 
\end{equation}
using the fact that the curve is closed. This variation trivializes when exponentiated and multiplied by $P(\gamma)$ as in the definition of the Wilson loop in Eq.~\eqref{eq:gauge_invariant_wilson}, verifying our claim.

\section{Orbifolding the Particle on a Circle}
\label{app:poc} 

In this appendix we discuss the Villain discretization of the particle on a circle with charge conjugation symmetry gauged, in other words a particle moving on $S^1/\ZZ_2$ where the $\ZZ_2$ is a reflection across a fixed axis. We start with the standard particle on a circle. The ingredients in the Villain description are a real scalar $\varphi \in \calC^0(\Gamma, \RR)$, and associated integer Villain field $w \in \calC^1(\Gamma, \ZZ)$, with gauge redundancy
\begin{equation} \label{eq:compactness} 
\varphi \to \varphi + 2\pi r\,, \  w \to w +dr\,,
\end{equation}
with $r \in \calC^0(\Gamma,\ZZ)$. Both fields are odd under charge conjugation, $\varphi \to \G \cup\varphi$ and $w \to \G \cup w$, which we take to be a global symmetry for now. The Villain action in the presence of charge conjugation backgrounds is
\begin{equation}
S_{\text{p.o.c.}}(\C) = \frac{\kappa}{2} \sum_{\text{links}} (\dC \varphi - 2\pi w)^2 + i\theta \sum_{\text{links}} w\,, 
\end{equation}
where $\kappa \ge 0$ and $\theta\sim \theta + 2\pi$ is a theta parameter associated with the winding of the compact scalar (only $\theta = 0,\pi$ respect $\C$). This theory has a generalized mixed anomaly between the $U(1)$ global shift symmetry and the periodicity of the $\theta$ parameter~\cite{Gaiotto:2017yup,Kikuchi:2017pcp,Cordova:2019jnf}. This anomaly persists on the lattice and can be seen explicitly by coupling to a gauge field $A \in \calC^1(\Gamma,\RR)$ and compact scalar $(\Sigma, \tilde N)$ with $\Sigma \in \calC^0(\Gamma,\RR)$ and $\tilde N \in \calC^1(\Gamma,\ZZ)$ and writing 
\begin{multline}\label{eq:poc_bgd} 
S_{\text{p.o.c.}}(\C,A,\Sigma) = \frac{\kappa}{2} \sum_{\text{links}} (\dC \varphi -A- 2\pi w)^2 + i\theta \sum_{\text{links}} w + i \sum_{\text{links}} w \cup_{\C} \Sigma \\
-i \sum_{\text{links}}  \varphi \cup_{\C} (\tilde N - \frac{1}{2\pi}\dC\Sigma)\,, 
\end{multline}
where we take the backgrounds to be $\C$ odd $A \to \G \cup A, \Sigma \to \G \cup \Sigma, \tilde N \to \G \cup \tilde N$ and the remaining background gauge transformations act as 
\begin{equation}
\begin{split}
A &\to A + \dC\Lambda + 2\pi M\,, \ \Sigma \to \Sigma + 2\pi \tilde M\,, \ \tilde N \to \tilde N + \dC \tilde M\,,  \\
\varphi &\to \varphi + \Lambda\,, \ w \to w - M\,,
\end{split}
\end{equation}
where $\Lambda \in \calC^0(\Gamma,\RR), M \in \calC^1(\Gamma,\ZZ), \tilde M \in \calC^0(\Gamma,\ZZ)$. Under this transformation the action shifts by
\begin{equation} \label{eq:poc_anomaly}
\Delta S_{\text{p.o.c.}} = - i \sum_{\text{links}}\Lambda \cup_{\C} (\tilde N -\frac{1}{2\pi}\dC \Sigma)  + M \cup_{\C} \Sigma\,,
\end{equation}
signaling an anomaly. In 3d $O(2)$ gauge theory we can use the orbifolded particle on a circle to cancel the anomaly inflow onto the twist vortex worldline, as discussed in Sec.~\ref{sec:3d_tv}. 

\smallskip
Let us now take $\c$ to be a dynamical gauge field. We can solve this theory exactly by rewriting the action in terms of an auxiliary field $z \in \calC^0(\Gamma,\RR)$, 
\begin{equation}
\frac{1}{2\kappa}\sum_{\text{sites}}  z^2 + i\sum_{\text{links}} z \cupc (\dc \varphi - 2\pi w) + \theta w\,.
\end{equation}
Summing over $w$ sets $-2\pi z + \theta = 2\pi \tilde w$, so the dual action becomes 
\begin{equation}
\tilde S = \frac{1}{2\kappa(2\pi)^2}\sum_{\text{sites}} \left(\tilde w - \frac{\theta}{2\pi}\right)^2 + i\sum_{\text{links}} \dc \tilde w \cupc  \varphi 
\end{equation}
The equation of motion of $\varphi$ sets $d\tilde w = 2c\cup \tilde w$. In other words, $\tilde w$ is a constant except in the presence of a $\c$ defect where it flips sign. On a spacetime circle of length $L$ with periodic boundary conditions there are two terms in the sum over $\c$ --- we either insert or don't insert a single $\c$ defect. The only configuration that contributes to the term with the defect is $\tilde w = 0$, and as a result the partition function is 
\begin{equation}
Z(\kappa,L,\theta) \sim \sum_{\tilde w \in \ZZ_{\ge 0}}e^{-\frac{L}{2(2\pi)^2\kappa} \left(\tilde w-\frac{\theta}{2\pi}\right)^2}\,.
\end{equation}

\section{Orbifolding the Compact Boson}
\label{app:orbifold}

The 2d compact boson has a symmetry- and anomaly-preserving Villain lattice discretization~\cite{Sulejmanpasic:2019ytl,Gorantla:2021svj} whose ingredients are a pair of real scalars $\varphi,\tilde\varphi \in \calC^0(\Sigma, \RR)$, and an integer Villain field $w \in \calC^1(\Sigma, \ZZ)$. Both scalars are compact, with gauge redundancy
\begin{equation} 
\varphi \to \varphi + 2\pi r\,, \  w \to w +\dC r\,, \tilde\varphi \to \tilde\varphi + 2\pi \tilde r\, .
\end{equation}
with $r,\tilde r \in \calC^0(\Sigma,\ZZ)$. We write the action in the presence of $\C$ backgrounds as
\begin{equation}
S_{\text{c.b.}}(\C) = \frac{\kappa}{2} \sum_{\text{links}} (\dC \varphi - 2\pi w)^2 + i \sum_{\text{plaq.}} \dC w \cup_{\C} \tilde\varphi \,, \quad \kappa \ge 0\,. 
\end{equation}
Coupling to background $U(1)$ gauge fields for the momentum and winding symmetries, described by pairs $(A,N)$ and $(\tilde A,\tilde N)$ where $A,\tilde A \in \calC^1(\Sigma, \RR)$ and $N,\tilde N \in \calC^2(\Sigma,\ZZ)$, we have
\begin{multline}
S_{\text{c.b.}}(\C,A,N,\tilde A,\tilde N) = \frac{\kappa}{2} \sum_{\text{links}} (\dC \varphi -A- 2\pi w)^2 + i \sum_{\text{plaq.}} (\dC w+N) \cup_{\C} \tilde\varphi + i \sum_{\text{plaq.}} w \cup_{\C} \tilde A \\
-i \sum_{\text{plaq.}} \varphi \cup_{\C} (\tilde N - \frac{1}{2\pi}\dC \tilde A)\,.
\end{multline}
All fields are $\C$ odd, $(A,N) \to (\G \cup A, \G \cup N)$, $(\tilde A, \tilde N) \to (\G \cup \tilde A, \G \cup \tilde N)$, and 
\begin{equation}
\begin{split}
A &\to A + \dC\Lambda + 2\pi M\,, \ N \to N + \dC M\,, \\
\tilde A &\to \tilde A + \dC\tilde \Lambda + 2\pi \tilde M\,,  \ \tilde N \to \tilde N + \dC \tilde M\,,  \\
\varphi &\to \varphi + \Lambda\,, \ w \to w - M\,, \ \tilde\varphi \to \tilde\varphi + \tilde\Lambda\,,
\end{split}
\end{equation}
where $\Lambda,\tilde\Lambda \in \calC^0(\Gamma,\RR), M,\tilde M \in \calC^1(\Gamma,\ZZ)$. The anomaly is
\begin{equation} \label{eq:cb_anomaly} 
\Delta S_{\text{c.b.}} = -i \sum_{\text{plaq.}} \Lambda \cup_{\C} (\tilde N - \frac{1}{2\pi}\dC \tilde A) -N \cup_{\C} \tilde \Lambda + M\cup_{\C}(\tilde A + \dC\tilde\Lambda)\,. 
\end{equation}
This gauge variation matches the anomaly inflow onto the twist vortex worldsheet in 4d $O(2)$ gauge theory, see  Sec.~\ref{sec:4d_tv}. 

\smallskip
If we now make $\c$ a dynamical field, we have
\begin{equation}
S_{\text{orbifold}} = \frac{\kappa}{2} \sum_{\text{links}} (\dc \varphi - 2\pi w)^2 + i \sum_{\text{plaq.}} \dC w \cup_{\c} \tilde\varphi + i\pi \sum_{\text{plaq.}} dc \cup v\,.
\end{equation}
This breaks the invertible momentum and winding symmetries down to their $\ZZ_2$ subgroups. The analog of the higher-group structure encountered in $O(2)$ gauge theory now becomes a standard group extension, i.e. $\ZZ_{2,m}^{(0)} \times \ZZ_{2,w}^{(0)}$ gets extended to $D_8$. As described in Ref.~\cite{Thorngren:2021yso}, this is captured by the twisted cocycle condition
\begin{equation}
dB_v = B_m \cup B_w
\end{equation}
which is the analog of Eq.~\eqref{eq:higher_group}. Moreover, this $D_8$ symmetry is anomaly-free (as explained in Appendix~A.2 of Ref.~\cite{Thorngren:2021yso}), much like the 2-group in $O(2)$ gauge theory in three dimensions. 

\smallskip
It is amusing to note that the local twist vortex operator $e^{i \pi v}$ is not gauge-invariant. Instead, the action transforms by a shift 
\begin{equation}
i \sum_{\text{plaq.}} \dc^2 r \cupc \tilde \varphi = -2i \sum_{\text{plaq.}} \overline{dc}\cup r \cup \tilde \varphi\,. 
\end{equation}
To deal with this, we can stack a `0d field theory' to the twist vortex which constrains $\tilde\varphi$ to be an integer multiple of $\pi$ at the location of the insertion, 
\begin{equation}
T(s)  = e^{i \pi v_s} \, \delta(\tilde\varphi_s \in \pi \ZZ) = \sum_{h \in\ZZ} e^{i \pi v_s} e^{2i h\tilde\varphi_s}\,. 
\end{equation}

\smallskip
The orbifold at a generic radius also has continuous non-invertible symmetries which are the analogs of the non-invertible electric and magnetic symmetries discussed in the main text. In particular, we can apply the construction from Section~\ref{sec:general_construction} to write down the symmetry operators 
\begin{equation}
\begin{split}
U_\theta(\tilde\gamma) &= P(\tilde\gamma^\vee) \sum_{\hat\g_* = \pm1} \prod_{\ell \in \star[\tilde\gamma]} \exp\left[-\frac{\kappa}{2}(\dc\varphi - \hat\g_* \eta(\gamma_{\ell,*}) \theta \star[\tilde\gamma]-2\pi w)_\ell^2 + \frac{\kappa}{2} (\dc \varphi - 2\pi w)_\ell^2 \right]\,, \\
V_\alpha(\gamma) &= P(\gamma)\sum_{\hat\g_* = \pm1} \exp\left(i\hat\g_*\alpha \sum_{\ell \in\gamma} \eta(\gamma_{\ell,*}) w_\ell \right)\,,
\end{split}
\end{equation}
for the non-invertible momentum and winding symmetries. 

\section{$O(2)$ vs. $Pin^{-}(2)$}
\label{app:pin2} 

\begin{figure}[h!] 
   \centering
   \includegraphics[width=\textwidth]{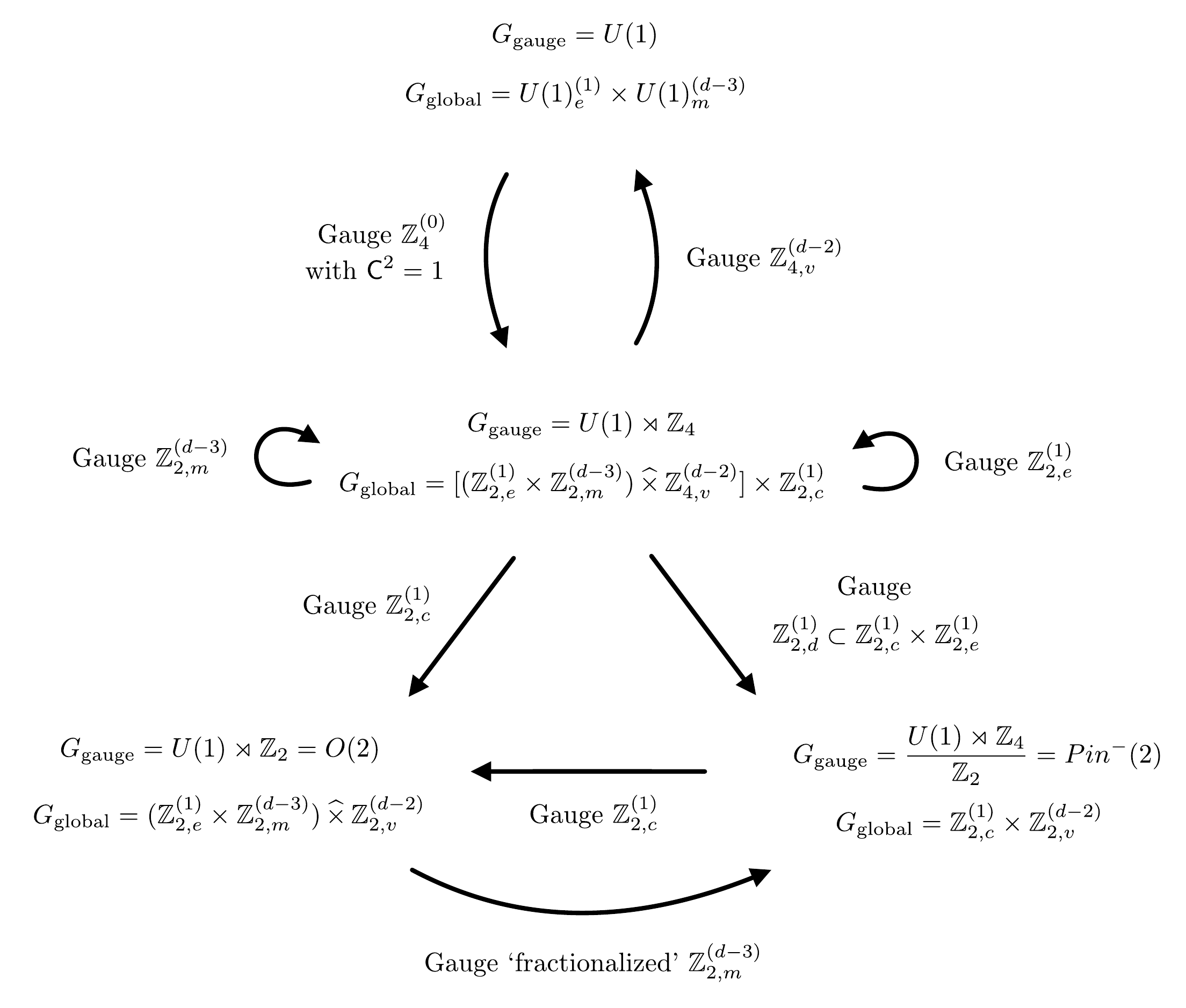} 
   \caption{Interrelations between $U(1), U(1)\rtimes \ZZ_4, O(2),$ and $Pin^{-}(2)$ gauge theories via generalized gauging of discrete higher-form symmetries. The group-like, invertible symmetries are distinct for each choice of gauge group. Note that in general, these equivalence hold only up to a rescaling of the $U(1)$ gauge coupling $\beta$. The notation $\widehat\times$ indicates a higher-group. }
   \label{fig:globalforms}
\end{figure}

In this Appendix we describe the global symmetries of, and relationships between, pure gauge theories with gauge groups $O(2) = U(1)\rtimes \ZZ_2, U(1)\rtimes \ZZ_4$, and $Pin^{-}(2) = \frac{U(1)\rtimes \ZZ_4}{\ZZ_2}$.\footnote{We thank T. Rudelius for helpful discussions regarding the contents of this section.} Each of these theories can be connected by generalized gauging of global symmetries, in a way which is summarized in Fig.~\ref{fig:globalforms}. 

\smallskip
Let us describe first in words how to get $Pin^{-}(2)$ gauge theory, keeping track of the (invertible) global symmetries. We start with $U(1)$ gauge theory enriched with a $\ZZ_4^{(0)}$ 0-form global symmetry whose $\ZZ_2$ subgroup acts trivially. In other words, we have codimension-1 operators $\C$ with $\ZZ_4$ fusion rules, such that $\C$ acts like charge conjugation on the $U(1)$ gauge theory and $\C^2$ has a trivial action. We then proceed to gauge $\ZZ_4^{(0)}$ to obtain $U(1) \rtimes \ZZ_4$ gauge theory. The $U(1)^{(1)}_e$ and $U(1)^{(d-3)}_m$ symmetries of $U(1)$ gauge theory are broken to the $\ZZ_{2,e}^{(1)}$ and $\ZZ_{2,m}^{(d-3)}$ subgroups which commute with the $\ZZ_4$ action. The electric and magnetic symmetries are involved in a non-trivial higher-group with the dual $\ZZ_{4,v}^{(d-2)}$ symmetry obtained from gauging. Furthermore, we have a $\ZZ_{2,c}^{(1)}$ 1-form symmetry from the fact that we gauged a non-effectively acting $\ZZ_2 \subset \ZZ_4$. 

\smallskip
We now contemplate gauging various symmetries. Clearly if we gauge $\ZZ_{4,v}^{(d-2)}$ we return back to the $U(1)$ gauge theory. It is more interesting to consider gauging different subgroups of the $\ZZ_{2,e}^{(1)}\times \ZZ_{2,c}^{(1)}$ 1-form `electric' symmetry.

\begin{enumerate}
\item If we gauge $\ZZ_{2,e}^{(1)}$, this effectively rescales the electric gauge field by $a \to a/2$. This does two things. First, $\ZZ_4$ shifts of the electric gauge field now become $\ZZ_2$ shifts which commute with charge conjugation, so we retain a $\ZZ_{2,e}^{(1)}$ symmetry. Second, we get a dual $\ZZ_2$ $(d-3)$-form symmetry from gauging which acts on the magnetic gauge field. So we end up with the same symmetries we started with --- in other words, up to a rescaling of the gauge coupling, the $U(1)\rtimes \ZZ_4$ theory is self-dual under gauging $\ZZ_{2,e}^{(1)}$.  

\item If we gauge $\ZZ_{2,c}^{(1)}$ we obtain $U(1)\rtimes \ZZ_2 = O(2)$. Indeed, this gauging breaks $\ZZ_{4,v}^{(d-2)} \to \ZZ_{2,v}^{(d-2)}$ due to the mixed anomaly between $\ZZ_{4,v}^{(d-2)}$ and $\ZZ_{2,c}^{(1)}$. There is no dual symmetry from gauging, because $\ZZ_{2,c}^{(1)}$ itself came from gauging a non-effectively-acting symmetry (the would-be charged object under the dual symmetry lives at the boundary of the codimension-2 $\ZZ_{2,c}^{(1)}$ symmetry operator, but no such boundary exists because the symmetry operator is charged under $\ZZ_{2,v}^{(d-2)}$). We are left with the symmetries discussed in detail in the main text. 

\item If we gauge the diagonal $\ZZ_{2,d}^{(1)} \subset \ZZ_{2,e}^{(1)} \times \ZZ_{2,c}^{(1)}$ we obtain $\frac{U(1)\rtimes \ZZ_4}{\ZZ_2} = Pin^{-}(2)$ gauge theory. As in the case above, this breaks $\ZZ_{4,v}^{(d-2)} \to \ZZ_{2,v}^{(d-2)}$, and again there is no dual magnetic symmetry from gauging since there is no gauge-invariant codimension-3 operator for the dual symmetry to act on. Since there is no magnetic symmetry, there is also no higher-group. The invertible symmetry is simply $\ZZ_{2}^{(1)} \times \ZZ_{2,v}^{(d-2)}$. 

\end{enumerate} 

We now turn to the lattice for a concrete realization of the above discussion. We work in 4d, and start with $U(1) \rtimes \ZZ_4$ gauge theory, 
\begin{equation}  \label{eq:U(1)Z_4}
S_{U(1)\rtimes\ZZ_4} = \frac{\beta}{2}\sum_{\text{plaq.}} (d_{\c^2}a - 2\pi n)^2 + i\sum_{\text{hyper-cu.}} (d_{\c^2}n)\cup_{\c^2}\tilde a + \frac{i\pi }{2} \sum_{\text{hyper-cu.}} dc\cup v\,.
\end{equation}
Here $\c = e^{\frac{2\pi i}{4}c} \in \calC^1(\Lambda,\ZZ_4)$ and what appears above is the charge-2 covariant derivative, $d_{\c^2}a = da + (\c^2 -\1)\cup a  = da - 2\overline{c} \cup a$. The $\ZZ_4$ gauge transformations act as
\begin{equation}
a \to \g^2 \cup a\,,\ n \to \g^2 \cup n\, , \ \tilde a \to \g^2 \cup \tilde a\, ,\  c \to c + dg\,,
\end{equation}
where $\g = e^{\frac{2\pi i}{4}g} \in \calC^0(\Lambda,\ZZ_4)$. There are further gauge redundancies 
\begin{equation}
a \to a + d_{\c^2}\lambda + 2\pi m \,,\ n \to n + d_{\c^2}m\,,\ \tilde a \to \tilde a + d_{\c^2}\tilde \lambda + 2\pi \tilde m\,,\ v \to v + dt\,,
\end{equation}
with $\lambda,\tilde \lambda \in \calC^0(\Lambda,\RR)$, $m,\tilde m \in \calC^1(\Lambda,\ZZ)$, and $t \in \calC^1(\Lambda, \ZZ_4)$. This theory has a $\ZZ_{2,e}^{(1)} \times \ZZ_{2,c}^{(1)}$ 1-form electric symmetry, where the factors act as $a \to a + \pi V$ and $c \to c + 2U$. There is also a $\ZZ_{2,m}^{(1)}$ magnetic symmetry taking $a \to a + \pi \tilde V$, and a $\ZZ_{4,v}^{(2)}$ generated by the $\c$ Wilson line $\eta$. As in the $O(2)$ case we have a 3-group captured by the twisted cocycle condition
\begin{equation}
dB_v = 2B_e \cup B_m \text{ mod } 4\,.
\end{equation}

\smallskip
If we gauge $\ZZ_{4,v}^{(2)}$, we add the coupling 
\begin{equation}
\frac{i\pi }{2} \sum_{\text{hyper-cu.}} (c\cup b_v + u \cup db_v) \,, \quad b_v \in \calC^3(\Lambda,\ZZ_4)\,,\ u \in\calC^0(\Lambda,\ZZ_4)
\end{equation}
and summing over $b_v$ sets $c=du$ to be pure gauge, leaving us with $U(1)$ gauge theory. If we gauge $\ZZ_{2,c}^{(1)}$ we effectively replace $\c^2 \to \c$ where $\c$ is a $\ZZ_2$ rather than $\ZZ_4$ gauge field. This lands us on the $O(2)$ theory. If we gauge $\ZZ_{2,e}^{(1)}$, we minimally substitute $n \to n + \frac{1}{2}b_e$ with $b_e \in \calC^1(\Lambda,\ZZ_2)$ constraint to be a flat $\ZZ_2$ gauge field by a Lagrange multiplier $u \in \calC^1(\Lambda,\ZZ_2)$,
\begin{multline}
S_{Pin^-(2)} = \frac{\beta}{2}\sum_{\text{plaq.}}(d_{\c^2}a -\pi b_e- 2\pi n)^2 + i \sum_{\text{hyper-cu.}}\left(d_{\c^2}n + \frac{1}{2}d_{\c^2}b_e \right)\cup_{\c^2}\tilde a \\
+ \frac{i\pi }{2} \sum_{\text{hyper-cu.}}dc\cup v + i \pi \sum_{\text{hyper-cu.}} db_e \cup u\,.
\end{multline}
For the action to be invariant under $\tilde a \to \tilde a + 2\pi \tilde m$, we must assign $u \to u + \tilde m$. This allows us to define a gauge-invariant, `fractional' 't Hooft line which is schematically $e^{\frac{i}{2} \sum_\gamma \tilde a + i \pi \sum_\gamma u}$. This line is the charged object under the dual magnetic symmetry. To compare to our starting point, we can rename $\hat n = b_e+2n$, $a = \frac{1}{2}\hat a$, and $\tilde a + 2\pi u = 2\hat{\tilde a}$. The action becomes Eq.~\eqref{eq:U(1)Z_4} with hatted variables and a rescaled gauge coupling $\hat\beta = \frac{1}{4}\beta$. Hence we return back to $U(1) \rtimes \ZZ_4$ but with a rescaled coupling.   

\smallskip
Finally, if gauge the diagonal $\ZZ_{2,d} \subset \ZZ_{2,e}^{(1)} \times \ZZ_{2,c}^{(1)}$, we get $Pin^{-}(2) = \frac{U(1)\rtimes\ZZ_4}{\ZZ_2}$. We can write the action as 
\begin{multline}
S_{Pin^-(2)} = \frac{\beta}{2}\sum_{\text{plaq.}}(d_{\c^2}a -\pi b_d- 2\pi n)^2 + i \sum_{\text{hyper-cu.}}\left(d_{\c^2}n + \frac{1}{2}d_{\c^2}b_d \right)\cup_{\c^2}\tilde a \\
+ \frac{i\pi }{2} \sum_{\text{hyper-cu.}}(dc-2b_d)\cup (v-du)\,,
\end{multline}
where $b \in \calC^2(\Lambda,\ZZ_2)$ and $u \in \calC^1(\Lambda,\ZZ_2)$ is a Lagrange multiplier setting $b$ to be a flat $\ZZ_2$ gauge field. The gauge redundancies now involve
\begin{equation}
a \to a + \pi v_d\,,\ n \to n - \overline{c}\cup v_d\,,\ c \to c + 2v_d\,, b_d \to b_d + dv_d\,,\ 
\end{equation}
as well as 
\begin{equation}
\tilde a \to \tilde a + 2\pi\tilde m \,,\ v \to v + dt\,,\ u \to u + \tilde m + t\,. 
\end{equation}
To rewrite the action in a more familiar form we again define $\hat n = b_d + 2n$,  $a =\frac{1}{2}\hat a$, and $\tilde a + 2\pi u = 2\hat{\tilde a}$. The action becomes
\begin{equation} \label{eq:pin^-(2)}
S_{Pin^-(2)} = \frac{\hat\beta}{2}\sum_{\text{plaq.}}(d_{\c^2}\hat a -2\pi \hat n)^2 + i \sum_{\text{hyper-cu.}} \left(d_{\c^2}\hat n \right)\cup_{\c^2}\hat{\tilde a} + \frac{i\pi }{2} \sum_{\text{hyper-cu.}}(dc-2\hat n)\cup v \,,
\end{equation}
where $\hat\beta = \frac{1}{4}\beta$. This resembles $O(2)$ gauge theory except that integrating out $v$ gives the constraint 
\begin{equation} \label{eq:pin_constraint} 
dc = 2\hat n \text{ mod } 4.
\end{equation}
The gauge transformation properties are also intertwined relative to the $O(2)$ case, 
\begin{equation}
\hat a \to \hat a + 2\pi m\,,\ \hat n\to \hat n + d_{\c^2}m\,,\ c \to c + 2m\,,\ \hat{\tilde a} \to \hat{\tilde a} + 2\pi \tilde m + \pi t, \ v \to v + dt\,.
\end{equation}
One can see that the genuine 't Hooft lines in the theory are still of the form $e^{2 i \sum_\gamma \hat{\tilde a}}$, while the `fractional' 't Hooft lines must be attached to surfaces built from $v$, of the form $e^{i \sum_{\partial\sigma} \hat{\tilde a} + i\pi \sum_{\sigma} v}$. We therefore \emph{do not} have a dual magnetic symmetry. This can also be seen from the constraint in Eq.~\eqref{eq:pin_constraint}, since the mod-2 magnetic charge
\begin{equation}
e^{i \pi \sum_{p \in S} \hat n_p} = e^{\frac{i\pi}{2} \sum_{p \in S} (dc)_p} = 1
\end{equation}
is always trivial.

\smallskip
The $Pin^-(2)$ theory does have a residual electric symmetry, generated by the charge-2 twist vortex. While the minimal twist vortex $e^{\frac{i\pi}{2} \sum_\Gamma v}$ sets $dc = \star[\Gamma^\vee] +2\hat n \text{ mod } 4$ and requires worldvolume degrees of freedom, the charge-2 twist vortex preserves flatness mod 2 which keeps the action gauge invariant. Moreover, this charge-2 twist vortex is topological, and generates the residual $\ZZ_{2}^{(1)}$ symmetry that acts on the gauge-invariant combination 
\begin{equation}
W_{1/2}(\gamma) e^{-i \frac{\pi}{2} \sum_{\gamma} c}\,. 
\end{equation}
Finally, the theory also has a $\ZZ_{2,v}^{(2)}$ symmetry generated by $e^{i \pi \sum_\gamma c}$ which counts twist vortices mod 2. 

\smallskip
We can also move between the $Pin^-(2)$ and $O(2)$ theories via generalized gauging. Starting with $Pin^-(2)$, we simply gauge the residual $\ZZ_{2}^{(1)}$ symmetry to get $O(2)$. Going in the other direction is more subtle --- we have to gauge the magnetic $\ZZ_{2,m}^{(d-3)}$ symmetry with a particular fractionalization class activated by $\ZZ_{2,v}^{(d-2)}$, namely $B_v = \frac{1}{2}dB_m$. This amounts to gauging the magnetic symmetry by adding the terms 
\begin{multline}
i\pi \sum_{\text{hyper-cu.}} n \cup b_m + \frac{i\pi}{2} \sum_{\text{hyper-cu.}} c \cup db_m + i\pi \sum_{\text{hyper-cu.}} h \cup db_m \\
= \frac{i\pi}{2}\sum_{\text{hyper-cu.}} (d(c+2h) -2n)\cup (b_m + 2v)\,,
\end{multline} 
with $h \in \calC^1(\Lambda,\ZZ_2)$ and $b_m \in \calC^2(\Lambda, \ZZ_2)$. Note that the combination $\hat c = c + 2h$ combines to an element of $\calC^2(\Lambda,\ZZ_4)$, so we have effectively extended the gauge field from $\ZZ_2 \to \ZZ_4$. Defining $\hat v = b_m + 2v$, the action becomes equivalent to Eq.~\eqref{eq:pin^-(2)}.





\newpage
\linespread{1}\selectfont
\bibliographystyle{utphys}
\bibliography{O2draft}

\end{document}